\documentclass[journal]{IEEEtran}

\usepackage[T1]{fontenc}
\usepackage{graphicx}
\usepackage{cite}
\usepackage{caption}
\usepackage{subcaption}
\usepackage{epstopdf}

\usepackage{overpic}
\usepackage{amssymb}
\usepackage{amsmath}
\usepackage{amsthm}
\usepackage{array}
\usepackage{color}
\usepackage{url}
\usepackage{bm}
\usepackage{tikz}
\usepackage[siunitx]{circuitikz}
\usetikzlibrary{decorations.pathreplacing,calligraphy}
\usepackage{relsize}

\IEEEoverridecommandlockouts

\theoremstyle{plain}

\newcommand{\vect}[1]{\mathbf{#1}}

\def\diag{\mathrm{diag}}

\def\tr{\mathrm{tr}}

\def\sinc{\mathrm{sinc}}
\def\Htran{\mbox{\tiny $\mathrm{H}$}}
\def\Ttran{\mbox{\tiny $\mathrm{T}$}}
\def\CN{\mathcal{N}_{\mathbb{C}}} 
\def\imagunit{\mathsf{j}} 
\newcommand{\B}[1]{\boldsymbol{#1}}
\usepackage{amsfonts}
\usepackage{amsmath}
\usepackage{commath}
\usepackage{amssymb}
\usepackage{cite}
\usepackage{dsfont}
\usepackage{latexsym}
\usepackage{subcaption}
\usepackage{times}
\usepackage{url}
\usepackage{verbatim}

\def\cV{{\mathcal{V}}}

\def\bee{{\mathbf{e}}}

\def\bmm{{\mathbf{m}}}

\def\bp{{\mathbf{p}}}

\def\bu{{\mathbf{u}}}

\def\bx{{\mathbf{x}}}

\def\bA{{\mathbf{A}}}

\def\bE{{\mathbf{E}}}

\def\bJ{{\mathbf{J}}}

\def\bM{{\mathbf{M}}}

\def\bQ{{\mathbf{Q}}}
\def\bR{{\mathbf{R}}}

\def\bT{{\mathbf{T}}}

\def\bur{{\bu_{r}}}
\def\bue{{\bu_{\theta}}}
\def\bua{{\bu_{\phi}}}

\def\arad{{\alpha_{\sf rad}}}
\def\aang{{\alpha_{\sf ang}}}

\def\e0{{\epsilon_0}}

\begin{document}

\title{Towards 6G MIMO: Massive Spatial Multiplexing, Dense Arrays, and Interplay Between Electromagnetics and Processing}

\author{
\IEEEauthorblockN{\noindent Emil Bj{\"o}rnson,~\IEEEmembership{Fellow,~IEEE}, Chan-Byoung Chae,~\IEEEmembership{Fellow,~IEEE}, Robert W. Heath Jr.,~\IEEEmembership{Fellow,~IEEE},\\ Thomas L. Marzetta,~\IEEEmembership{Life Fellow,~IEEE}, Amine Mezghani,~\IEEEmembership{Member,~IEEE}, Luca  Sanguinetti,~\IEEEmembership{Senior Member,~IEEE}, Fredrik 
Rusek,~\IEEEmembership{Member,~IEEE}, Miguel R. Castellanos,~\IEEEmembership{Member,~IEEE},\\ Dongsoo Jun,~\IEEEmembership{Graduate Student Member,~IEEE}, and \"Ozlem Tu\u{g}fe Demir,~\IEEEmembership{Member,~IEEE}}%
\thanks{E.~Bj\"ornson is with the School of Electrical Engineering and Computer Science, KTH Royal Institute of Technology, Stockholm, Sweden. E-mail: emilbjo@kth.se.\\
C.-B. Chae and D. Jun are with the School of Integrated Technology, Yonsei University, Seoul 03722, South Korea. E-mail: \{cbchae, dongsoo.jun\}@yonsei.ac.kr. \\
R. W. Heath, Jr. is with the Department of Electrical and Computer Engineering, University of California, San Diego, La Jolla, CA 92161, USA. E-mail: rwheathjr@ucsd.edu. \\
T. L. Marzetta is with the Department of Electrical and Computer Engineering, New York University, Brooklyn, NY 11201, USA. E-mail: tom.marzetta@nyu.edu. \\
A. Mezghani is with the Department of
Electrical and Computer Engineering, University of Manitoba, Winnipeg, MB R3T 5V6, Canada. E-mail: amine.mezghani@umanitoba.ca. \\
L. Sanguinetti is with the Dipartimento di Ingegneria dell’Informazione, University of Pisa, Pisa, Italy. E-mail: luca.sanguinetti@unipi.it. \\
F. Rusek is with Sony Europe B.V., Lund, Sweden. E-mail: fredrik.x.rusek@sony.com. \\
M. R. Castellanos is with the Department of Electrical and Computer Engineering, North Carolina State University, 
Raleigh, NC 27695, USA. E-mail: mrcastel@ncsu.edu. \\
\"O. T. Demir is with the TOBB University of Economics and Technology, Ankara, T\"urkiye. E-mail: ozlemtugfedemir@etu.edu.tr. 
}
\thanks{The work of E.~Bj\"ornson is supported by Swedish Research Council and Swedish Foundation for Strategic Research. 
The work of C.-B. Chae and D. S. Jun is in part supported by the Institute of Information and Communication Technology Promotion (IITP) grant funded by the Ministry of Science and ICT (MSIT), Korea (2021-0-00486, 2021-0-02208). 
The work of R. W. Heath, Jr. was supported in part by the National Science Foundation under grant nos. NSF-ECCS-2153698, NSF-CCF-2225555, NSF-CNS-2147955  and is supported in part by funds from federal agency and industry partners as specified in the Resilient \& Intelligent NextG Systems (RINGS) program, as well as by support from Nokia, Samsung and Qualcomm. 
The work of T. L. Marzetta was supported by NYU WIRELESS.
The work of A. Mezghani was in part supported by the Natural Sciences and Engineering Research Council of Canada as well as Research Manitoba. 
The work of L. Sanguinetti was partially supported by the Italian Ministry of Education and Research (MUR) in the framework of the FoReLab project (Departments of Excellence).
The work of M. R. Castellanos was supported in part by Nokia, Motorola, and Samsung.
The work of \"O. T. Demir was supported by 2232-B International Fellowship for Early Stage Researchers Programme funded by the Scientific and Technological Research Council of T\"urkiye.
}
}

\maketitle

\begin{abstract} 
The increasing demand for wireless data transfer has been the driving force behind the widespread adoption of Massive MIMO (multiple-input multiple-output) technology in 5G. The next-generation MIMO technology is now being developed to cater to the new data traffic and performance expectations generated by new user devices and services in the next decade. The evolution towards ``ultra-massive MIMO (UM-MIMO)'' is not only about adding more antennas but will also uncover new propagation and hardware phenomena that can only be treated by jointly utilizing insights from the communication, electromagnetic (EM), and circuit theory areas. This article offers a comprehensive overview of the key benefits of the UM-MIMO technology and the associated challenges. It explores massive multiplexing facilitated by radiative near-field effects, characterizes the spatial degrees-of-freedom, and practical channel estimation schemes tailored for massive arrays. Moreover, we provide a tutorial on EM theory and circuit theory, and how it is used to obtain physically consistent antenna and channel models. Subsequently, the article describes different ways to implement massive and dense antenna arrays, and how to co-design antennas with signal processing. The main open research challenges are identified at the end.
\end{abstract}


\IEEEpeerreviewmaketitle

\section{Introduction}

\IEEEPARstart{T}{he} need for equipping transmitters and receivers with multiple antennas in wireless communication systems has been recognized for over a century. The first observed benefit was the adaptive directivity achievable by controlling the constructive and destructive superposition of electromagnetic (EM) signals using an antenna array \cite{Alexanderson1919a,Bondyopadhyay2000a}. The transmitter can use this feature, traditionally referred to as \emph{beamforming}, to focus a transmitted signal at the desired receiver while avoiding interference at specific locations. Similarly, the receiver can amplify signals impinging from a particular direction using multiple antennas while suppressing undesired interference.
The second observed benefit was the higher robustness against channel fading achieved by using multiple antennas \cite{Peterson1931a,Friis1937a,Alamouti1998a,Tarokh1999a}, as it becomes less likely that all transmit-receive antenna pairs experience deep fades simultaneously as we increase the number of antennas and the array size.
This feature is called \emph{spatial diversity} and \emph{channel hardening} \cite{Hochwald2004a}.
The third and most recently discovered benefit is multiple-input multiple-output (MIMO) communications~\cite{Winters1987a,Swales1990a,Foschini1998a,Telatar1999a,Gesbert2008}, where antenna arrays are used to spatially multiplex many layers of data at the same time and frequency. This can be done in multi-user MIMO mode, where a multiple-antenna base station (BS) communicates with multiple user equipments (UEs) simultaneously. This is enabled using adaptive beamforming: the BS gives each transmitted signal a different spatial directivity, has the ability to amplify signals received from UEs in different directions, and can filter out interference in both transmission directions. 
There is also the single-user MIMO mode, where a multi-antenna BS and multi-antenna UE exchange multiple data layers simultaneously by beamforming through different propagation paths.

The MIMO technology was first introduced in cellular and WiFi networks as a premium feature but is nowadays a mainstream technology. The 5G technology was built around the \emph{Massive MIMO} concept \cite{Marzetta2010a} of having a surplus of antennas at the BS compared to the UE side, which makes it practically feasible to protect the data layers from mutual interference through spatial filtering, even under imperfect channel state information (CSI) and hardware impairments \cite{Bjornson2016b,massivemimobook}. A typical 5G BS in 2023 had 64 antenna ports and can support up to 16 data layers, such as 8 UEs assigned with two layers each.
The driving force behind the MIMO adoption is the rapidly increasing demand for data traffic in cellular networks, currently growing by 40\% per year \cite{EricssonMobility2023}. 
The 5G MIMO technology, particularly in the 3.5\,GHz band, can supply current BS sites with significantly higher capacity than in 4G, to support higher speeds per device and accommodate more simultaneously served devices.
If the data traffic continues to grow at the current pace over the next ten years, 6G technology must deliver $1.4^{10} \approx 30$ times higher capacity than current networks. 
New emerging user devices (e.g., for augmented reality) and services (e.g., federated learning, hyper-reliable and low-latency communication) might create an even faster wireless traffic growth; thus, the next-generation MIMO technology should be developed to at least support 100 times higher capacity than in current networks. A portion of that can be achieved by expanding the bandwidth. However, since spectrum is scarce in bands suitable for wide-area coverage, the focus should be on increasing the sum spectral efficiency (SSE) [bit/s/Hz]. The SSE is the total data transmitted per second and per Hertz among all the spatially multiplexed layers.

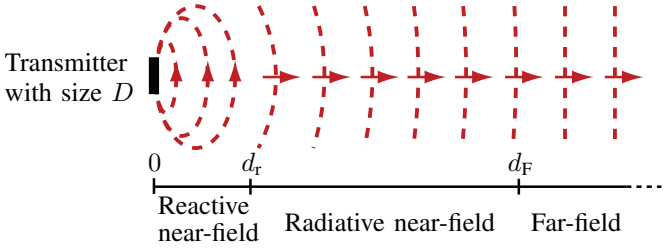
\begin{figure}[t!]
	\centering 
	\begin{overpic}[width=\columnwidth,tics=10]{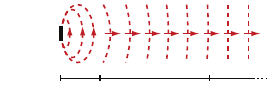}
	 \put (0,24) {Transmitter}
	 \put (0,19.5) {with size $D$}
  \put (23,2.5) {Reactive}
  \put (23,-1.5) {near-field}
  \put (42,0) {Radiative near-field}
  \put (21.5,9) {$0$}  
  \put (35.5,9) {$d_\textrm{r}$}  
  \put (75.5,9) {$d_\textrm{F}$}
  \put (79,0) {Far-field}
\end{overpic} \vspace{0.1mm}
	\caption{The EM field looks different depending on the distance from the transmitting aperture antenna. The wavefront is almost planar in the far-field, while the spherical curvature is clearly noticeable in the radiative near-field but not reactive effects such as inductive coupling and evanescent waves.}
	\label{fig:three-fields}
\end{figure}

The spectral efficiency (SE) per spatial layer is fundamentally limited by the signal-to-noise ratio (SNR), as expressed by the Shannon formula $\log_2(1+\textrm{SNR})$ bit/s/Hz \cite{Shannon1949b}. The logarithmic nature of this function places a fundamental constraint on massive improvements through the use of multiple antennas, except for UEs experiencing very low SNRs. However, the spectral efficiency (SSE) of $\nu$ data layers of this kind is upper-bounded by $\nu \cdot \log_2(1+\textrm{SNR})$, a linearly increasing function of $\nu$. To enhance SSE in 6G, the logical approach is to employ ``more MIMO''--increasing spatial multiplexing with more layers to serve additional UEs. On the BS side, this involves using larger antenna arrays, either physically larger or relative to the wavelength (if the carrier frequency is increased compared to 5G). At first glance, this evolution may seem like an engineering challenge: assembling more antennas using current technology and applying the same algorithms found in textbooks such as \cite{Marzetta2016a,massivemimobook} with larger-dimensional matrices.

However, the reality is vastly different on multiple levels. Antenna array design in 6G requires not only a drastic increase in the number of antennas but also fundamental changes in EM properties. Despite the numerous antennas in 5G Massive MIMO systems, the aperture size is small enough to neglect near-field effects, focusing primarily on the far-field. In 6G, the significantly increased number of antennas and higher frequency bands will expand the near-field region, especially with dense BS deployments in urban and indoor environments. Consequently, MIMO systems with extremely larger antenna arrays, referred to as X-MIMO in industry terminology, are gaining attention. Various terms, such as extremely large aperture arrays (ELAA) \cite{Bjornson2019c}, extremely large-scale MIMO (XL-MIMO) \cite{a, b, c, cui2022channel}, and \emph{Ultra-Massive MIMO (UM-MIMO)}, have been suggested in academic literature. In this paper, we use the latter terminology.

In addition to near-field propagation becoming more dominant, we approach fundamental limits on spatial degrees-of-freedom; more physically accurate models of channels, antennas, and hardware effects are necessary; and new implementation challenges emerge. These aspects will be further described in the remainder of this paper.

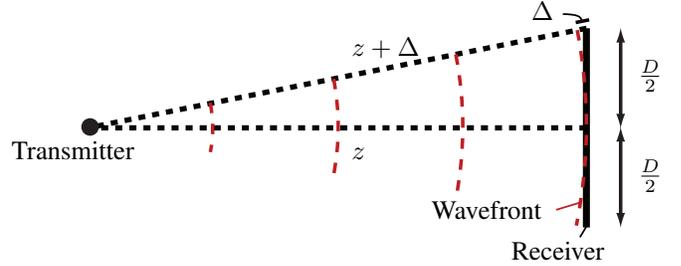
\begin{figure}[!t]
	\centering
	\begin{overpic}[width=\columnwidth,tics=10]{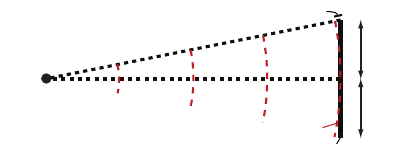}
	 \put (0,15) {Transmitter}
	 \put (75,0) {Receiver}
	 \put (63,6) {Wavefront}
	 \put (51,30.5) {$z+\Delta$}
	 \put (51,15) {$z$}
	 \put (78,36) {$\Delta$}
	 \put (94,11.5) {$\frac{D}{2}$}
	 \put (94,26.5) {$\frac{D}{2}$}
\end{overpic} 
	\caption{The curvature of an impinging spherical wave creates a delay $\frac{\Delta}{c}$ between the center of the receiver and the edge. The delay turns into a phase-shift of $2\pi f_c\frac{\Delta}{c}=\frac{2\pi}{\lambda}\Delta$.}
	\label{fig:fraunhofer} 
\end{figure}

The remainder of the paper is organized as follows. Section~\ref{sec:near-field} provides an overview of radiative near-field propagation effects and how they can be exploited for  finite-depth beamforming and massive spatial multiplexing.
Section~\ref{sec:spat_deg_freedom} introduces the spatial degrees-of-freedom concept, which is the physical limit on the multiplexing capability of an array.
We then describe efficient channel estimation techniques in Section~\ref{section:estimation}, with a focus on exploiting array geometry and propagation characteristics. 
Having introduced the basic concepts, the paper then provides a tutorial on the underlying theory.
Section~\ref{sec:EMI} provides a linear system approach to EM theory. Section~\ref{sec:physically-consistent-channel} expands on this approach by using circuit theory to obtain a physically consistent end-to-end MIMO channel representation. Section~\ref{sec:polarization-modeling} describes how to account for realistic antenna properties, such as mutual coupling, polarization, and near-field propagation for MIMO array modeling. Next, Section~\ref{sec:antenna-array-design} describes four antenna array architectures that might be used in 6G UM-MIMO systems.
Section~\ref{sec:array_implementation} takes a closer look at how the hybrid array architecture can be optimized jointly with the signal processing algorithms.
Finally, the paper is concluded in Section~\ref{sec:conclusions} by describing some open research challenges.

\section{MIMO Communication in the Radiative Near-Field}
\label{sec:near-field}

The behavior of a wireless channel is governed by Maxwell's equations, which can be solved to determine the electric and magnetic field distributions that a transmission creates at different locations in a specific environment.
This fundamental approach often results in very complex expressions, but they can be fortunately simplified and tailored to the specific scenario, allowing for practical use in system design and optimization. As contemporary MIMO systems expand in size, the once-accurate simplified propagation models must be enriched to account for ``new'' phenomena that were always present but previously considered negligible.
This section provides a tutorial on such properties, starting from the near- and far-fields of antennas and arrays, and subsequently describing the impact they have on future UM-MIMO communication systems. We will go deeper into many of these aspects in later sections.

We begin by considering the transmission from a point source. The electric field observed at a distance $z$ from the source, in any direction perpendicular to the
propagation direction, is proportional to \cite{Dardari2020a}
\begin{equation} \label{eq:Green-function-polarization}
 \frac{ -\imagunit \eta e^{-\imagunit \frac{2\pi}{\lambda} z }  }{2 \lambda z} \left( 1 + \frac{\imagunit}{2\pi z / \lambda} - \frac{1}{(2\pi z / \lambda)^2}  \right),
\end{equation} 
where $\eta$ denotes the impedance of free space and $\lambda$ is the signal's wavelength. 
The first term in \eqref{eq:Green-function-polarization} has a squared magnitude that decays proportionally to $1/z^2$, consistent with the classical pathloss behavior for free-space propagation \cite{friis1946note}.
The second factor in \eqref{eq:Green-function-polarization} is often overlooked in communications---for good reasons---because 
\begin{equation} \label{eq:Green-function-polarization2}
\left| 1 + \frac{\imagunit}{2\pi z / \lambda} - \frac{1}{(2\pi z / \lambda)^2} \right|^2 = 
1 -  \frac{1}{(2\pi z / \lambda)^2}  +  \frac{1}{(2\pi z / \lambda)^4},
\end{equation}
which quickly approaches $1$ when the propagation distance $z$ increases. Already at a distance of $2\lambda$ from the transmitter, \eqref{eq:Green-function-polarization2} equals $0.99$. 
The region $z \geq 2 \lambda$ where \eqref{eq:Green-function-polarization} can be approximated as
$ \frac{-\imagunit \eta e^{-\imagunit \frac{2\pi}{\lambda} z }  }{2 \lambda z}$ is known as the far-field, while $z < 2\lambda$ represents the near-field region.
There is no strict boundary between these regions since \eqref{eq:Green-function-polarization2} approaches $1$ in a continuous manner.

If we replace the point source with a transmitting aperture antenna having a maximum length $D$ that is larger than the wavelength, then the distances that characterize the near- and far-field change as well.
In particular, the near-field can be divided into two parts: the \emph{reactive} and \emph{radiative} near-field.
For most antenna types, terms similar to the second factor in \eqref{eq:Green-function-polarization} must be taken into account at distances $z \leq d_\textrm{r} = 0.62 \sqrt{D^3/\lambda}$ \cite{Selvan2017a,balanis2015antenna}. These terms represent EM components that remain around the transmitter instead of being radiated; for example, related to inductive coupling and evanescent fields. It is in the reactive near-field that the induction-based near-field communication (NFC) technology operates and it is commonly used for tags and keycards. This paper does not consider such technologies. 
A transmitting aperture antenna has similar far-field behavior as a point source when observed at a distance $z > d_\textrm{F} = \frac{2D^2}{\lambda}$, where $d_\textrm{F}$ is greater than $d_\textrm{r}$ for $D$ larger than $\lambda$. This boundary is known as the \emph{Fraunhofer distance} or Rayleigh distance. The far-field region is also known as the \emph{Fraunhofer} region.
In between the mentioned distance boundaries is a range $d_\textrm{r}< z \leq d_\textrm{F}$ that is called the \emph{radiative near-field} or \emph{Fresnel} region.
Fig.~\ref{fig:three-fields} illustrates these different regions and emphasizes the core difference between the radiative near-field and conventional far-field: the curvature of the wavefront. 
It is spherical in both cases, which means that the received signal power decays with the distance $z$ proportionally to $1/z^2$, but the curvature is only noticeable in the radiative near-field.

The implications of the strongly curved wavefront in the radiative near-field are easier to comprehend when considering an aperture antenna with the length $D$ that receives a signal from a transmitting point source. This setup is shown in Fig.~\ref{fig:fraunhofer} for a transmitter at the distance $z$ in the broadside direction. When the wavefront reaches the receiver's center, a distance $\Delta$ remains until it reaches the edge. The extra distance can be calculated as
\begin{equation}
\Delta = \sqrt{z^2+\frac{D^2}{4}} - z =  z\sqrt{1+\frac{D^2}{4z^2}} - z \approx \frac{D^2}{8z},
\end{equation}
where the last expression follows from the first-order Taylor approximation $\sqrt{1+x} \approx 1+\frac{x}{2}$, that is accurate for small $x$ (i.e., when $z\gg\frac{D}{2}$). 
The Fraunhofer distance $z= d_\textrm{F}$, where $d_\textrm{F}\gg \frac{D}{2}$ so that we can use the above approximation, gives rise to the phase-shift
\begin{equation}
2\pi f_c\frac{\Delta}{c}=\frac{2\pi}{\lambda} \Delta \approx \frac{2\pi}{\lambda} \frac{D^2}{8d_\textrm{F}} = \frac{\pi}{8},
\end{equation}
where $f_c$ is the carrier frequency and $c$ is the speed of the EM radiation. There is nothing deeper behind the Fraunhofer distance than the fact that $\cos(\pi/8) \approx 0.92 \approx 1$ so that the spherical curvature creates a tiny phase variation over the antenna \cite{Sherman1962a}.

The spherical curvature also gives rise to power variations over the receive antenna. 
Since the squared magnitude of the EM field is proportional to $1/z^2$, the power difference is
\begin{equation} \label{eq:power-difference}
\frac{z^2}{(z+\Delta)^2} \approx \frac{z^2}{\left(z+ \frac{D^2}{8z}\right)^2}
\end{equation}
 between the center and edge.
The power variations are negligible if $z = d_\textrm{B} = 2D$ \cite{Bjornson2020a} so that the propagation distance is twice as large as the surface, in which case \eqref{eq:power-difference} becomes $0.94$.
We notice that $d_\textrm{F} = d_\textrm{B} \frac{D}{\lambda}$, which means that the phase-variations are more prevalent when considering large antennas.

The aforementioned approximations are conventionally made in wireless communications without further discussion, but there are ways to be more precise.
The gain of a receiving aperture antenna can be computed by taking an integral of the impinging field over the aperture. For example, if the antenna spans the interval $x \in [-a/2,a/2], y \in [-b/2,b/2]$ in the $xy$-plane and the electric field is denoted as $E(x,y)$, then the gain (relative to an isotropic reference antenna in the far-field) can be computed as \cite{Kay1960a,Bjornson2021b}
\begin{equation} \label{eq:antennagain}
G =  \frac{ \left| \int_{-a/2}^{a/2} \int_{-b/2}^{b/2}  E(x,y) d x \, d y \right|^2}{A_\textrm{iso}  \int_{-a/2}^{a/2} \int_{-b/2}^{b/2} \left| E(x,y) \right|^2 d x \, d y  },
\end{equation}
where $A_\textrm{iso} = \frac{\lambda^2}{4\pi}$ is the effective area of an isotropic antenna.
Looking at this expression, it might seem logical that a larger antenna has a higher gain because of the expanded integration intervals, but this is not a necessity because phase variations in $E(x,y)$ over the antenna impact the numerator in \eqref{eq:antennagain}.
This phenomenon occurs even in the far-field when the impinging signal arrives from a non-broadside direction, creating linear phase variation similar to that of a plane wave. However, the effect is particularly dominant in the radiative near-field due to the wave's spherical curvature. 
For example, for a transmitting point source in the broadside direction $(0,0,z)$, with $z > \max(d_\textrm{r}, d_\textrm{B})$, the electric field observed at the receiver can be expressed as
\begin{equation} \label{eq:electric-field}
E(x,y) = \frac{E_0}{\sqrt{4\pi} z} e^{-\imagunit \frac{2\pi}{\lambda} \sqrt{x^2+y^2+z^2}},
\end{equation}
where $E_0$ is a scaling factor for the electric intensity and
$\sqrt{x^2+y^2+z^2}$ is the distance from the transmitter to the location $(x,y,0)$ at the receiver. This expression assumes that the power variations over the antenna are negligible, but there are still phase variations across the antenna aperture when $\max(d_\textrm{r}, d_\textrm{B})<z\leq d_\textrm{F}$. By substituting \eqref{eq:electric-field} into \eqref{eq:antennagain}, the gain simplifies to
\begin{equation} \label{eq:antennagain2}
G =  \frac{ \left| \int_{-a/2}^{a/2} \int_{-b/2}^{b/2}  e^{-\imagunit \frac{2\pi}{\lambda} \sqrt{x^2+y^2+z^2}} d x \, d y \right|^2}{A_\textrm{iso} ab },
\end{equation}
which becomes $ab/ A_\textrm{iso}$ if the phase in the numerator is approximately constant over the antenna aperture. This is a classical approximation that is valid for $z>d_\textrm{F}$ because the electric field can then be approximately as $E(x,y)\approx\frac{E_0}{\sqrt{4\pi}z}e^{-\imagunit \frac{2\pi}{\lambda}z}$. 
The ratio $ab/ A_\textrm{iso}$ between antenna areas is how the antenna gain is normally computed in the far-field.
By contrast, a strictly smaller gain is obtained for $z < d_\textrm{F}=\frac{2(a^2+b^2)}{\lambda}$ (with $D = \sqrt{a^2+b^2}$ being the antenna's diagonal length) because of the noticeable spherical phase variations that make some parts of the impinging field cancel other parts.
The way to circumvent that effect is to use an array of many small antennas instead of one large antenna so that each antenna achieves its physically maximum gain and then the gains are superimposed through receiver processing.
In other words, we want to use small transmit and receive antennas that are in each others' far-field, but build large antenna arrays that can observe radiative near-field effects across the array.

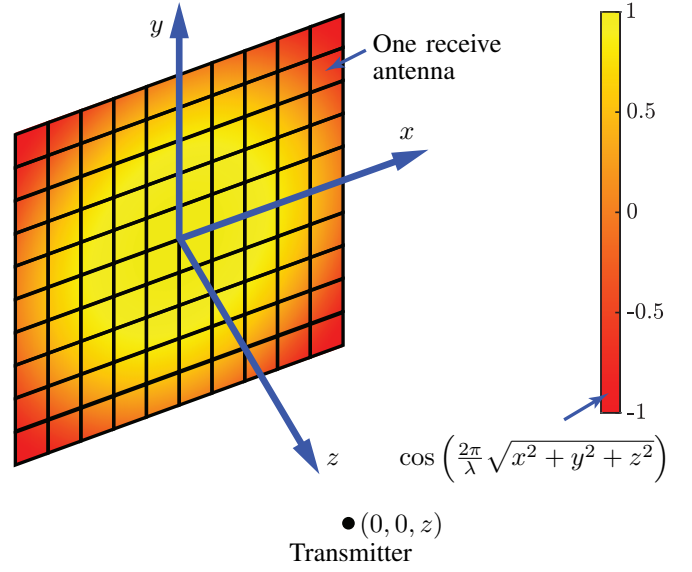
\begin{figure}[t!]
	\centering 
	\begin{overpic}[width=\columnwidth,tics=10]{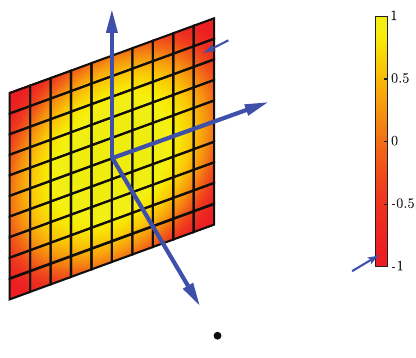}
  \put (60,63) {$x$}
  \put (22.5,79) {$y$}
  \put (49,14) {$z$}
  \put (54,4) {$(0,0,z)$}
  \put (43.5,-0.5) {Transmitter}
  \put (56,76) {One receive}
  \put (56,72) {antenna}
  \put (60,14) {$\cos \left( \frac{2\pi}{\lambda} \sqrt{x^2+y^2+z^2}\right)$}
\end{overpic} 
	\caption{The (normalized) real part of the electric field in \eqref{eq:electric-field} is shown for an antenna array deployed in the $xy$-plane. The transmitter is located in a broadside direction in the radiative near-field, leading to spherical phase variations.}
	\label{fig:phase-variations}
\end{figure}

Fig.~\ref{fig:phase-variations} illustrates this situation for an array of $10 \times 10$ $\lambda/2$-spaced antennas in the $xy$-plane, and the coloring shows the (normalized) real part $\cos(2\pi \sqrt{x^2+y^2+z^2} / \lambda)$ of the impinging electric field when the transmitter is $8\lambda$ from the receiver. The spherical-shaped phase variations result in values between $+1$ to $-1$. If there were only one big antenna covering the entire surface, the red parts would cancel the yellow parts when computing \eqref{eq:antennagain2}, leading to a large loss in antenna gain. The gain of this large antenna would be only 35\% of the maximum gain $ab/A_\textrm{iso}$ when $a=b=10\frac{\lambda}{2}=5\lambda$, and the percentage shrinks as the antenna size increases. However, in Fig.~\ref{fig:phase-variations}, the surface is divided into 100 antennas so that the phase variation is small across each individual antenna. This division results in a negligible gain loss per antenna so that at least 95\% of the maximum gain is achieved. This comes at the expense of requiring receiver hardware that can combine the antenna signals, which we anyway need for beamforming in mobile scenarios.

When the array is large, there can also be power variations between the antennas deployed at the surface, caused by having substantially different distances (and angles) to the transmitter. In contrast to the phase-shifts, this geometric pathloss effect cannot be mitigated through signal processing or making the antenna elements smaller. Fortunately, the effect is negligible in many practical situations. If we let $D$ denote the largest dimension of the array (i.e., the diagonal in Fig.~\ref{fig:phase-variations}), then it follows that $d_\textrm{F} = d_\textrm{B} \frac{D}{\lambda} \gg d_\textrm{B}$ for large arrays with $D \gg \lambda$. This implies that there is a wide distance range $d_\textrm{B} < z \leq d_\textrm{F}$ where the power variations over the antenna array are negligible but not the phase variations. We will call this the \emph{Fresnel region}. As an example, consider a half-wavelength-spaced BS array with the size $1 \times 0.5$\,m, and operating at $30$\,GHz (i.e., $\lambda=0.01$\,m). This array contains $5000$ antennas in the configuration $100 \times 50$. It then follows that $d_\textrm{B}=2D=2\sqrt{1^2+0.5^2}\approx 2.24$\,m so most UEs will be located beyond that distance, while they might be closer than the Fraunhofer distance that becomes $d_\textrm{F}=\frac{2D^2}{\lambda}=\frac{2(1^2+0.5^2)}{0.01}=250$\,m.
We will focus on the Fresnel region range $d_\textrm{B} < z \leq d_\textrm{F}$ in the remainder of this section, but note that there is one particular situation when the power variations over the antenna array cannot be neglected: when studying the asymptotic limits of large arrays because otherwise one obtains implausible results where more power is received than was transmitted \cite{Bjornson2020a}. One must also accommodate for such effects when modeling non-line-of-sight propagation environments \cite{Gao2015b}, where different parts of a large array might see different scattering objects.

\subsection{Beamfocusing in the radiative near-field of arrays}

Suppose the antenna array is used to transmit a signal to a particular receiver. When $M$ antennas transmit the signal with phase-shifts that create constructive interference at the receiver's location, the received power becomes $M$ times larger than when transmitting with the same power from a single antenna. This is called the \emph{array gain} or \emph{beamforming gain} and is achievable in line-of-sight scenarios at any propagation distance $z > d_\textrm{B} = 2D$ (i.e., there are no pathloss variations over the array). The gain is the same in the far-field and radiative near-field, but there is an essential difference in how the radiated signal behaves at other locations, such as the shape of the focus area around the receiver. These geometric properties will be analyzed in this section.

Consider a uniform square array with $M$ antennas arranged as $N \times N$ antennas, where $N = \sqrt{M}$ is an integer and $\Delta$ denotes the antenna spacing. 
We let $n \in \{1,\ldots,N\}$ be the antenna index in the $x$ dimension and $m \in \{1,\ldots,N\}$ be the index in the $y$ dimension.
The antenna with index $(n,m)$ is centered at the point $(\bar{x}_n,\bar{y}_m,0)$, where
\begin{align}
\bar{x}_n &= \left(n-\frac{N+1}{2} \right) \Delta, \\
\bar{y}_m &= \left(m-\frac{N+1}{2} \right) \Delta.
\end{align}
We consider an isotropic broadside receiver at the location $(0,0,z)$.
If each transmit antenna has the gain $G$ towards the receiver, the channel coefficient becomes
\begin{align} \label{h_nm_exact}
h_{n,m} &= \frac{\lambda \sqrt{G}}{4\pi z} e^{-\imagunit \frac{2\pi}{\lambda} \sqrt{\bar{x}_n^2+\bar{y}_m^2+z^2}} \\
& \approx  \frac{\lambda \sqrt{G}}{4\pi z} e^{-\imagunit \frac{2\pi}{\lambda} \left(z  + \frac{\bar{x}_n^2}{2z}  + \frac{\bar{y}_m^2}{2z} \right)}, \label{h_nm_approx}
\end{align}
where \eqref{h_nm_exact} is the exact expression and the first-order Taylor approximation $\sqrt{1+x} \approx 1+\frac{x}{2}$ can be utilized to simplify it to \eqref{h_nm_approx} since $z$ is substantially larger than $\bar{x}_n$ and $\bar{y}_m$ when $z > d_\textrm{B}$.
This is known as the Fresnel approximation \cite{Polk1956a}.

If the data symbol $s \in \mathbb{C}$ is transmitted with power $p$, the received signal is
\begin{equation}
r = \sum_{n=1}^{N} \sum_{m=1}^{N} h_{n,m} \frac{e^{\imagunit \psi_{n,m}}}{\sqrt{M}} s + w_{n,m}, 
\end{equation}
where $w_{n,m} \sim \CN(0,\sigma^2)$ is independent complex Gaussian receiver noise, $\psi_{n,m}$ is the phase-shift assigned at antenna $(n,m)$, and $1/\sqrt{M}$ divides the total power equally among the transmit antennas.
The SNR becomes
\begin{align} \label{eq:SNR-formula} \nonumber
\mathrm{SNR} &= 
 \frac{p}{\sigma^2} \left|   \sum_{n=1}^{N} \sum_{m=1}^{N}  h_{n,m} \frac{e^{\imagunit \psi_{n,m}}}{\sqrt{M}} \right|^2 \\
 &= \frac{p}{\sigma^2} 
  \frac{\lambda^2 G }{(4\pi z)^2} \underbrace{\frac{1}{M}
  \left|   \sum_{n=1}^{N} \sum_{m=1}^{N}  e^{-\imagunit \frac{2\pi}{\lambda} \sqrt{\bar{x}_n^2+\bar{y}_m^2+z^2}} e^{\imagunit \psi_{n,m}} \right|^2 }_{=\mathrm{AG}},
\end{align}
where $\mathrm{AG}$ denotes the array gain. It becomes $N^4/M = M$ when $\psi_{n,m} = \frac{2\pi}{\lambda} \sqrt{\bar{x}_n^2+\bar{y}_m^2+z^2}$ so that the transmitter cancels all the phase-shifts. This is the maximum array gain.

Suppose the transmitter instead focuses the signal on a point at the distance $z$ but with some other small angle $\varphi$ measured from the boresight in the horizontal plane. The focus point is at $(z \sin(\varphi),0,z \cos(\varphi))$ and is obtained when the transmitter assigns the phase-shifts
\begin{align} \nonumber
\psi_{n,m} &= \frac{2\pi}{\lambda} \sqrt{(\bar{x}_n-z\sin(\varphi))^2+\bar{y}_m^2+z^2 \cos^2(\varphi)} \\
& = \frac{2\pi}{\lambda} \sqrt{\bar{x}_n^2+\bar{y}_m^2+z^2 -2z \bar{x}_n \sin(\varphi)}.
\end{align}
The array gain at the original receiver can then be computed, using the Fresnel approximation, as
\begin{align} \nonumber
&\frac{1}{M}
  \left|   \sum_{n=1}^{N} \sum_{m=1}^{N}  e^{-\imagunit \frac{2\pi}{\lambda} \sqrt{\bar{x}_n^2+\bar{y}_m^2+z^2}} e^{\imagunit \frac{2\pi}{\lambda} \sqrt{\bar{x}_n^2+\bar{y}_m^2+z^2 -2z \bar{x}_n \sin(\varphi)}}  \right|^2 \\ \nonumber
 & \approx \frac{1}{M}
  \left|   \sum_{n=1}^{N} \sum_{m=1}^{N}  e^{-\imagunit \frac{2\pi}{\lambda}  \left(z  + \frac{\bar{x}_n^2}{2z}  + \frac{\bar{y}_m^2}{2z} \right) } e^{\imagunit \frac{2\pi}{\lambda} \left(
  z + \frac{\bar{x}_n^2}{2z}  + \frac{\bar{y}_m^2}{2z} -  \bar{x}_n \sin(\varphi)  \right)
   }  \right|^2 \\ \nonumber
&=   \frac{N^2}{M}   \left|   \sum_{n=1}^{N}  e^{-\imagunit \frac{2\pi}{\lambda} \left(n-\frac{N+1}{2} \right) \Delta \sin(\varphi) 
   }  \right|^2 \\
 & \approx  \frac{N^2}{M}   \left|   \int_{0}^{N}  e^{-\imagunit \frac{2\pi}{\lambda} n  \Delta \sin(\varphi) 
   }  d n \right|^2 = M \sinc^2 \left( \frac{1}{\lambda} N \Delta \sin(\varphi) \right), \label{eq:beamforminggain}
\end{align}
where we also approximated the summation over many antennas by the corresponding integral. This approximation is tight when the antenna spacing is small, similar to how a Riemann sum approaches a Riemann integral.
This array gain expression depends on the angle $\varphi$ but is independent of the propagation distance, which implies that it is the same in the radiative near-field as in the far-field. 
The squared sinc-function determines how the array gain observed at the receiver tapers off when the transmitter aims the signal at a different location at the same distance. Since $\sinc^2(0.443) \approx 0.5$, half the array gain is achieved at $\varphi = \pm \arcsin ( \frac{0.443 \lambda}{N \Delta} ) \approx \pm\frac{0.443 \lambda}{N \Delta}$. The half-power angular beamwidth then becomes
\begin{equation} \label{eq:half-power}
\textrm{BW}_{3\,\textrm{dB}} \approx \frac{0.886 \lambda}{N \Delta} \, \textrm{rad},
\end{equation}
which is inversely proportional to the width $N \Delta$ of the array and proportional to the wavelength.
Although the angular beamwidth is the same at all propagation distance for which $z > d_\textrm{B}$, the physical beamwidths (in meters) is approximately $\textrm{BW}_{3\,\textrm{dB}} z$; thus, the width of the beam around the receiver is proportional to the distance to the receiver and much smaller in the radiative near-field than in the far-field.

When the antenna spacing is $\Delta = \lambda/2$, the beamwidth expression simplifies to $\textrm{BW}_{3\,\textrm{dB}} = \frac{1.772}{N}$. It is clear that more antennas per dimension leads to a narrower beamwidth.

The radiated signal from an array is often illustrated as a cone with an angular width of $\textrm{BW}_{3\,\textrm{dB}} $. However, this description is incomplete because the array gain also tapers off in the depth domain.
We can characterize this phenomenon similarly to the beamwidth analysis by supposing that the transmitting array focuses the emitted signal on another point $(0,0,F)$ in the same direction but at a distance $F \neq z$. 
Let us define the focal point deviation
\begin{equation}
z_\textrm{eff} = \left| \frac{1}{F} - \frac{1}{z} \right|^{-1} = \frac{Fz}{|F-z|}.
\end{equation}
The array gain at the original receiver can then be computed, using the Fresnel approximation, as \cite{Bjornson2021b}
\begin{align} \nonumber
&\frac{1}{M}
  \left|   \sum_{n=1}^{N} \sum_{m=1}^{N}  e^{-\imagunit \frac{2\pi}{\lambda} \sqrt{\bar{x}_n^2+\bar{y}_m^2+z^2}} e^{\imagunit \frac{2\pi}{\lambda} \sqrt{\bar{x}_n^2+\bar{y}_m^2+F^2 }}  \right|^2 \\ \nonumber
 & \approx \frac{1}{M}
  \left|   \sum_{n=1}^{N} \sum_{m=1}^{N}  e^{-\imagunit \frac{2\pi}{\lambda}  \left(z  + \frac{\bar{x}_n^2}{2z}  + \frac{\bar{y}_m^2}{2z} \right) } e^{\imagunit \frac{2\pi}{\lambda} \left(
  F + \frac{\bar{x}_n^2}{2F}  + \frac{\bar{y}_m^2}{2F}  \right)
   }  \right|^2 \\ \nonumber
 & \approx  \frac{1}{M}   \left|   \int_{-N/2}^{N/2} 
 e^{\imagunit \frac{\pi}{\lambda} 
   \frac{n^2 \Delta^2}{ z_\textrm{eff}}  } d n
  \int_{-N/2}^{N/2}  e^{\imagunit \frac{\pi}{\lambda} 
   \frac{m^2 \Delta^2}{ z_\textrm{eff}}  }  d m \right|^2  \\
  & = M \left( \frac{8 z_\textrm{eff}}{d_\textrm{F}} \right)^2 \left( C^2 \left( \sqrt{\frac{d_\textrm{F} }{8 z_\textrm{eff}} }\right) \!+\! S^2 \left( \sqrt{\frac{d_\textrm{F} }{8 z_\textrm{eff}} } \right) \right)^2, \label{eq:AF-depth}
\end{align}
where $d_\textrm{F} = 4 N^2 \Delta^2 / \lambda$ is the Fraunhofer distance of the considered array, while
 $C(x) = \int_{0}^{x} \cos(\pi t^2/2) dt$ and $S(x) = \int_{0}^{x} \sin(\pi t^2/2) dt$ denote the Fresnel integrals.

The array gain expression in \eqref{eq:AF-depth} has the structure  $A(x) = (C^2(\sqrt{x}) + S^2(\sqrt{x}) )^2 /x^2$, where $x = d_\textrm{F} / (8 z_\textrm{eff} ) $. This is a decreasing function for $x \in [0,2]$ with $A(0) =1$ and $A(1.25) \approx 0.5$.  Hence, half the array gain is achieved when 
\begin{equation}
1.25 = \frac{d_\textrm{F}}{8 z_\textrm{eff}} = \frac{d_\textrm{F} |F-z|}{8 Fz}  \,\, \rightarrow \,\, z = \frac{d_\textrm{F} F  }{d_\textrm{F}   \pm 10 F}.
\end{equation}
This implies that when the transmitter focuses a signal on the point $(0,0,F)$, the array gain will only be large at some of the potential locations $(0,0,z)$ in the same direction.
The smallest solution is  $z = \frac{d_\textrm{F} F  }{d_\textrm{F}   + 10 F}< F$, which is the starting point of the beam in the depth domain. If $F < d_\textrm{F} /10$, we also have the solution $z= \frac{d_\textrm{F} F  }{d_\textrm{F}  - 10 F} $, which marks the end of the beam \cite{Bjornson2021b}.
Hence, for transmission to receivers at locations  $d_\textrm{F} < F < d_\textrm{F}/10$ in the radiative near-field, the transmitted signal behaves as a beam with finite half-power beamdepth:
\begin{equation}
    \textrm{BD}_{3\,\textrm{dB}} = \frac{d_\textrm{F} F  }{d_\textrm{F}  - 10 F}-\frac{d_\textrm{F} F  }{d_\textrm{F}  + 10 F} = \frac{20 d_\textrm{F} F^2}{d_\textrm{F}^2 - 100F^2}.
\end{equation}
This is a unique phenomenon for the beamdepth because for transmission to points $F > d_\textrm{F}/10$ (which includes the far-field), the beam continues all the way to infinity.
Interestingly, as $F \to \infty$ so that the beam is focused at a faraway location, the lower limit $\frac{d_\textrm{F} F  }{d_\textrm{F}   + 10 F}$ converges to $d_\textrm{F} /10$. This is an alternative boundary between the near-field and far-field when considering the beamdepth.

\begin{figure}[t!]
	\centering 
	\begin{overpic}[width=\columnwidth,tics=10]{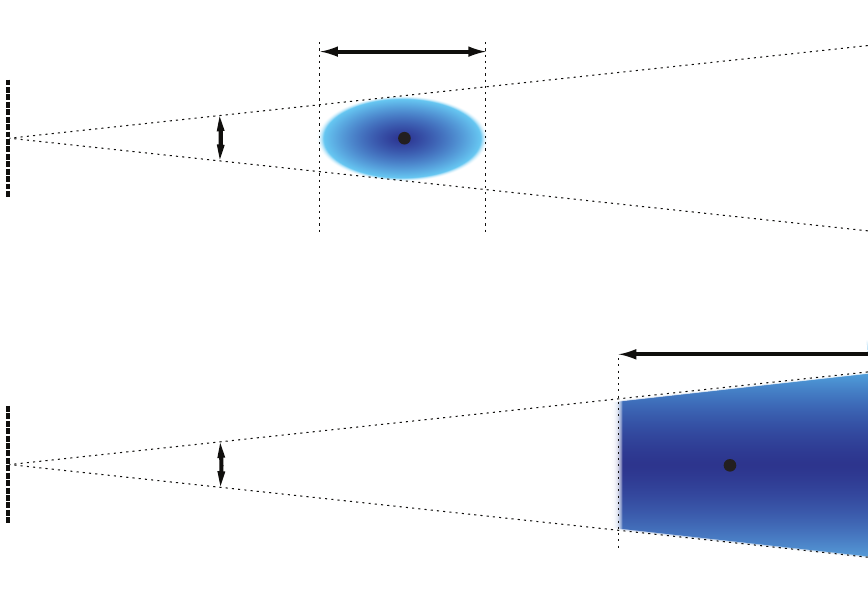}
  \put (15,37) {(a) Beamforming in the radiative near-field}
  \put (20,0) {(b) Beamforming in the far-field}
  \put (37,65) {Finite depth}
  \put (75,30) {Infinite depth}
  \put (67,28) {$\frac{d_\textrm{F}}{10}$}
  \put (26.5,52) {$\textrm{BW}_{3\,\textrm{dB}}$}
  \put (26.5,14.5) {$\textrm{BW}_{3\,\textrm{dB}}$}
\end{overpic} 
	\caption{Beamforming leads to a limited beamwidth regardless of whether the signal is focused on a receiver in the near-field or far-field. However, if the focus point is at a closer distance than $d_\textrm{F}/10$, the beamdepth will be finite. This is not the case when the focus point is beyond $d_\textrm{F}/10$, because then the beam continues until infinity.}
	\label{fig:beamdepth_nearfield}
\end{figure}
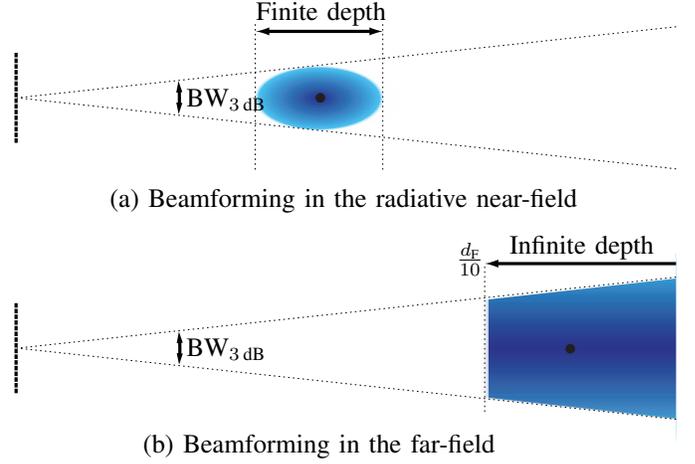

Fig.~\ref{fig:beamdepth_nearfield} illustrates the differences between beamforming in the radiative near-field and far-field, which in this case ends at $d_\textrm{F} /10$. The beamdepth is finite in the near-field but semi-infinite in the far-field, meaning that it starts at $d_\textrm{F} /10$ and then continues to infinity. These phenomena resemble the depth of focus of camera lenses, which is finite when focusing on a nearby object (leading to a blurry background) and semi-infinite when the object is far away~\cite{lens16,lens18,lens_p_22}.

The array processing that underpins near-field beamfocusing has been studied for decades, starting with the works \cite{PirzDesignWidebandConstantBeamwidth1979,RyanGoubranNearFieldBeamformingMicrophone1997,NordholmEtAlChebyshevOptimizationDesignBroadband1998} on microphone arrays.
It is the applications within long-range wireless communication networks that are novel \cite{MyersHeathInfocus2022,g,FriedlanderLocalizationSignalsNearField2019,Phang2018,DeshpandeEtAlWidebandGeneralizationNearField2023} and considered for the 6G era.

\subsection{Near-field spatial multiplexing}

The fact that signals transmitted toward receivers in the near-field have a smaller focus area (both in width and depth) means a drastically reduced risk of causing interference between concurrent signal transmissions. This is by itself an enabler of the emerging \emph{massive spatial multiplexing} paradigm, where we are not serving tens of UEs per BS as in 5G but hundreds or a thousand. The classical SE formulas and transmit/receiver signal processing schemes (see \cite{Marzetta2016a,massivemimobook}) can be utilized, but they will result in substantially higher values thanks to the more favorable propagation conditions obtained in the radiative near-field compared to the far-field.

To exemplify these effects, we begin by considering an uplink single-cell multi-user MIMO setup with $K$ single-antenna UEs. 
The channel between the $M$-antenna BS and UE $k$ is denoted by $\mathbf{h}_k \in \mathbb{C}^M$.
The received signal $\mathbf{y}\in\mathbb{C}^M$ is modeled as
\begin{equation} \label{eq:SIMO-MU-MIMO}
  \mathbf{y}=\sum_{k=1}^{K}\mathbf{h}_k s_k + \mathbf{n}, 
\end{equation}
where $s_k$ is the data signal transmitted by UE $k$ with power $p_k$ and
$\mathbf{n} \sim \mathcal{N}_\mathbb{C}\left(\mathbf{0},\sigma^2 \mathbf{I}_M\right)$ is circularly symmetric complex Gaussian noise with variance $\sigma^2$ and $\mathbf{I}_M$ is the $M$-dimensional identity matrix.
The BS can apply a receive combining vector $\mathbf{v}_k\in\mathbb{C}^M$ to the received signal in \eqref{eq:SIMO-MU-MIMO} as $\mathbf{v}_k^{\Htran}\mathbf{y}$ to detect the signal  $s_k$. By treating the co-user interference as noise, the SE for UE $k$ becomes
\begin{align}\label{eq:uplinkSE}
  &\log_2\left(1+ \frac{p_k \left|\mathbf{v}_k^{\Htran} \mathbf{h}_k\right|^2}
  { \sum\limits_{i=1,i \neq k}^{K}{p_i \left|\mathbf{v}_k^{\Htran} \mathbf{h}_i\right|^2} + \sigma^2\left\|\mathbf{v}_k\right\|^2} \right) \\
  &\leq \log_2\left(1+ p_k \mathbf{h}_k^{\Htran} \left( \sum_{i=1,i \neq k}^{K}{{p_i \mathbf{h}_i \mathbf{h}_i^{\Htran} }} + {\sigma^2}\mathbf{I}_M\right)^{-1} \mathbf{h}_k \right),
\end{align}
where the upper bound is achieved by using the linear minimum mean-squared error (LMMSE) receive combining vector:
\begin{align}\label{eq:mmse}
  \mathbf{v}_k^{\rm LMMSE} = p_k \left( \sum_{i=1}^{K}{{p_i \mathbf{h}_i \mathbf{h}_i^{\Htran}}} + {\sigma^2}\mathbf{I}_M\right)^{-1} \mathbf{h}_k.
\end{align}
This combining scheme finds the SE-maximizing tradeoff between amplifying the signal power by aligning the receiver to $\mathbf{h}_k$ and rejecting interference by spatial whitening of the received signal using the inverse of $\mathbb{E}\{ \vect{y} \vect{y}^{\Htran} \}$.
The provided formulas are typical ones for uplink multi-user MIMO because the difference lies in the channels.

\begin{figure}[t!]
       \centering
       \begin{subfigure}[b]{\columnwidth} \centering 
	\begin{overpic}[trim={5mm 0mm 15mm 10mm},width=\columnwidth,tics=10]{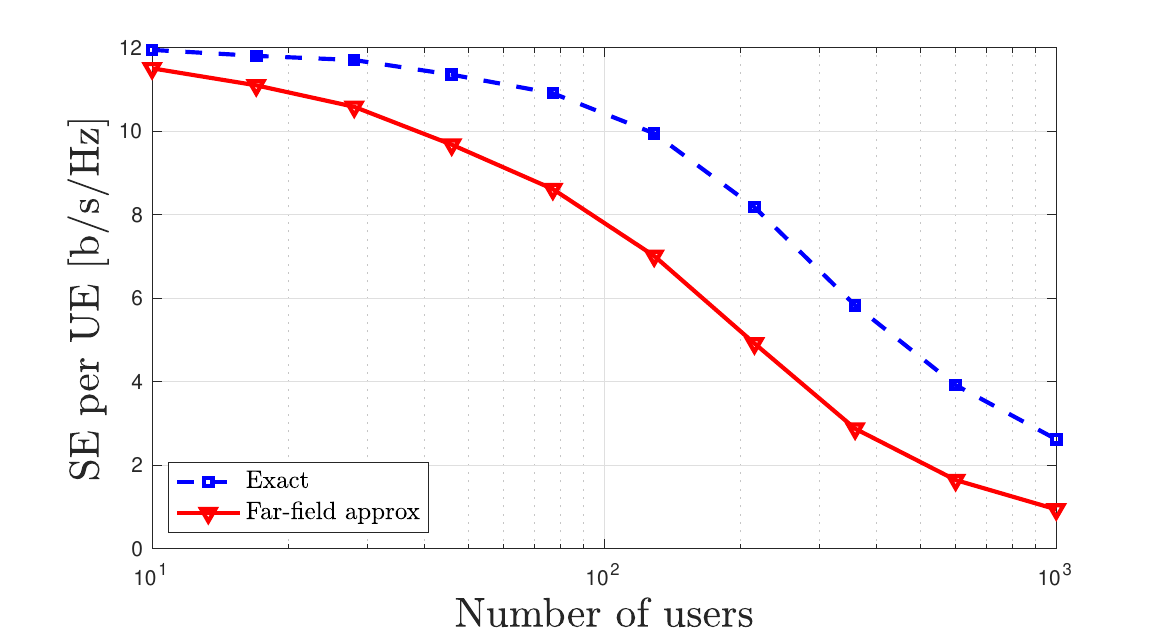}
\end{overpic}   
               \caption{The average SE per UE.} 
       \end{subfigure}\\
       \begin{subfigure}[b]{\columnwidth} \centering  \vspace{+10mm}
	\begin{overpic}[trim={5mm 0mm 15mm 10mm},width=\columnwidth,tics=10]{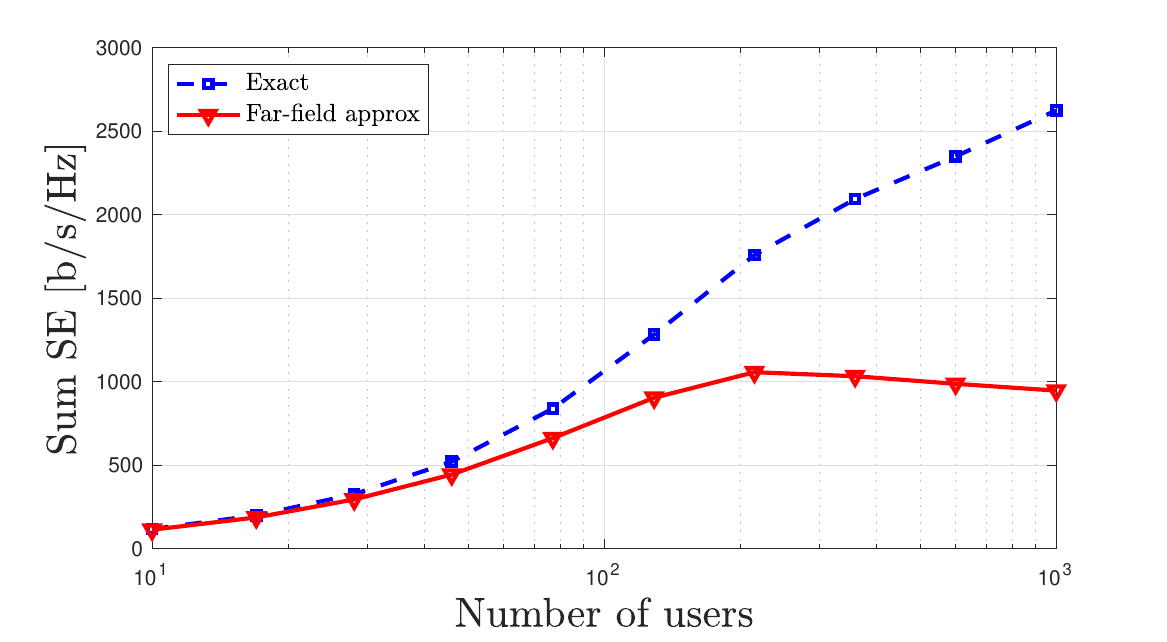}
\end{overpic}  
               \caption{The average sum SE for all UEs.} 
       \end{subfigure} 
       \caption{The average uplink SE per UE and sum SE in a setup with $5000$ antennas. The exact near-field propagation model leads to much higher values than in an identical setup where a mismatched far-field approximation is utilized.} 
       \label{figure_MUMIMO}
\end{figure}

Fig.~\ref{figure_MUMIMO} shows the uplink SE achieved in a single-cell multi-user MIMO setup operating at $30$\,GHz. The BS has a half-wavelength-spaced array of $1 \times 0.5$\,m, containing $5000$ antennas in the configuration $100 \times 50$. The single-antenna UEs are uniformly distributed in the horizontal plane in the angular sector $\varphi \in \left[-\pi/3, +\pi/3\right]$ and in the distance range $15$--$500$\,m. The Fraunhofer distance is $d_\textrm{F}=250$\,m in this setup, so some of the UEs are located in the radiative near-field and others in the far-field. The propagation parameters are otherwise the same as in \cite{Bacci2023a}.

The figure contains one curve obtained using an exact line-of-sight channel model and one curve obtained using the conventional far-field approximation, which is mismatched for the UEs actually located in the radiative near-field. Fig.~\ref{figure_MUMIMO}(a) shows the average SE per UE, as a function of the number of UEs. A logarithmic scale is used on the horizontal axis since we consider the range $10$--$1000$ UEs. The SE per UE reduces as more UEs are added to the setup, due to the increased interference.  The optimal LMMSE receive combining from \eqref{eq:mmse} is utilized. There is a substantial gap between the curves where the exact model consistently provides better results. The far-field approximation basically moves near-field UEs outwards to the far-field and thereby makes the resulting beamfocus areas wider and deeper, leading to increased interference.
This showcases how the ability to utilize the depth domain to distinguish between UE channels makes it easier to deal with interference. 

Fig.~\ref{figure_MUMIMO}(b) shows the SSE, which is the SE values from Fig.~\ref{figure_MUMIMO}(a) multiplied by the respective number of UEs. The massive difference between using the exact and mismatched far-field model becomes evident in this case: With the exact model, the SSE keeps growing with the number of UEs while it saturates at around $200$ UEs with the mismatched model.
From a network operational perspective, we want to transmit as much data as possible, and the new propagation phenomena observed in the radiative near-field enable higher SEs per UE and efficient spatial multiplexing of many more UEs than if the same MIMO system would operate in the far-field. It is through the massive spatial multiplexing of $1000$ UEs that one can reach groundbreaking SSE levels in 6G.

While the depth perception facilitated the spatial multiplexing of many UEs, the tiny beamwidth can enable multiple data streams to be transmitted to a single UE---even in line-of-sight scenarios where we are used to only support a single spatial layer in the far-field.
For the sake of argument, suppose the BS and UE are both equipped with uniform linear arrays (ULAs) with $M$ antennas. The UE has half-wavelength spacing since it is supposed to be a compact device, while the spacing $\Delta$ at the BS can be optimized if its ULA is deployed on the facade of a building.
When the BS transmits, the half-power beamwidth is given in \eqref{eq:half-power} and is inversely proportional to $\Delta$.
If we make it sufficiently small, we can beamform a different signal to each antenna in the receiver array.

We denote the distance between the transmitter and receiver as  $d$, and let the ULAs be deployed perpendicularly to the propagation direction. The BS then sees two adjacent UE antennas with an angular difference of $\sin(\varphi) \approx (\lambda/2)/d$.
We can select $\Delta$ so that the array gain is zero at the adjacent antenna. The expression in \eqref{eq:beamforminggain} is zero when $\varphi$ is such that the argument of the sinc-function is $1$. By solving for $\Delta$, we obtain
\begin{align}
    \frac{1}{\lambda} N \Delta \sin(\varphi) = 1 \quad \Rightarrow \quad \Delta = \frac{\lambda}{N \sin(\varphi)} \approx \frac{2d}{N}.
\end{align}
Hence, the total length of the ULA should be $N\Delta = 2d$, which makes it longer than the propagation distance.
More practical deployment scenarios can be achieved by tuning the antenna spacing in the ULAs at both the transmitter and receiver. If the spacings are $\Delta_\textrm{t}$ and $\Delta_\textrm{r}$, respectively, the same result can be achieved if \cite{Bohagen2007a,multi2011,Do2021a}
\begin{equation} \label{eq:MIMO-antenna-spacing}
    \Delta_\textrm{t}\Delta_\textrm{r} = \frac{\lambda d}{M}.
\end{equation}
The main message is that one can benefit from increasing the antenna spacing in single-user MIMO systems if that pushes the propagation into the radiative near-field, where the beamwidth can be smaller than the array.

\begin{figure}[t!]
	\centering 
	\begin{overpic}[trim={5mm 0mm 15mm 10mm},width=\columnwidth,tics=10]{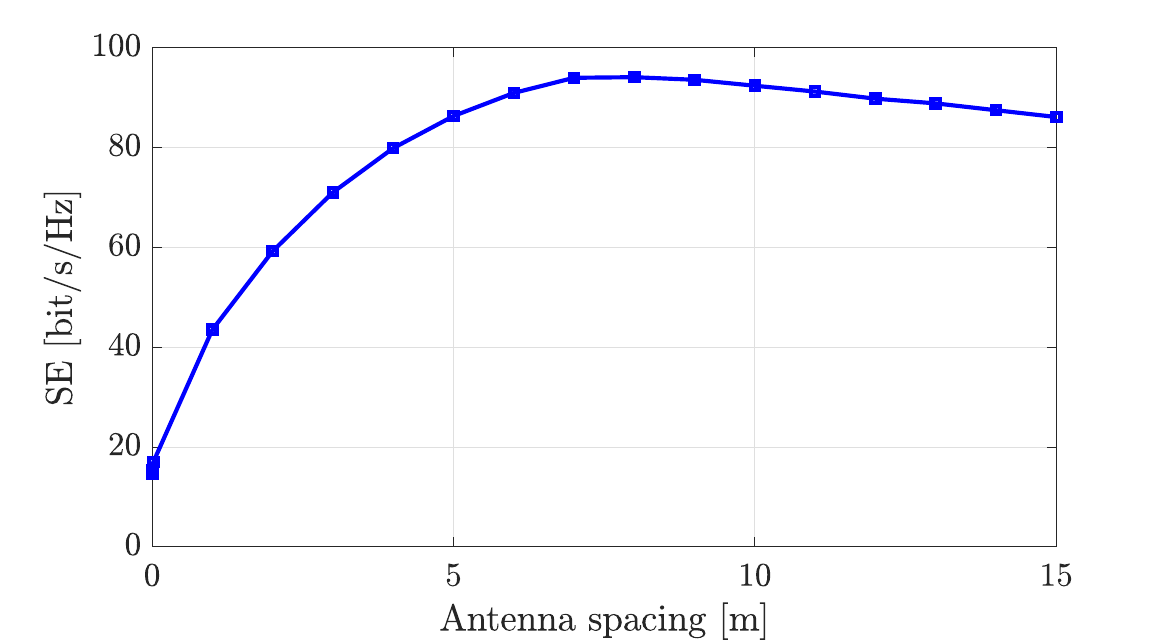}
\end{overpic} 
	\caption{The SE of single-user MIMO channel with $16$ antennas at each side depends on the antenna spacing. The UE has half-wavelength spacing while the BS has a varying spacing, which can be fine-tuned to maximize the SE.}
	\label{fig:SU_MIMO}
\end{figure}

Fig.~\ref{fig:SU_MIMO} shows how the SE over a single-user MIMO channel with $16$ antennas at the transmitter and receiver. The propagation distance is $d=50$\,m, the frequency is $30$\,GHz, and the SNR for a single layer is $20$\,dB. The UE has a ULA with half-wavelength spacing while the antenna spacing at the BS is varied on the horizontal axis. The first point on the curve represents half-wavelength spacing and the figure shows that drastically higher SEs can be achieved by increasing the spacing. The reason is that the MIMO channel matrix transitions from having one non-zero singular value in the beginning to $16$ equally large singular values at the peak value at around $8$\,m. At that point, the BS can transmit a different beam toward each receive antenna, and it will only be ``heard'' by the designated antennas thanks to the narrow beamwidth. Beyond that antenna spacing, the SE begins to decay again due to sidelobe effects. 

The main point of this example is that large SEs can also be achieved for a single UE in the radiative near-field, since we can make the beamwidth so narrow that we can beamform different signals toward different parts of the receiver.

\section{Spatial degrees-of-freedom}
\label{sec:spat_deg_freedom}
A key reason for increasing the number of antennas in future UM-MIMO systems is to enable more spatial layers, as demonstrated in the last section. We will now take a closer look at the maximum number of spatial layers that can be transmitted efficiently over a given channel, which is called the \emph{spatial degrees-of-freedom (DoF)}. As a background to this concept, we begin by reviewing the concept of \emph{spectral DoF}.

\subsection{Spectral degrees-of-freedom}

Wireless channels are also known as \emph{waveform channels} because they take a transmitted analog waveform/signal as the input and produce a received analog waveform at the output. Waveforms can be represented either as time-domain signals or as spectra in the frequency domain, and the Fourier transform serves as the bridge between these identical representations. Let us consider a  waveform channel that accepts complex-valued continuous-time signals $s(t)$ that is approximately limited to the $T$-length time interval $[-T/2, T/2]$ and strictly band-limited to the spectral interval  $[-B/2, B/2]$. A band-limited signal cannot be entirely time-limited, but according to the Shannon-Nyquist sampling theorem~\cite{Shannon1949b}, we can approximate $s(t)$ as
\begin{align} \label{eq:sampling_theorem}
s(t) \approx \sum_{n = -TB/2}^{TB/2-1} s \left(\frac{n}{B}\right) \sinc \left(B t - n\right),
\end{align}
where the approximation error becomes negligible as $TB \rightarrow \infty$. 
Notably, \eqref{eq:sampling_theorem} is a band-limited orthonormal series expansion characterized by the finite set of coefficients	 
\begin{align}
\left\{s \left(\frac{n}{B}\right): n = -\frac{TB}{2}, \ldots ,\frac{TB}{2}-1\right\}.
\end{align}
The cardinality of this set is
\begin{align} \label{eq:DoF_waveform}
\eta = TB 
\end{align}
and is called the \emph{dimension} or DoF of the waveform.
Since the signal $s(t)$ is completely determined by $B$ complex-valued equal-spaced samples per second, we can call these the spectral DoF.

In communication systems, we want to design the waveform $s(t)$ to carry data over the channel. A practical system might operate over a real-valued passband channel with bandwidth $B$ around some carrier frequency $f_c$; that is, it accepts real-valued waveforms that are band-limited to the spectral interval $[f_c-B/2,f_c+B/2]$.
This channel can be identically represented as a complex baseband waveform channel of the kind described above~\cite{Tse2005a}.\footnote{Due to Doppler spread and use of other pulses than the sinc-function, the bandwidth can be slightly larger than $B$.}
Hence, the data signal is completely determined by $B$ complex-valued equal-spaced samples per second; these are the spectral DoF available for shaping the communication signal at the transmitter. 
For example, the transmitter can place information into these samples using a $16$-QAM (Quadrature Amplitude Modulation) scheme, which is a complex constellation with $16$ different states. Since each sample represents $\log_2 (16) = 4$ bits, the transmitter can convey four bits per spectral DoF. If the channel has the bandwidth $B=10$\,MHz, which leads to $10 \cdot 10^6$ DoF per second, the data rate becomes $4 \cdot 10 \cdot 10^6 = 40$\,Mbps.
More information can be conveyed by increasing the constellation size, but it is essential that the data rate is below the channel capacity, so the receiver can decode the data without error. The SE is the theoretical limit on the number of bits to convey per sample. We previously said that its unit is bit/s/Hz, but it can be equivalently expressed as bit/sample or bit/DoF.

In 5G, the typical bandwidth in the $3.5$\,GHz band is $100$\,MHz, and the maximum constellation size is $1024$-QAM. This corresponds to $100 \cdot 10^6$ DoF/s and $10$ bit/DoF. Consequently, the maximum data rate is $1$\,Gbps and it is determined by two variables: the spectral DoF (i.e., bandwidth) and SE.
It is not realistic to increase the SE beyond $10$ bit/DoF in future networks, because one needs an SNR greater than $30$\,dB to reach that number. When moving beyond that SNR value, the system is typically limited by hardware fidelity rather than noise.
We also cannot expect to increase the bandwidth by more than a factor of $10$ compared to 5G.
Hence, the important question is: \emph{How can we increase the data rate by a factor of $100\times$ or $1000\times$?} The answer is that we need to create new signal dimensions through MIMO instead.

\begin{figure}[t!]
	\centering 
	\begin{overpic}[width=\columnwidth,tics=10]{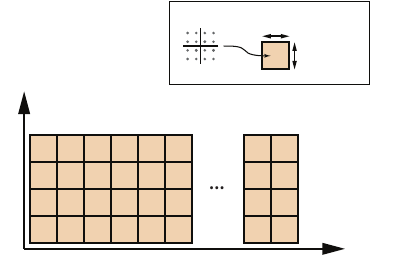}
   \put (42.5,46.5) {16-QAM}
   \put (78,8) {Spectral DoF}
   \put (42.5,62) {One DoF:}
   \put (0,45) {Spatial DoF}
   \put (74,51) {1 antenna}
   \put (64,59) {1 Hz}
   \put (67,51) {1}
\end{overpic} 
	\caption{The basic building blocks of wireless waveforms are called DoF, which is the number of complex-valued coefficients that describes the signal per second. There are both spectral and spatial DoF and their product is the total number of DoF. Each DoF can represent an amount of data that equals the SE of the channel, for example, represented using a 16-QAM constellation.}
	\label{fig:DOF1}
\end{figure}
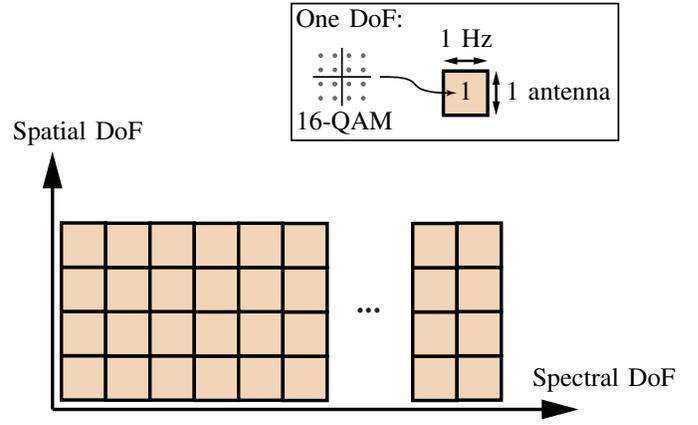

\subsection{(Massive) MIMO: An information-theoretic perspective}

Two types of MIMO communication systems were introduced earlier in this paper: single-user MIMO involves a multi-antenna BS and a multi-antenna UE, while multi-user MIMO involves a multi-antenna BS and multiple UEs. In both cases, the expansion of the signal space into the spatial (antenna) domain creates multiple parallel spatial channels representing the \emph{spatial DoF}. The simultaneous transmission of independent data signals over these spatial channels, as opposed to different time slots and/or frequency subbands, results in a traffic multiplier or \emph{spatial multiplexing gain}, as long as effective measures are taken to mitigate interference between the transmitted signals. 
Fig.~\ref{fig:DOF1} illustrates this as a two-dimensional DoF plane, where the spatial and spectral dimensions are orthogonal. The basic building blocks of a signal is $1$ DoF spanning over one Hertz and one antenna dimension. Although we might always have fewer spatial DoF (e.g., hundreds to thousands) than spectral DoF (e.g., millions to billions) in wireless systems, the total number of DoF is their product.
We will now take a closer look at the spatial DoF delivered by the two MIMO categories.

The capacity of single-user MIMO was first established in~\cite{Foschini1998a,Telatar1999a}. Such a channel with $M_\textrm{t}$ transmit antennas and $M_\textrm{r}$ receive antennas is represented by the MIMO channel matrix $\vect{H} \in \mathbb{C}^{M_\textrm{r} \times M_\textrm{t}}$, which can be modeled in various ways depending on propagation environment. The channel capacity is generally determined by the $M_{\min} = \min(M_\textrm{r}, M_\textrm{t})$ singular values $\mu_1,\ldots,\mu_{M_{\min}}$ of $\vect{H}$, such that \cite[Sec.~7.1]{Tse2005a}
\begin{align}
C = \sum_{i=1}^{M_{\min}} \log_2 \left(1 + \frac{p_i \mu_i^2}{\sigma^2} \right),
\end{align}
where the summation is over the different spatial layers and the corresponding transmit powers are $p_1,\ldots,p_{M_{\min}}$. For an ideal MIMO channel where all the singular values are equal and each entry of $\vect{H}$ has an equal squared magnitude $\beta$, the capacity becomes 
\begin{align}
C = M_{\min} \log_2 \left( 1 + \mathrm{SNR} \frac{M_\textrm{r} M_\textrm{t}}{M_{\min}^2} \right),
\end{align}
where $\mathrm{SNR} = p \beta / \sigma^2$, $p$ is the total transmit power, and $\sigma^2$ is the noise variance. This is the kind of MIMO channel that was obtained by optimizing the antenna spacing according to \eqref{eq:MIMO-antenna-spacing}.
The number of signals that are spatially multiplexed, or equivalently, the number of spatial DoFs, increases with the minimum of $M_\textrm{r}$ and $M_\textrm{t}$. Hence, the total DoF of a single-user MIMO system with bandwidth $B$ is $\min(M_\textrm{r}, M_\textrm{t}) \times B$ per second. With $M_\textrm{t}=M_\textrm{r}=64$, there is a potential for a $64$-fold increase in capacity compared to a single-antenna system.

The capacity of multi-user MIMO systems was characterized in the early $2000$s~\cite{Caire2003a,Goldsmith2003a,Viswanath2003a}. The capacity requires non-linear signal processing at the transmitter or receiver, which is challenging to implement in practical systems. Hence, there is a variety of alternative SE expressions for specific linear processing schemes, such as the ones provided in \eqref{eq:uplinkSE}.
We will consider an uplink expression from \cite[Table 3.2]{Marzetta2016a} for the case of i.i.d. Rayleigh fading channels, $K$ single-antenna UEs and $M$ BS antennas. If all UEs have the same average SNR, the SSE becomes
\begin{align}
K \log_2 \left(1 + \frac{M \, \mathrm{SNR}}{K \,\mathrm{SNR} + 1} \right).
\end{align}
The multiplicative factor $K$ preceding the logarithm indicates the availability of $K$ spatial DoFs. What is more intriguing is that by increasing both $M$ and $K$ simultaneously (with some fixed ratio $M/K$ that is preferably large), we can maintain a nearly constant SE per UE while providing a $K$-fold increase in SSE.
If we could somehow allow the UEs to collaborate in a multi-user MIMO system, we obtain a single-user MIMO setup whose capacity must be equal or higher. This indicates that the spatial DoFs cannot surpass $\min(M,K)$ in a multi-user MIMO system.

The basic principles of MIMO outlined above were established for specific idealistic channel models and perfect CSI. However, the spatial multiplexing capability of single-user MIMO is traditionally hampered by having a low-rank channel matrix, while imperfect CSI acquisition is a main limiting factor for multi-user MIMO. The Massive MIMO concept was introduced in \cite{Marzetta2010a} to address these challenges. Firstly, it relies on deploying a significantly larger number of BS antennas than spatially multiplexed devices (e.g., $M/K \ge 8$).
In addition to the reason described above, there are two further practical advantages: it reduces interference between UEs and provides the so-called \emph{channel hardening}, which ensures minimal SNR fluctuations after the precoding/combining has been applied. Secondly, the protocols were designed for time-division duplexing (TDD) to enable CSI acquisition for arbitrarily many BS antennas through uplink pilot transmission~\cite{Marzetta2016a,massivemimobook}. We will return to the channel estimation challenge in Section~\ref{section:estimation}.

Current 5G BSs are built using the Massive MIMO principle with digital planar arrays of $M = 32$ or $M = 64$ antennas. When increasing these numbers in future networks to hundreds or thousands, it would be desirable to pack the antennas more closely. However, this will make the channel coefficients more similar (e.g., statistically correlated) and we cannot increase the capacity indefinitely by doing so. This is analogous to the waveform channel where, given the bandwidth constraint $B$ and the transmission interval $T$, increasing the number of time samples will also not increase the spectral DoF indefinitely.
This is called temporal oversampling.
The available spectral DoF in the given time interval is always limited to $TB$.  A fundamental question arises: \emph{given an area limitation for the antenna array, what is the intrinsic number of DoF available in the channel?} 
To answer this question, it is necessary to extend the Shannon-Nyquist sampling theorem to spatially bandlimited fields.

\subsection{Shannon-Nyquist sampling theorem for electric fields}

While a single receive antenna takes samples of an impinging waveform at different times, an antenna array can take many spatially separated samples of the waveform at the same time. We can apply the classical Shannon-Nyquist sampling theorem also in this situation, with the only difference being that the samples are collected over a limited spatial interval rather than a limited time interval.
When transforming a spatial signal into the frequency domain, we obtain the spatial frequencies, also known as \emph{wavenumbers}.
Before we can apply the sampling theorem to this situation, we need an accurate description of electric fields in the wavenumber domain, which is tied to its physical nature and geometry of the antenna array, as it imposes different boundary conditions on the set of Maxwell's equations. In both the EM and communications literature, various methodologies for the treatment of deterministic and stochastic spatial fields have been explored. A non-exhaustive list of relevant publications in this area is~\cite{Bucci_1987,Bucci_1998,Sayeed2002a,poon2005degrees,Kennedy_2007,Hanlen_2007,Franceschetti2017a,Hu2018a,Pizzo2020a,Pizzo_2022,Pizzo_2022a}. 

\begin{figure}[t!]
         \centering\includegraphics[width=0.28\textwidth]{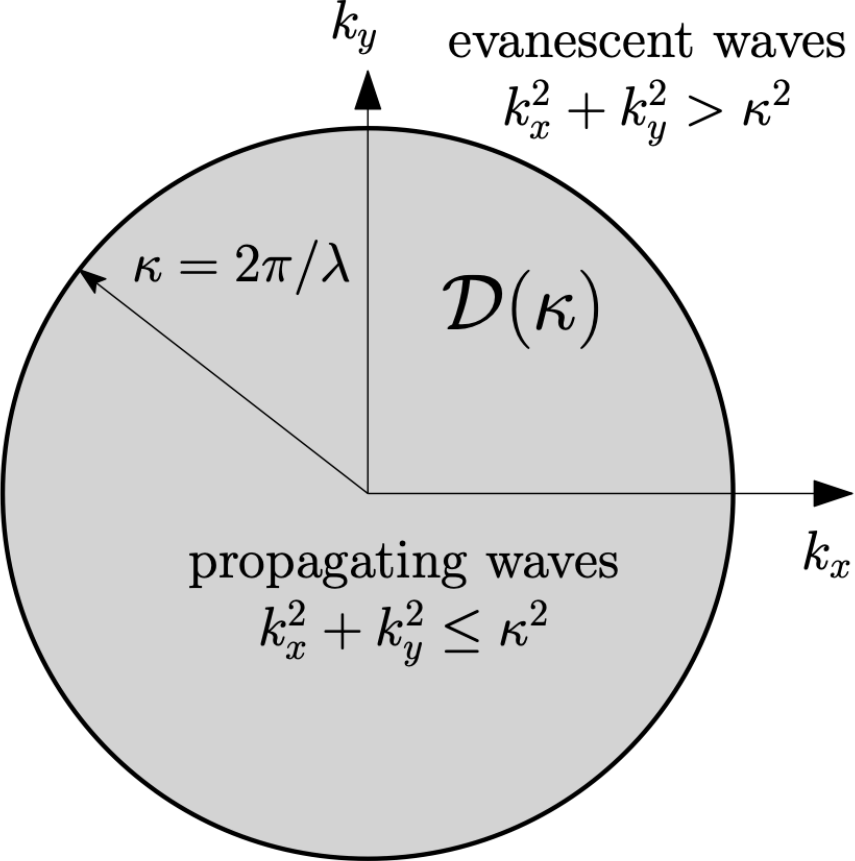}
         \caption{Wavenumber support of the  electric field $s(x, y, z)$.}
         \label{fig:wavenumber_support}
\end{figure}

To understand the basic principles, we consider a scenario where EM waves propagate through a homogeneous, isotropic, source-free, and scattered infinite medium. Monochromatic waves with no polarization then behave as acoustic waves and the electric field $\{s(x,y,z): (x,y,z) \in \mathbb{R}^3\}$ satisfies the scalar Helmholtz equation in the frequency domain~\cite[Eq.~(1.2.17)]{ChewBook} 
\begin{equation} \label{eq:Helmholtz_frequency_mono} 
\left(\nabla^2 + \kappa^2  \right) s(x,y,z)  =  0,
\end{equation}
where $\kappa = \frac{2 \pi}{\lambda}$ is the angular wavenumber of the considered signal (i.e., the angular variation in radians per unit of length). The  solution to \eqref{eq:Helmholtz_frequency_mono} takes the form
\begin{equation} \label{eq:solution}
s(x,y,z) = \alpha e^{\imagunit (k_x x + k_y y + k_z z)} 
\end{equation}
where $\alpha\in \mathbb{C}$ is an unknown complex scaling factor and $(k_x,k_y,k_z)\in\mathbb{R}^3$ are three wavenumber coefficients that characterize the solution. These represent the wavenumbers observed in the $x$, $y$, and $z$ dimensions, respectively.
Plugging \eqref{eq:solution} into \eqref{eq:Helmholtz_frequency_mono} yields the condition
\begin{equation} \label{eq:wavenumberConstraint}
k_x^2+k_y^2+k_z^2 = \kappa^2.
\end{equation}
This implies that the wavenumber $\kappa$ of the original wave is divided between the three dimensions. We can observe the value ranges $k_x,k_y,k_z \in [-\kappa,\kappa]$, but if we increase the magnitude of one wavenumber, the others must reduce accordingly.
In fact, the constraint in \eqref{eq:wavenumberConstraint} allows us to eliminate one of the three coefficients. 
Specifically, we consider the half-space where $k_z$ is positive such that 
\begin{equation} \label{eq:kappa_z}
k_{z}=\sqrt{\kappa^2 - k_x^2 - k_y^2}.
\end{equation} 
It follows that $(k_x,k_y)$ must have compact support given by
\begin{equation} \label{eq:disk_wavenumber}
\mathcal{D}(\kappa) = \left\{(k_x,k_y)\in\mathbb{R}^2 : k_x^2 + k_y^2 \le \kappa^2\right\},
\end{equation}
which is a disk of radius $\kappa$ centered on the origin, as illustrated in Fig. \ref{fig:wavenumber_support}. 
From \eqref{eq:solution} and \eqref{eq:kappa_z}, we thus have that 
\begin{equation} \label{eq:plane_wave}
s(x,y,z) = \alpha e^{\imagunit \left(k_x x + k_y y + \sqrt{\kappa^2 - k_x^2 - k_y^2} z\right)},
\end{equation}
which is the equation of an incident \emph{plane-wave} impinging on the spatial point $(x,y,z)$.
Note that, by imposing the condition $(k_x,k_y,k_z)\in \mathbb{R}^3$, we exclude the \emph{evanescent waves} from the analysis and consider only \emph{propagating waves}. 
This is consistent with our previous assumption of studying the radiative near-field and far-field.
If no directionality is enforced on the EM waves (corresponding to an isotropic scattering environment), then the support of $s(x,y,z)$ is limited to $\left|\mathcal{D}(\kappa)\right| = \pi \kappa^2$, revealing the spatially band-limited nature of EM fields. Scattering mechanisms act as a filter limiting further the field to a smaller support \cite{Pizzo_2022}. Compared to the classical definition of a band-limited signal in the frequency domain, here the notion of band-limited support applies in the wavenumber domain.

The above findings are crucial for calculating the spatial DoF of electric fields through the generalization of the sampling theorem. Assume for example that $s(x,y,z)$ is observed over a one-dimensional (1D) line segment of length $L_x$ along the $x$-axis. Then, the spatial DoF that characterize all possible fields that can be observed are given by \cite{Franceschetti2017a,Hu2018a,Pizzo2020a,Pizzo_2022}
\begin{align} \label{eq:sampling_theorem_space}
\eta_{\rm 1D} = \frac{2}{\lambda}L_x = \frac{\kappa}{\pi}L_x.
\end{align}
The first expression is the product between the length $L_x$ of the spatial interval where the field is observed, while $ \frac{2}{\lambda}$ is the length of the interval $[-1/\lambda,1/\lambda]$ of spatial frequencies (i.e., non-angular wavenumbers) that the field might contain. This is the spatial-wavenumber counterpart to the product $TB$ between time duration $T$ and bandwidth $B$ in \eqref{eq:DoF_waveform}.
The second expression in \eqref{eq:sampling_theorem_space} expresses the same relation using the angular wavenumber $\kappa$.

Suppose that the electric field $s(x, y, z)$ is instead observed over a two-dimensional (2D) rectangle with side lengths $L_x$ and $L_y$. In this case, the spatial DoF are given by \cite{Franceschetti2017a,Hu2018a,Pizzo2020a,Pizzo_2022}
\begin{align} \label{eq:sampling_theorem_space_LHLV}
\eta_{\rm 2D} = \frac{\pi}{\lambda^2}L_xL_y,
\end{align}
which are proportional to the surface area $L_xL_y$ normalized by the squared wavelength. Hence, each portion of the array aperture with area $\lambda^2$ can observe $\pi$ DoF from the impinging electric field.

One may expect that the spatial DoFs of a two-dimensional array would be the product of the DoFs that can observed horizontally and vertically, using the one-dimensional formula in \eqref{eq:sampling_theorem_space}.
That computation leads to 
\begin{align} \label{eq:sampling_theorem_space_LHLV_1}
\tilde{\eta} = \left( \frac{2}{\lambda}L_x  \right) \left( \frac{2}{\lambda}L_y  \right) =  \frac{4}{\lambda^2}L_xL_y, 
\end{align}
which is different from \eqref{eq:sampling_theorem_space_LHLV}. Specifically $\eta_{\rm 2D} / \tilde{\eta} = \pi/4 \approx 0.79$, so the correct spatial DoFs is smaller. The difference is exactly the ratio between the areas of the disk $D(\kappa)$ and the square circumscribing it, regardless of the dimensions of the rectangular aperture. 
The intuition is that the wavenumbers observed horizontally and vertically in a planar array are correlated. For example, a wave that arrives from a large elevation angle (i.e., near the $y$-axis) can give rise to rather small horizontal variations, similar to how there are smaller distances when moving around the Earth near the North Pole compared to the equator.

\subsection{Implications for MIMO Systems}
\label{subsec:MIMO-DOF}

The signal transmission in wireless communications generates electric fields at the transmitter and samples them at the receiver. The spatial DoF determine how many coefficients are required to characterize these electric fields, not in general but from the perspective of a particular antenna array; if two different electric fields look the same to the array, then we cannot use their difference to carry any additional data. If we are given a particular deployment area of $L_x \times L_y$ meters to deploy an array, \emph{how should we deploy the antennas to obtain all the available spatial DoF?}

The common practice in array design has been to use uniform planar arrays (UPAs) with the spacing $\Delta=\lambda/2$ both horizontally and vertically. We can fit
\begin{equation}
M = \frac{L_x}{\Delta} \frac{L_y}{\Delta} = \frac{4}{\lambda^2}L_xL_y
\end{equation}
antennas into this area. This value matches with \eqref{eq:sampling_theorem_space_LHLV_1} and is, thus, larger than the available spatial DoF that are given by \eqref{eq:sampling_theorem_space_LHLV}. Consequently, the conventional approach to array design is sufficient to capture all the available spatial DoF.

\begin{figure}[t!]
     \begin{subfigure}[b]{0.47\textwidth}
         \centering
         \includegraphics[width=\textwidth]{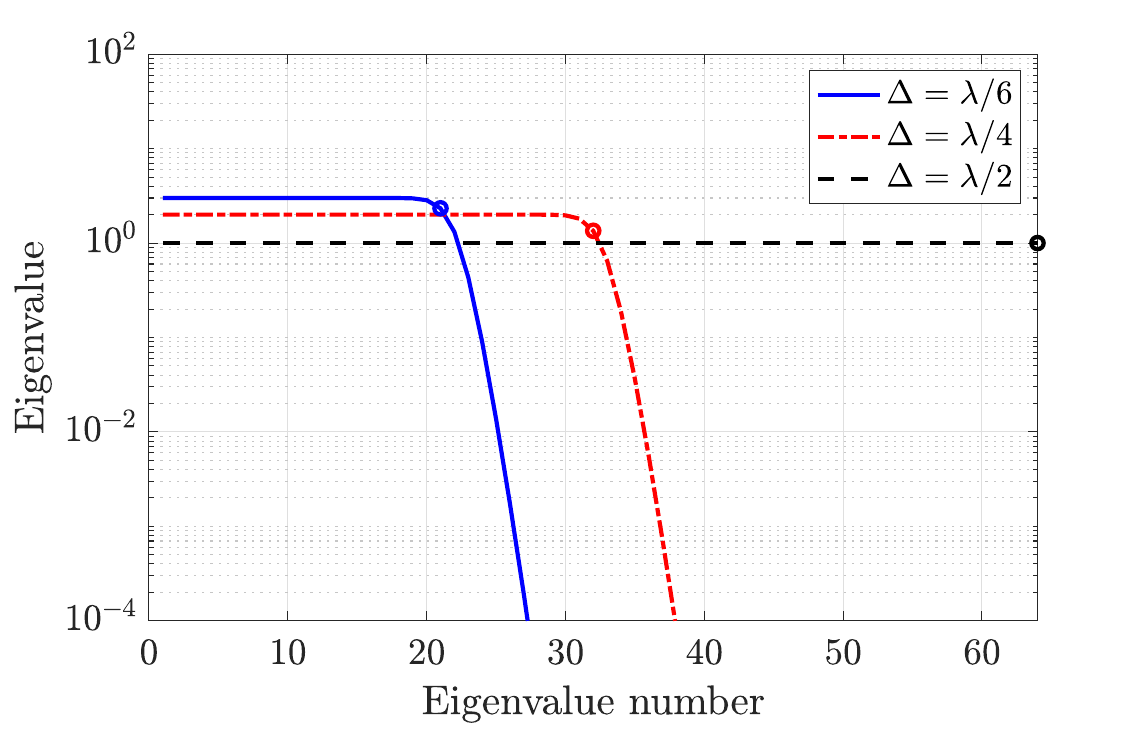}
         \caption{Uniform linear array with $64$ antennas.}
         \label{fig:SE_K2_c}
     \end{subfigure}
     \hfill
     \begin{subfigure}[b]{0.47\textwidth}
         \centering
         \includegraphics[width=\textwidth]{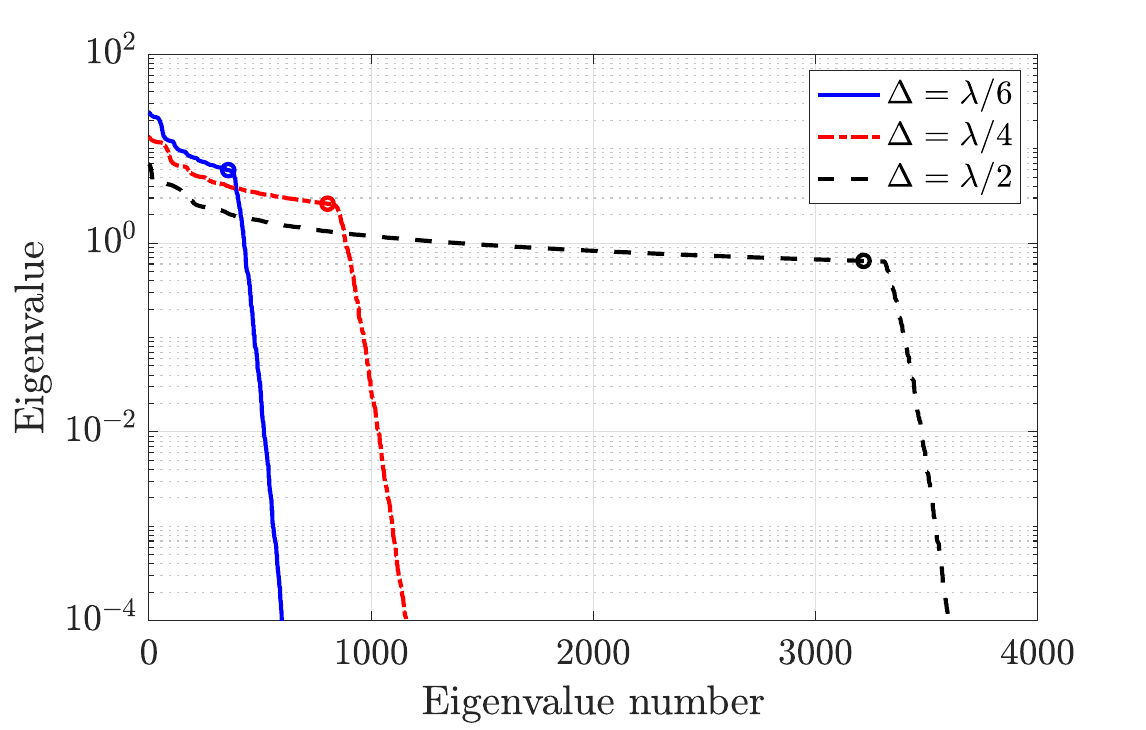}
         \caption{Uniform planar array with $64 \times 64$ antennas.}
     \end{subfigure}
        \caption{The normalized eigenvalues of the spatial correlation matrix $\bf R$ in decreasing order. We consider an isotropic scattering environment with different array geometries and antenna spacings. The numbers of large eigenvalues are determined by the spatial DoF captured by the particular array type. The theoretical values from \eqref{eq:sampling_theorem_space} and \eqref{eq:sampling_theorem_space_LHLV} are shown using circles.} \label{fig:eigenvalue}
\end{figure}

Fig.~\ref{fig:eigenvalue} illustrates what would happen if we reduce the antenna spacing to $\Delta < \lambda/2$. We consider an isotropic scattering environment where plane waves can arrive at the $M$ receive antennas from any direction with equal probability. In this situation, one can compute the spatial correlation matrix $\vect{R} = \mathbb{E}\{ \vect{h} \vect{h}^{\Htran} \}$ of the channel vector $\vect{h} \in \mathbb{C}^M$ and study its eigenvalues. The number of large eigenvalues represents the number of spatial DoF that the array observes, and this is the maximum value since an isotropic environment excites all possible channel dimensions. Fig.~\ref{fig:eigenvalue}(a) considers a uniform linear array (ULA) with $M=64$ antennas and varying antenna spacings. In the case of $\Delta=\lambda/2$, the array covers a spatial interval of $M \lambda/2$ meters, and the spatial DoF in \eqref{eq:sampling_theorem_space} then becomes $M$, which results in all eigenvalues being equally large.
However, if we decrease the antenna spacing to $\lambda/4$ or $\lambda/6$, the number of spatial DoF respectively reduces to $M/2$ and $M/3$. These numbers are illustrated with circles in the figure and clearly predict the number of large eigenvalues; that is, the number of channel dimensions. Fig.~\ref{fig:eigenvalue}(b) considers the case of a UPA with $64 \times 64$ antennas. We see the same general trends as in the case of a ULA, with the main difference that even the case with $\lambda/2$ spacing leads to only 79\% large eigenvalues.

The main conclusion from this section is that we should continue using arrays with $\lambda/2$-spacing in future UM-MIMO systems. We have seen earlier in the paper (e.g., in Fig.~\ref{fig:SU_MIMO}) that one can possibly benefit from increasing the antenna spacing beyond that limit, to observe near-field propagation effects that improve the channel conditions. 
However, one should not reduce the antenna spacing in the hope of increasing the spatial DoF because that is not possible---it corresponds to spatial oversampling.
If the array dimensions are limited, there are still some good reasons for filling it with more antennas than what is needed to capture all the available spatial DoF. Small antennas have a more isotropic-like radiation pattern, which improves the array's ability to transmit and receive signals in any direction. But if we shrink the antenna sizes, we should compensate by adding more antennas into the aperture to keep the total antenna area fixed.


\section{Channel Estimation for Large MIMO Arrays}
\label{section:estimation}

CSI is necessary to make efficient use of a large number of antennas, so that the transmitted signals can be beamformed to the intended location in the downlink, and the received signal can be combined coherently in the uplink. The basic way of acquiring CSI is to transmit a predefined pilot sequence and estimate the channel coefficients at the receiver side. The length of the pilot sequence must equal the number of transmit antennas, while arbitrarily many receive antennas can collect channel estimates simultaneously. 
New communication systems are nowadays built using TDD spectrum, where the same band is used in both uplink and downlink. Hence, we have the liberty to choose in which direction to transmit pilots.
Since the 5G Massive MIMO technology builds on serving tens of UEs with a large number of BS antennas, the pilot sequences (called sounding reference symbols in 5G) are transmitted in the uplink where the pilot sequence length equals the number of UE antennas \cite{Marzetta2010a}. 

The same procedure can be used in the next generation of MIMO technology, but if we increase the number of antennas per UE and the number of spatially multiplexed UEs, the pilot sequence length will increase accordingly. In this section, we will outline how the pilot sequence can be reduced depending on what prior information exists regarding the channel properties.

For brevity in presentation, we focus on the channel estimation from an $M$-antenna UE to one of the many antennas at the BS. The approach described in this section can be applied separately for each BS antenna.
The considered channel vector is denoted by $\vect{h}\in \mathbb{C}^M$. The channel is made of a superposition of multipath components. In the far-field, it can be described as a discrete summation of plane waves arriving from different directions. In the radiative near-field,
the channel can be expressed as a continuous summation of plane waves \cite{Sayeed2002a}, which is also an accurate representation of spherical waves. Hence, we may write the channel vector as
\begin{equation} \label{eq:channel1}
\vect{h} =   \iint_{-\pi/2}^{\pi/2} g(\varphi,\theta) \vect{s}(\varphi,\theta) d\theta d\varphi 
\end{equation}
where $ \vect{s}(\varphi,\theta)$ is the far-field array response vector for the azimuth angle $\varphi$ and elevation angle $\theta$, while the \emph{angular spreading function} $g(\varphi,\theta)$ specifies the gain and phase-shift from each direction.
The integration limits for the azimuth angle are limited to $\varphi \in [-\frac{\pi}{2},\frac{\pi}{2}]$ because waves can only arrive from the front of the array.

The channel realization is determined by the angular spreading function. 
Since small-scale UE mobility is hard to model accurately and the interaction with the surrounding multipath environment is complex, it is customary to model small-scale variations stochastically. In this section, we consider the block fading model, where the channel vector $\vect{h}$ is constant within one block of time-frequency resources and takes independent realization across blocks from a stationary stochastic distribution.
We will model $g(\varphi,\theta)$ as a spatially uncorrelated circularly symmetric Gaussian stochastic process with cross-correlation
\begin{equation} \label{eq:scattering-correlation-model}
\mathbb{E} \{ g(\varphi,\theta) g^*(\varphi',\theta') \} = \beta f(\varphi,\theta) \delta(\varphi-\varphi')  \delta(\theta-\theta'),
\end{equation}
where  $\beta$ denotes the average channel gain and $f(\varphi,\theta)$ is the normalized \emph{spatial scattering function} \cite{Sayeed2002a}.  This is a probability density function (PDF) that provides a statistical representation of the multipath environment in terms of how likely it is for signals to arrive from different directions. As with any PDF, it holds that
$\iint f(\varphi,\theta) d\theta d\varphi  = 1$. 
Based on these assumptions, we obtain the classical correlated Rayleigh fading channel distribution
\begin{equation} \label{eq:corr-Rayleigh}
\vect{h} \sim \CN(\vect{0},\vect{R}),
\end{equation}
but the key reason for the aforementioned assumptions is that we can give the spatial correlation matrix a geometric model:
\begin{equation} \label{eq:spatial-correlation}
\vect{R} = \mathbb{E}\{ \vect{h} \vect{h}^{\Htran} \} = \beta  \iint_{-\pi/2}^{\pi/2} f(\varphi,\theta) \vect{s}(\varphi,\theta) \vect{s}^{\Htran}(\varphi,\theta) d \theta d \varphi 
\end{equation}
which follows from the property in \eqref{eq:scattering-correlation-model}. 
The average gain of the channel is $\mathbb{E}\{ \| \vect{h} \|^2 \} = \tr(\vect{R}) = M\beta$.

We will consider different ways of estimating the realizations of $\vect{h}$ depending on which parts of the statistical characterization are known. 
In all these cases, the estimation is based on transmitting a predefined pilot sequence of some length $\tau_p \leq M$. It is desirable to make this sequence as short as possible to not spend unnecessarily many signal resources (i.e., DoF) on channel estimation, and we will later show that different estimators allow for different pilot lengths. 
We let $\boldsymbol{\Phi}\in \mathbb{C}^{\tau_p \times M}$ denote the pilot sequence matrix, where the $(n,m)$th entry represents the pilot symbol transmitted from antenna $m$ of the UE at time instance $n$ in the $\tau_p$-length sequence. 
We assume that the average pilot power is normalized to one, in the sense that 
\begin{align} \label{eq:avg-power-pilot}
\tr\left(\boldsymbol{\Phi}^{\Htran}\boldsymbol{\Phi}\right)=\tau_p.
\end{align}
 If we collect all the $\tau_p$ received symbols at the BS antenna in a vector $\vect{y}\in\mathbb{C}^{\tau_p}$, we can express it as
\begin{align}
    \vect{y} = \sqrt{p}\vect{\Phi}\vect{h}+ \vect{n} ,\label{eq:received_signal_pilot}
\end{align}
where $p>0$ is the pilot power and $\vect{n}\sim \CN\left(\vect{0}, \sigma^2\vect{I}_{\tau_p} \right)$ is the independent noise. 

\subsection{Least-squares estimation}

The simplest channel estimation method is least-squares (LS) estimation which does not require any statistical information regarding the channel vector. This estimator finds the solution to the \emph{least squares} problem
\begin{align} \label{eq:LS-problem}
\underset{\widehat{\vect{h}} \in \mathbb{C}^M}{\text{minimize}} \quad    \left \Vert\vect{y}-\sqrt{p}\vect{\Phi}\widehat{\vect{h}}\right\Vert^2
\end{align}
between the actual received signal $\vect{y}$ and the potential received signal $\sqrt{p}\vect{\Phi}\widehat{\vect{h}}$ in the absence of noise.
If the pilot length is  $\tau_p =M$ so that the pilot matrix $\boldsymbol{\Phi}$ can be selected to be invertible, the solution to \eqref{eq:LS-problem} is obtained as
\begin{align} \label{eq:LS-estimate}
   \widehat{ \vect{h}}_{\rm LS} = \frac{\boldsymbol{\Phi}^{-1}\vect{y}}{\sqrt{p}} = \vect{h} + \underbrace{\frac{\boldsymbol{\Phi}^{-1}\vect{n}}{\sqrt{p}}}_{=\widetilde{\vect{h}}_{\rm LS}},
    \end{align}
    where $\widetilde{\vect{h}}_{\rm LS}\in \mathbb{C}^M$ is the channel estimation error. This choice makes the term inside the square in \eqref{eq:LS-problem} zero.   
    The channel estimation quality can be quantified by the average power of the channel estimation error, which is called the mean-squared error (MSE). The MSE of the LS estimator is computed as
    \begin{align}
\text{MSE}_{\rm LS}&=\mathbb{E}\left\{\left\Vert\vect{h}-\widehat{ \vect{h}}_{\rm LS}\right\Vert^2\right\}= \mathbb{E}\left\{\left\Vert\widetilde{\vect{h}}_{\rm LS}\right\Vert^2\right\} \nonumber\\
& = \frac{\sigma^2}{p}\tr\left((\boldsymbol{\Phi}^{\Htran}\boldsymbol{\Phi})^{-1}\right)
\end{align}
by utilizing the fact that $\mathbb{E}\left\{\vect{n}\vect{n}^{\Htran}\right\}=\sigma^2\vect{I}_M$.
This expression reveals that the channel estimation quality depends on the selection of the pilot matrix. It can be shown that selecting $\boldsymbol{\Phi}$ as any unitary matrix minimizes the MSE, under the average power constraint in \eqref{eq:avg-power-pilot}. The resulting minimum MSE is 
\begin{align}\label{eq:MSE-LS-minimum}
    \text{MSE}_{\rm LS}^{\star} = \frac{M\sigma^2}{p }
\end{align}
which is proportional to the number of UE antennas (i.e., the number of unknowns) and a linearly decreasing function of the pilot SNR, $\frac{p}{\sigma^2}$. The key observation is that the optimal pilot matrix is obtained by treating each dimension of the $M$-dimensional vector space equally by allocating the same amount of power to all of them.

The main drawback of the LS estimator is that the pilot length must equal the number of UE antennas, otherwise, we cannot invert the pilot matrix. To reduce the pilot length and obtain better-quality channel estimates, we need to exploit more about the structure of the channel, as elaborated in the following part.

\subsection{Minimum mean-squared error estimation}

The considered channel $\vect{h}$ follows the correlated Rayleigh fading model in \eqref{eq:corr-Rayleigh} with the spatial correlation matrix $\vect{R}$ defined in \eqref{eq:spatial-correlation}. If this correlation matrix is completely known at the BS, the minimum MSE (MMSE) estimate can be computed as \cite{kay1993a}
\begin{align}\label{eq:MMSE-estimate}
\widehat{\vect{h}}_{\rm MMSE} = \sqrt{p}\vect{R}\vect{\Phi}^{\Htran}\left(p\vect{\Phi}\vect{R}\vect{\Phi}^{\Htran}+\sigma^2\vect{I}_M\right)^{-1}\vect{y}.
\end{align}
As the name suggests, the MMSE estimator minimizes the MSE among all conceivable estimators that have access to the statistical characterization.
The true channel $\vect{h}$ can be decomposed as $\vect{h} = \widehat{\vect{h}}_{\rm MMSE}+\widetilde{\vect{h}}_{\rm MMSE}$, where the estimation error  $\widetilde{\vect{h}}_{\rm MMSE}$ is independent of the estimate $\widehat{\vect{h}}_{\rm MMSE}$. Consequently, the MSE can be computed as
\begin{align}
    \text{MSE}_{\rm MMSE}&= \tr\left(   \mathbb{E}\left\{\widetilde{\vect{h}}_{\rm MMSE}\widetilde{\vect{h}}_{\rm MMSE}^{\Htran}\right\}      \right) \nonumber \\
    &\hspace{-12mm}=\tr\left(   \mathbb{E}\left\{\vect{h}\vect{h}^{\Htran}\right\}      \right)- \tr\left(   \mathbb{E}\left\{\widehat{\vect{h}}_{\rm MMSE}\widehat{\vect{h}}_{\rm MMSE}^{\Htran}\right\}      \right) \nonumber \\
    &\hspace{-12mm} = M\beta -\tr\left(p\vect{R}\vect{\Phi}^{\Htran}\left(p\vect{\Phi}\vect{R}\vect{\Phi}^{\Htran}+\sigma^2\vect{I}_M\right)^{-1}\vect{\Phi}\vect{R}\right). \label{eq:MMSE-MSE}
\end{align}
This MSE depends on the pilot matrix $\vect{\Phi}$ and, thus, it can be minimized by properly designing the pilot matrix.
We let $\vect{R}=\vect{U}\vect{\Lambda}\vect{U}^{\Htran}$ denote the eigendecomposition of the spatial correlation matrix $\vect{R}$, where the unitary matrix $\vect{U}\in \mathbb{C}^{M \times M}$ contains the eigenvectors as its columns, and the corresponding eigenvalues are located in decreasing order along the diagonal of the matrix $\vect{\Lambda} = \diag(\lambda_1,\ldots,\lambda_M)$.
It can be shown that the MSE is minimized by selecting the $\tau_p \times M$ pilot matrix as $\vect{\Phi}=\vect{D}\vect{U}^{\Htran}$ \cite{Kotecha2004a}, where $\vect{D} \in \mathbb{C}^{\tau_p \times M}$ is a rectangular diagonal matrix where the diagonal entries are 
\begin{align}  \label{eq:poweralloc_pilot}
    d_m = \sqrt{\max\left(0,\mu-\frac{\sigma^2}{p\lambda_m}\right)}, \quad m=1,\ldots,\tau_p,
 \end{align}
 where $\mu>0$ is selected such that 
 \begin{align} \sum_{m=1}^{\tau_p}\max\left(0,\mu-\frac{\sigma^2}{p \lambda_m}\right)=\tau_p.
 \end{align}
This pilot matrix is matched to the eigendecomposition of the spatial correlation matrix since $\vect{U}$ is utilized, while $d_m^2$ determines how much power is allocated to estimating the channel components along the $m$th eigenvector.
The power allocation in \eqref{eq:poweralloc_pilot} has a  water-filling structure, where more power is allocated to the channel directions with larger eigenvalues.

If we substitute $\vect{\Phi}=\vect{D}\vect{U}^{\Htran}$ into
\eqref{eq:MMSE-estimate}, we can simplify the estimator as 
\begin{align}
\widehat{\vect{h}}_{\rm MMSE}&=\vect{U}\vect{A}\vect{y} \label{eq:MMSE-estimate2}
\end{align}
where $\vect{A}=\sqrt{p}\vect{\Lambda}\vect{D}^{\Ttran}\left(p \vect{D}\vect{\Lambda}\vect{D}^{\Ttran}+\sigma^2\vect{I}_{\tau_p}\right)^{-1}$ is a diagonal matrix.  
This MMSE estimator carries out two operations. First, it computes the MMSE estimates of the channel components in the $\tau_p$ strongest eigendirections as $\vect{A}\vect{y}$. Second, it brings this estimate back to the original channel space by multiplying it with the eigenvector matrix $ \vect{U}$.

The core difference between the MMSE and LS estimator is that the former knows the statistical strength of the channel in different in each eigendirection. It can therefore fine-tune the estimator and allocate more pilot power to stronger eigendirections, thereby reducing the MSE. 
While the LS estimator necessitates $\tau_p = M$, the MMSE estimator can be applied with any $\tau_p$ and will then only transmit pilots along the $\tau_p$ strongest eigendirections. 
To highlight the practical importance of this, we will consider a scenario where the rank of $\vect{R}$ is strictly smaller than $M$, as previously illustrated in Fig.~\ref{fig:eigenvalue}. 

\begin{figure}[t!]
	\centering 
	\begin{overpic}[trim={15mm 0mm 15mm 10mm},width=0.9\columnwidth,tics=10]{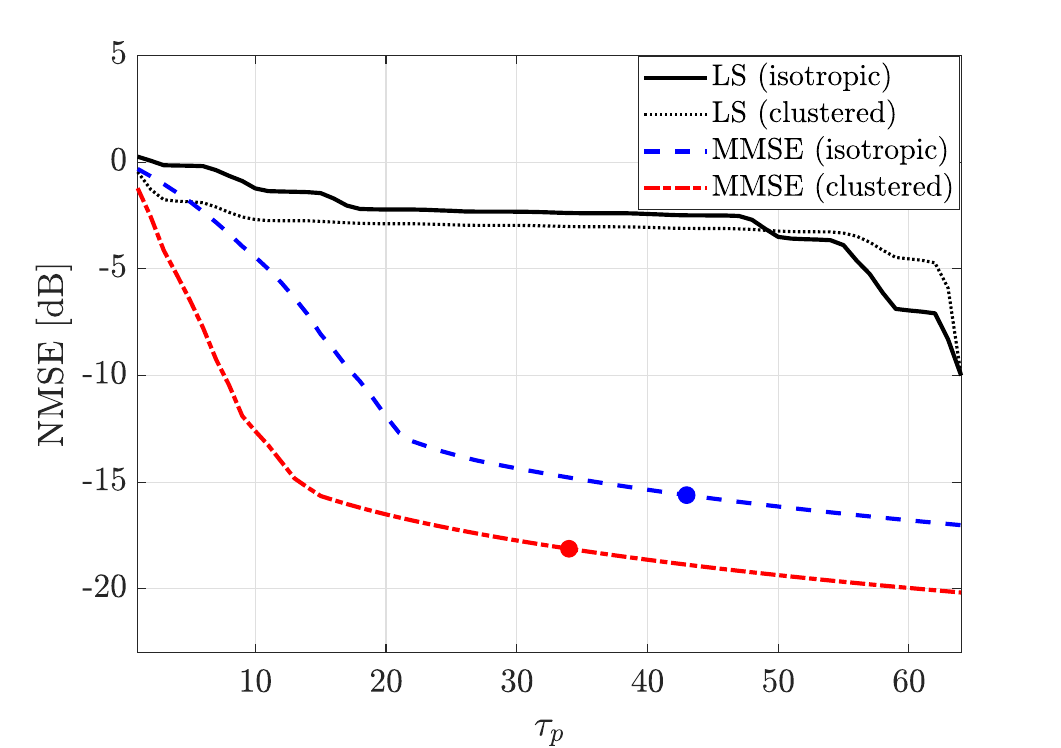}
\end{overpic} 
	\caption{The NMSE versus the pilot length $\tau_p$ for the LS and MMSE estimators in isotropic and clustered scattering environments. The dots show the NMSE performance when $\tau_p$ is equal to the rank of the corresponding spatial correlation matrix.}\vspace{2mm}
	\label{fig:LS_MMSE}
\end{figure}

In Fig.~\ref{fig:LS_MMSE}, we plot the normalized mean square error (NMSE), obtained by dividing the MSE by $\mathbb{E} \{\vect{h}\vect{h}^{\Htran} \}=\tr(\vect{R})$ for the LS and MMSE estimators in different propagation environments. We consider an $8\times 8$ UPA with the antenna spacing $\Delta=\lambda/4$. Two different propagation environments are considered: isotropic and clustered scattering. The isotropic environment assumes that the multipath components are equally strong in all directions (as in Fig.~\ref{fig:eigenvalue}). The clustered environment is generated using a model from \cite{demir2022channel} where there are three scattering clusters located in the azimuth directions $0$, $\pi/4$, and $-\pi/4$ and each having a $10^{\circ}$ angular standard deviation. The effective signal-to-noise ratio (SNR) is $p\tr(\vect{R})/(M\sigma^2)=10$\,dB.

The figure shows the NMSE as a function of the pilot length $\tau_p$. When $\tau_p<M$, the LS estimator is applied with the pseudoinverse of the pilot matrix instead of the true inverse. This leads to poor estimation performance, which is why we ruled out this situation previously.
The MMSE estimator outperforms the LS estimator since it exploits the spatial correlation matrix, but it particularly enables good estimation quality at significantly smaller pilot lengths.
The dots on the plot represent the MMSE estimator's performance when $\tau_p$ equals the rank of the spatial correlation matrices. In the case of isotropic scattering, the value matches the spatial DoF.
For clustered scattering, the rank is smaller since the scattering environment only excites a subset of the possible channel dimensions.
If we increase the pilot length beyond the dot-marked number, the pilot matrix will not explore any new channel dimensions but only benefit from increased pilot energy.
The MMSE estimator provides lower MSEs when the propagation environment has a structure, such as clustered scattering because it can then focus the pilot power into a few important eigendirections.

\subsection{Reduced-subspace least-square estimation}

In the preceding section, we demonstrated how the MMSE estimator effectively utilizes the spatial correlation matrix to minimize the MSE. This $M \times M$ matrix can be estimated in practice by collecting many channel realizations and forming a sample covariance matrix \cite{massivemimobook}, but much more than $M$ observations are needed to obtain an accurate estimate.
Hence, it is challenging to acquire the per-UE spatial correlation matrix information in practice, particularly in scenarios with a larger antenna number, rapid UE mobility that changes the statistics, or during short data packet transmissions \cite{demir2022channel}. 

In such situations, an alternative approach is to only exploit the spatial correlations induced by the array geometry and general characteristics of the propagation environment. For instance, it may be known that multipath components can only be observed within certain angular intervals. This implies that any plausible channel vector lies in a lower-dimensional subspace of $\mathbb{C}^M$. If we know the basis vectors of this subspace, the channel estimation can be performed exclusively within this subspace. This approach is called \emph{reduced-subspace least squares (RS-LS)} estimation \cite{demir2022channel}.

We collect the orthonormal basis vectors of the subspace as columns of the matrix $\overline{\vect{U}}\in \mathbb{C}^{M\times \overline{r}}$, where $\overline{r}\leq \tau_p<M$ is the dimension of the subspace.
Hence, any UE channel can be expressed as $\vect{h}=\overline{\vect{U}}\vect{v}$ for some $\vect{v}\in\mathbb{C}^{\overline{r}}$. 
 The RS-LS estimator builds on estimating $\vect{v}$ and consists of two steps.
 First, we compute the LS estimate of $\vect{v}$ in the subspace spanned by the columns of $\overline{\vect{U}}$ as
\begin{align} \label{eq:RS-LS-estimate0}
    \widehat{\vect{v}}_{\rm LS} = \frac{1}{{\sqrt{p}}}\left(\overline{\vect{U}}^{\Htran}\vect{\Phi}^{\Htran}\vect{\Phi}\overline{\vect{U}}\right)^{-1}\overline{\vect{U}}^{\Htran}\vect{\Phi}^{\Htran}\vect{y},
\end{align}
where we assumed $\tau_p\geq \overline{r}$ so that the inverse is well-defined.
The matrix $\left(\overline{\vect{U}}^{\Htran}\vect{\Phi}^{\Htran}\vect{\Phi}\overline{\vect{U}}\right)^{-1}\overline{\vect{U}}^{\Htran}\vect{\Phi}^{\Htran}$ in \eqref{eq:RS-LS-estimate0} gives the orthogonal projection of the channel onto the considered subspace.
Next, we return this estimate to the original $M$-dimensional space by multiplying $\widehat{\vect{v}}_{\rm LS}$ by $\overline{\vect{U}}$: 
\begin{align} \label{eq:RS-LS-estimate}
    \widehat{\vect{h}}_{\rm RS-LS} &= \overline{\vect{U}}\widehat{\vect{v}}_{\rm LS} = \frac{1}{{\sqrt{p}}}\overline{\vect{U}}\left(\overline{\vect{U}}^{\Htran}\vect{\Phi}^{\Htran}\vect{\Phi}\overline{\vect{U}}\right)^{-1}\overline{\vect{U}}^{\Htran}\vect{\Phi}^{\Htran}\vect{y} \nonumber\\
&=\underbrace{\overline{\vect{U}}\vect{v}}_{=\vect{h}}+\frac{1}{\sqrt{p}}\overline{\vect{U}}\left(\overline{\vect{U}}^{\Htran}\vect{\Phi}^{\Htran}\vect{\Phi}\overline{\vect{U}}\right)^{-1}\overline{\vect{U}}^{\Htran}\vect{\Phi}^{\Htran}\vect{n}.
\end{align}

The RS-LS estimator effectively eliminates noise from all unused channel dimensions when $\overline{r}<M$, and the required pilot length is reduced to $\tau_p\geq \overline{r}$ (compared to $\tau_p\geq M$ for the original LS estimator).
The remaining question is how to determine the basis vectors for the reduced subspace.
One option is to collect many different spatial correlation matrices over time and then compute the union of their span \cite[Lem.~3]{demir2022channel}. Alternatively, one can consider the worst-case scenario of isotropic scattering, where all conceivable channel dimensions might exist in the channel vector. Fig.~\ref{fig:eigenvalue} previously showed that such channels have a low-rank behavior when using UPAs or when the antenna spacing is less than $\lambda/2$
This property is not UE-channel-specific but rather array-dependent, thereby enabling the removal of noise from unused directions.

The MSE of the RS-LS estimator in \eqref{eq:RS-LS-estimate} is
\begin{align} \label{eq:mse-rsls}
    \text{MSE}_{\rm RS-LS}&= \frac{\sigma^2}{p}\tr\left(\overline{\vect{U}}\left(\overline{\vect{U}}^{\Htran}\vect{\Phi}^{\Htran}\vect{\Phi}\overline{\vect{U}}\right)^{-1}\overline{\vect{U}}^{\Htran}\right)\nonumber \\
    &=\frac{\sigma^2}{p}\tr\left(\left(\overline{\vect{U}}^{\Htran}\vect{\Phi}^{\Htran}\vect{\Phi}\overline{\vect{U}}\right)^{-1}\right)
\end{align}
and depends on the pilot matrix $\vect{\Phi}$. 
The MSE is minimized by selecting the pilot matrix as
\begin{align} \label{eq:optimal-phase-shift}
    \vect{\Phi}^{\star} = \sqrt{\frac{\tau_p}{\overline{r}}}\vect{S}\overline{\vect{U}}^{\Htran}
\end{align}
where $\vect{S}\in \mathbb{C}^{\tau_p\times \overline{r}}$ is an arbitrary matrix with orthonormal columns \cite{demir2022exploiting}. Substituting the optimal pilot matrix into the MSE expression in \eqref{eq:mse-rsls}, we obtain 
\begin{align}
    \text{MSE}_{\rm RS-LS}^{\star} = \frac{\overline{r}^2\sigma^2}{\tau_p p},
\end{align}
which is $(M/\overline{r})^2$ times smaller than the MSE achieved by the LS estimator in \eqref{eq:MSE-LS-minimum} when $\tau_p=M$.

\begin{figure}[t!]
	\centering 
	\begin{overpic}[trim={15mm 0mm 15mm 10mm},width=0.9\columnwidth,tics=10]{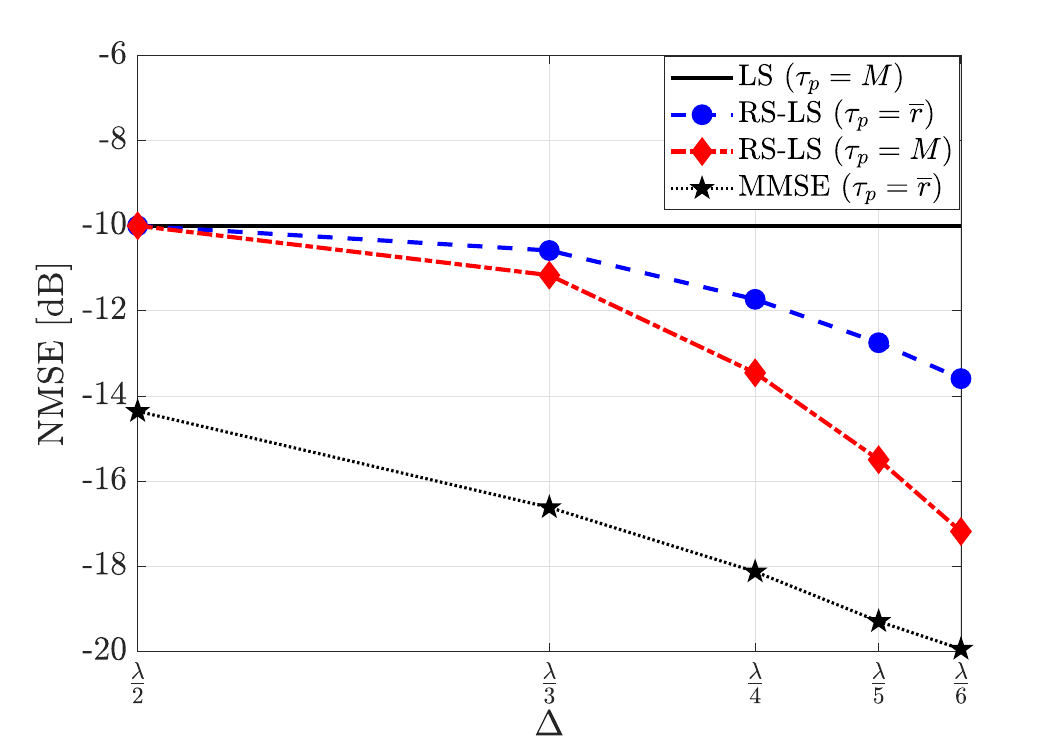}
\end{overpic} 
	\caption{The NMSE versus the antenna spacing $\Delta$ for the LS, RS-LS, and MMSE estimators in a clustered scattering environment.}
	\label{fig:LS_RSLS}
\end{figure}

Fig.~\ref{fig:LS_RSLS} shows the NMSE as a function of the antenna spacing $\Delta$ for the $8\times 8$ UPA considered previously for the same channel with clustered scattering. The LS estimator serves as a reference with $\tau_p=M$
and is unaffected by the antenna spacing, as evident from the MSE expression in \eqref{eq:MSE-LS-minimum}. By contrast, the other estimators exhibit smaller NMSE values as the antenna spacing decreases, because of the higher spatial correlation that these estimators exploit to varying degrees.

The isotropic spatial correlation matrix is utilized to construct the reduced subspace in the RS-LS estimator. Two RS-LS results are presented: one with $\tau_p=M$, and the other with a pilot length equal to the dimension of the reduced subspace. Since subspace size $\overline{r}$ decreases when $\Delta$ becomes smaller, the pilot length used by the RS-LS estimator and MMSE estimator with $\tau_p=\overline{r}$ also decreases. Despite this, the NMSE decreases as $\Delta$ reduces thanks to the increased spatial correlation.
We note that the NMSE with the RS-LS is better with $\tau_p=M$ than with $\tau_p=\overline{r}$, but the latter might still be preferable in practice since fewer pilot resources are required. 

In summary, the RS-LS estimator is a meaningful alternative to the conventional LS estimator, as both methods do not require UE-specific statistical information.

\subsection{Compressed-sensing-based channel estimation}

There is a middle ground between the MMSE estimator, which requires the complete UE-specific channel statistics, and the RS-LS estimator which only utilizes channel statistics at the UE population level.
If the BS knows that the UE channel features multipath propagation caused by only a small number of scattering clusters, this information can be utilized by the estimator. The goal is then to jointly sense the locations of the clusters and estimate their related parameters.
In such cases, channel estimation based on compressed sensing methods can be effective in achieving a good estimation quality with a relatively small pilot overhead \cite{gao2018compressive}.

\begin{figure}[t!]
	\centering 
	\begin{overpic}[trim={15mm 0mm 15mm 10mm},width=0.9\columnwidth,tics=10]{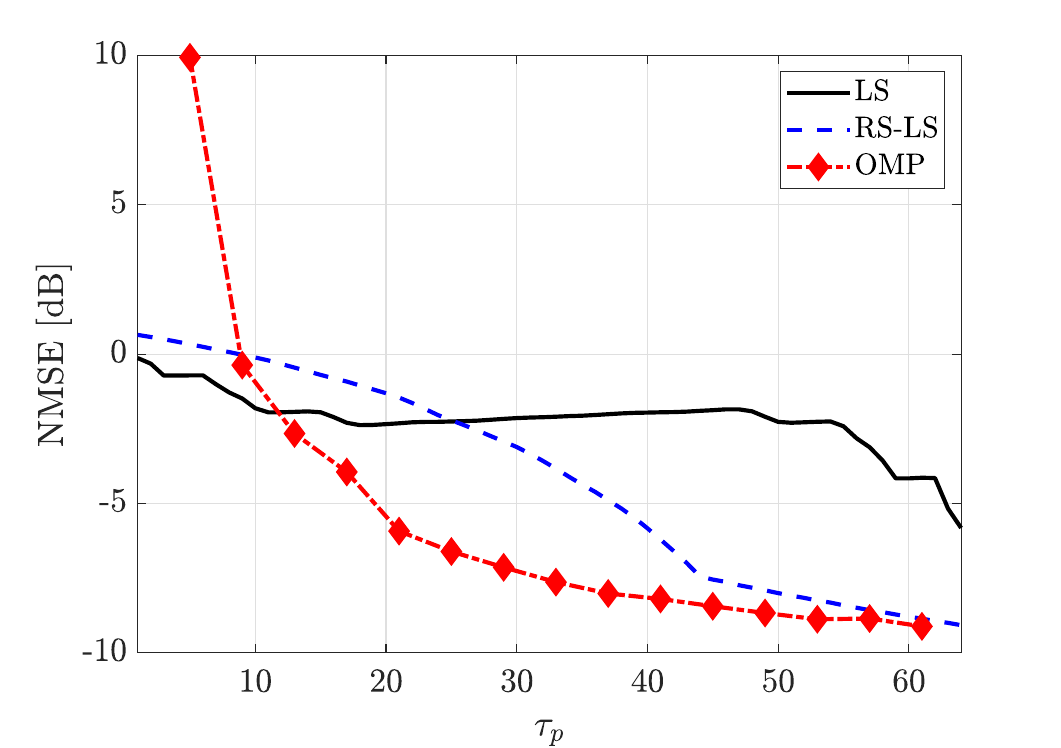}
\end{overpic} 
	\caption{The NMSE versus the pilot length ($\tau_p$) for the LS, RS-LS, and OMP-based channel estimators in a propagation environment with sparse scattering.}
	\label{fig:OMP}
\end{figure}

The first step in compressed-sensing-based channel estimation is to create a dictionary of vectors representing the channel from scattering clusters at different plausible locations. The goal is then to identify a linear combination of a small number of these dictionary vectors that results in a channel vector that resembles the one observed during the pilot transmission.
An example of this is shown in Fig.~\ref{fig:OMP}, where we consider a channel comprising $L=3$ paths from clusters located in the far-field and a UPA at the UE. The UE antennas are deployed as $8\times 8$ UPA with $\lambda/4$ antenna spacing.
Each path is represented by an array response vector, scaled by a complex channel gain. Hence, the dictionary contains many such array response vectors with uniform sampling of the plausible azimuth and elevation angles. The azimuth angular grid $\Psi=\sin(\varphi)\cos(\theta)$ and the elevation angular grid $\Omega=\sin(\theta)$ are sampled with a period of $1/40$, ensuring that all dictionary angle pairs satisfy the condition $\Psi^2+\Omega^2\leq 1$. This leads to a dictionary size of $5019$.

In this simulation, the paths are assumed equally strong on the average, and the pilot SNR is $10$\,dB. 
Multiple random channel realizations are considered for each pilot length, with azimuth and elevation angles chosen from the range $[-\frac{0.9\cdot \pi}{2},\frac{0.9\cdot\pi}{2}]$. The classical orthogonal matching pursuit (OMP) algorithm \cite{gao2018compressive} is utilized for the compressed-sensing-based estimation. 
Fig.~\ref{fig:OMP} shows the resulting estimation performance, compared with the LS and RS-LS estimators, for which all inverses are replaced by pseudo-inverses when matrix inversion issues arise. 
The results show that when $\tau_p\geq 10$, the OMP-based estimator significantly outperforms the conventional LS estimator and provides the lowest NMSE. $\tau_p=\overline{r}=44$ is the point where the RS-LS estimator has sufficient pilots to explore all the dimensions of the reduced subspace, after which there is a slight performance gap between the RS-LS and OMP algorithms. The OMP algorithm experiences a performance floor for $\tau_p> 50$, which stems from the limited dictionary size and on-grid channel estimation. 
More advanced compressed sensing-based algorithms may provide better performance, but they come with a significantly higher computational complexity (e.g., caused by increasing the dictionary size). Hence, compressed-sensing-based estimation methods are attractive to reduce the pilot length when estimating channels that are known to feature sparse scattering, but where the exact channel statistics are unknown. However, for less sparse channels or when we can afford longer pilots, the more computationally-friendly RS-LS estimator might be a better choice.

The goal of the dictionary design is that the channel path to any scattering cluster should be well represented by one of the dictionary vectors, in terms of having a large inner product (in the magnitude sense).
The simulation example considered far-field channels because we considered the channel between a single antenna of the BS and a relatively small array of the UE. In this scenario, the assumption of having a dictionary of far-field array response vectors chosen with uniform sampling of the azimuth and elevation angular domains makes good sense, because all scattering clusters will be in the far-field of the UE.
However, when considering the complete MIMO channel between the BS and UE, it becomes imperative to also account for radiative near-field effects. All the considered estimation algorithms are also applicable in this situation. When it comes to compressed sensing-based channel estimation, we must revise the dictionary design to also account for scattering clusters located in the radiative near-field of transmitter or receiver \cite{you2023near, demir2023new}. 
Recent proposals advocate for using a polar-domain grid where the dictionary vectors represent channels to points in the different angles and depths  \cite{cui2022channel, wu2023multiple, demir2023new}. 
This design is more challenging compared to the far-field case due to the new depth dimension, and the fact that the shape of a beam depends on both depth and angle. One basically needs to find a reasonably small collection of finite-depth and far-field beams that jointly cover all possible locations in the coverage area, in the sense of guaranteeing a large array gain anywhere. Once this dictionary has been designed, similar compressed sensing-based estimation algorithms can be applied as in the far-field  \cite{cui2022channel}.


\section{A Linear System Approach to Electromagnetic Theory}

\label{sec:EMI}

Highly simplified models for the function of antennas and the propagation of EM waves have taken us a long way in designing wireless communication systems, culminating in Massive MIMO---the most spectrally efficient scheme yet devised.
However, to further push the limits of the MIMO technology, we must not only consider radiative near-field effects (as earlier in this paper) but also thoroughly model the interactions among the antennas in an array. 
It seems necessary to create a close union of EM theory and communication theory. Important characteristics such as polarization must be taken into account, whose analysis is required to rigorously assess the spatial DoF \cite{yuan2021electromagnetic}.  We have to abandon the common assumptions of i.i.d. Rayleigh fading, independence of interference among receivers, and the neglect of mutual coupling \cite{sun2022characteristics}. 
None of these classic assumptions hold when considering large and densely spaced arrays.
Revising the basic system models is a daunting prospect for most communication theorists and signal processors. The very mention of EM theory evokes non-physical scalar and vector potentials, black-box finite-element simulations, complicated partial differential equations in cylindrical or spherical coordinates, and unfamiliar Bessel and Hankel functions.

In fact, tractable physics-based communication-theoretic models are obtainable quite simply from three fundamental principles \cite{Marzetta_Larsson_Hansen}:
1) Any system of antennas, operating in a fixed propagation medium, is completely characterized by an impedance matrix, which quantifies the interactions among all antennas in the system. 2) Maxwell's equations describe a linear space/time-invariant system. 3) The external EM field due to any space/time distribution of electrical currents can be exactly represented as a superposition of outgoing plane waves, both ordinary and evanescent.

In this section, we will provide a linear system approach to EM theory that is distinct from the conventional physicist's approach, which is based on vector and scalar potentials and the method of separation of variables. The linear system approach is ideally suited for the needs of the communications and signal processing communities when developing future wireless technologies.

\subsection{Impedance matrix description of antennas}

A resistor, capacitor, inductor, or an antenna is a \emph{ported device}, having a pair of wires carrying equal and opposite currents, across which is a voltage. A system of $M$ antennas is an \emph{$M$-port network}\cite{Ivrlac_Nossek}, \cite{Youla}. If operating in a linear, time-invariant medium, the relation between the $M$ voltages and the $M$ currents constitutes a linear time-invariant (LTI) system having a real-valued causal $M \times M$ matrix impulse response. The Fourier transform of the matrix impulse response is equal to the impedance matrix, which leads to the relation
\begin{equation} \label{ImpMatrix}
\mathbf{V}(\omega) = \mathbf{Z}(\omega) \mathbf{I}(\omega),
\end{equation}
where $\mathbf{V}(\omega)\in \mathbb{C}^M$ denotes the vector of voltages, $\mathbf{I}(\omega)\in \mathbb{C}^M$ denotes the vector of currents, and $\omega$ is the angular frequency. 
The diagonal elements of the impedance matrix $\mathbf{Z}(\omega)\in \mathbb{C}^{M\times M}$ are the self-impedances, while the off-diagonal elements are the mutual impedances.
A valid impedance matrix must satisfy four properties:
\begin{itemize}
    \item Conjugate-symmetry in frequency, $\mathbf{Z}(- \omega) = \mathbf{Z}^*(\omega)$;
    \item Causality: the real and imaginary parts of the $(n,m)$th entry $Z_{nm}( \omega)$ satisfy the Kramers-Kronig relations;
    \item Reciprocity: non-conjugate transpose symmetry, $\mathbf{Z}^{\Ttran}(\omega) = \mathbf{Z}(\omega)$;
    \item Conservation of energy: $\mathrm{Re} \left(\mathbf{Z}(\omega) \right)$ is nonnegative-definite.
\end{itemize}
A system comprising a set of transmit antennas and a set of receive antennas has a partitioned impedance matrix, such that
\begin{equation} \label{ImpMatrixPart}
\left[ \begin{array}{c} \mathbf{V}_{\mathrm{T}}(\omega) \\ \mathbf{V}_{\mathrm{R}}(\omega)
\end{array} \right] = 
\left[ \begin{array}{cc} \mathbf{Z}_{\mathrm{T}}(\omega) &  \mathbf{Z}_{\mathrm{TR}}(\omega) \\
\mathbf{Z}_{\mathrm{RT}}(\omega) & \mathbf{Z}_{\mathrm{R}}(\omega) \end{array} \right]
\left[ \begin{array}{c} \mathbf{I}_{\mathrm{T}}(\omega) \\ \mathbf{I}_{\mathrm{R}}(\omega)
\end{array} \right] ,
\end{equation}
where $\mathbf{Z}_{\mathrm{TR}}(\omega) = \mathbf{Z}_{\mathrm{RT}}^{\Ttran}(\omega)$ due to the reciprocity property.
The instantaneous transmitted sum-power is 
\begin{equation} \label{P_inst}
    p_{\mathrm{T}}(t) = \mathbf{i}_{\mathrm{T}}^{\Ttran}(t) \mathbf{v}_{\mathrm{T}}(t) .
\end{equation}
For a time-harmonic source, $\mathbf{i}_{\mathrm{T}}(t) = \mathrm{Re} \left( \mathbf{I}_{\mathrm{T}}(\omega) e^{- \imagunit \omega t}  \right)$ \footnote{A slight abuse of notation: elsewhere we represent the Fourier transform by $\mathbf{I}_{\mathrm{T}}(\omega)$.} and the time-average transmitted power is
\begin{align} \label{P_ave} 
\bar{P}_{\mathrm{T}} = &\frac{1}{2} \mathbf{I}_{\mathrm{T}}^{\Htran}(\omega) \mathrm{Re} \left(  \mathbf{Z}_{\mathrm{T}}(\omega) \right) \mathbf{I}_{\mathrm{T}}(\omega) \nonumber \\
&+ \frac{1}{2} \mathrm{Re} \left(  \mathbf{I}_{\mathrm{T}}^{\Htran}(\omega) \mathbf{Z}_{\mathrm{TR}}(\omega) \mathbf{I}_{\mathrm{R}}(\omega) \right).
\end{align}
If the receive currents are equal to zero (implying that open-circuit voltages are measured) or if the receive antennas are sufficiently decoupled from the transmit antennas (the typical cellular scenario) then the receive currents do not contribute to the transmitted power. In the communication literature, the power is typically computed by summing the squared magnitudes of the currents, but this is an inaccurate procedure whenever there is significant mutual coupling, i.e., the off-diagonal entries of $\mathbf{Z}_{\mathrm{T}}(\omega)$ are non-negligible.

We again stress that the impedance matrix is an \emph{exact} description of any system of antennas operating in an LTI propagation medium. In the context of wireless communication theory, the sole purpose of EM theory is to populate the entries of the impedance matrix.

\subsection{Space/Time Fourier solutions to Maxwell's equations}

We will now derive general expressions for the EM fields that can appear in communication systems, and thereby be used for carrying data. The electric and magnetic fields are obtained as solutions to Maxwell's equations. In a homogeneous and isotropic medium, Maxwell's equations become 
\begin{align} \label{Maxwelltxyz}
\nabla \times \mathbf{E}(t,\mathbf{p}) &= - \mu_0 \frac{\partial \mathbf{H}(t,\mathbf{p})}{\partial t} \nonumber \\
\nabla \times \mathbf{H}(t,\mathbf{p}) &= \epsilon_0 \frac{\partial \mathbf{E}(t,\mathbf{p})}{\partial t} + \mathbf{J}(t,\mathbf{p}) \nonumber \\
\epsilon_0 \nabla \cdot \mathbf{E}(t,\mathbf{p}) &= \rho(t,\mathbf{p}) \nonumber \\
\mu_0 \nabla \cdot \mathbf{H}(t,\mathbf{p}) &= 0 ,
\end{align}
where $\mathbf{p} = [x,y,z]^{\Ttran}$ contains the Cartesian coordinates.
The EM medium is characterized by the dielectric permittivity $\epsilon_0$ ($\frac{\text{A}\cdot\text{s}}{\text{V}\cdot \text{m}}$) and the magnetic permeability $\mu_0$ ($\frac{\text{V}\cdot\text{s}}{\text{A}\cdot \text{m}}$). The field quantities are defined as follows: the electric field intensity, $\mathbf{E}$ (V/m), the magnetic field intensity, $\mathbf{H}$ (A/m), and the electric charge density, $\rho$ ($ \text{A}\cdot \mathrm{s}/\text{m}^3$). Maxwell's equations are driven, in general, by a space/time distributed electric current density, $\mathbf{J}$ (A/m$^2$).

\subsubsection{Linear space/time-invariant system}
Maxwell's equations describe a system whose inputs are the three components of the vector-valued electric current density, and whose outputs are the six components of the vector-valued electric and magnetic fields.\footnote{The divergence of the second Maxwell equation, combined with the third, yields the charge density in terms of the current density, $\frac{ \partial \rho}{\partial t} = - \nabla \cdot \mathbf{J}$.} If the electric current density is displaced in time and space, the corresponding electric and magnetic fields are displaced in the same way. We are dealing with a linear space/time-invariant system for which there is a $6 \times 3$ space/time impulse response (called the \emph{Green's function}), $\mathbf{G}(t,\mathbf{p})$, such that
\begin{equation} \label{Green's}
    \left[ \begin{array}{c} \mathbf{E}(t,\mathbf{p}) \\  \mathbf{H}(t,\mathbf{p}) \end{array}   \right] = \mathbf{G}(t,\mathbf{p})* \mathbf{J}(t,\mathbf{p}),
\end{equation}
where $*$ denotes space/time convolution. Applying the space/time Fourier transform, $\int \int \int \int  \left( \cdot \right) e^{\imagunit \left( \omega t - \mathbf{k}^{\Ttran} \mathbf{p} \right)}d t \, d x \, d y \, d z$ to \eqref{Green's} yields\footnote{We use \emph{mixed notation}: the same symbol is used for the space/time function and its transform.}
\begin{equation}
    \left[ \begin{array}{c} \mathbf{E}(\omega,\mathbf{k}) \\  \mathbf{H}(\omega,\mathbf{k}) \end{array}   \right] =  \mathbf{G}(\omega,\mathbf{k}) \mathbf{J}(\omega,\mathbf{k}),
\end{equation}
where the convolution becomes a multiplication, and $\mathbf{k} = \left[ k_x,k_y,k_z \right]^{\Ttran}$ is the wavenumber. As previously,
\begin{equation}
\kappa = \omega \sqrt{\epsilon_0 \mu_0} = \frac{\omega}{c} = \frac{2\pi}{\lambda}.
\end{equation}

We can take the space/time Fourier transform of both sides of the four Maxwell's equations (\ref{Maxwelltxyz}) and then algebraically obtain a remarkably simple analytical solution for the electric and magnetic fields due to the source distribution:
\begin{align} \label{Maxsoln}
\left[ \begin{array}{c} \mathbf{E}(\omega,\mathbf{k}) \\ \mathbf{H}(\omega,\mathbf{k})  \end{array} \right] = \frac{1}{\mathbf{k}^{\Ttran} \mathbf{k} - \kappa^2 } \left[ \begin{array}{c} \left( \frac{\kappa^2 \mathbf{I}_3 - \mathbf{k} \mathbf{k}^{\Ttran}}{-\imagunit\omega \epsilon_0} \right) \\
\left( \imagunit\mathbf{k} \times \right)
\end{array} \right] \mathbf{J}(\omega,\mathbf{k}),
\end{align}
where
\begin{equation} \label{kcross}
    \mathbf{k} \times = \left[ \begin{array}{ccc}
    0 & -k_z & k_y \\
    k_z & 0 & -k_x \\
    -k_y & k_x & 0 \end{array} \right].
\end{equation} 

\subsubsection{Plane-wave solution}
All the action of the wave equation is embodied in the denominator polynomial of (\ref{Maxsoln}) which constitutes two simple poles in any of the three wavenumbers:
\begin{align} \label{2pole}
    \mathbf{k}^{\Ttran} \mathbf{k} - \kappa^2 &= \left( k_z - \gamma \right) \left( k_z + \gamma \right),
\end{align}
where
\begin{align}
    \gamma(\omega, k_x, k_y) &= \sqrt{ \kappa^2 - k_x^2 - k_y^2} .
\end{align}
The Sommerfeld rule for choosing the sign of the square root is that the imaginary part of the square root should be non-negative, and when the imaginary part is zero, the real part should be non-negative. Expressed as functions of $\left\{ \omega, k_x, k_y, z \right\}$, the electric and magnetic fields satisfy an ordinary second-order differential equation in $z$. Suppose that the electric current distribution is confined to the slab $|z| \leq z_0$ (e.g., $\mathbf{J}(\omega,\mathbf{p}) = \mathbf{0}, \ \forall |z| > z_0$). Then for $|z| > z_0$, the field has to satisfy the 1D homogeneous Helmholtz equation, 
\begin{equation} \label{1DHelm}
   \left[ \frac{\partial ^2}{\partial z^2} + \left( \kappa^2 - k_x^2 - k_y^2 \right) \right] \left[ \begin{array}{c} \mathbf{E}(\omega,k_x,k_y,z ) \\ \mathbf{H}(\omega, k_x, k_y, z)  \end{array} \right] = 0, 
\end{equation}
with $z$-dependence given by
\begin{equation} \label{PWsol}
     \left[ \begin{array}{c} \mathbf{E}(\omega,k_x,k_y,z ) \\ \mathbf{H}(\omega, k_x, k_y, z)  \end{array} \right] \propto e^{\imagunit \gamma(\omega, k_x, k_y) |z|  }, \ |z| > z_0 ,
\end{equation}
which implies that the field constitutes outgoing plane-waves on either side of the source distribution.

\begin{figure}[t!]
	\centering 
	\begin{overpic}[trim={5mm 0mm 15mm 10mm},width=0.8\columnwidth,tics=5]{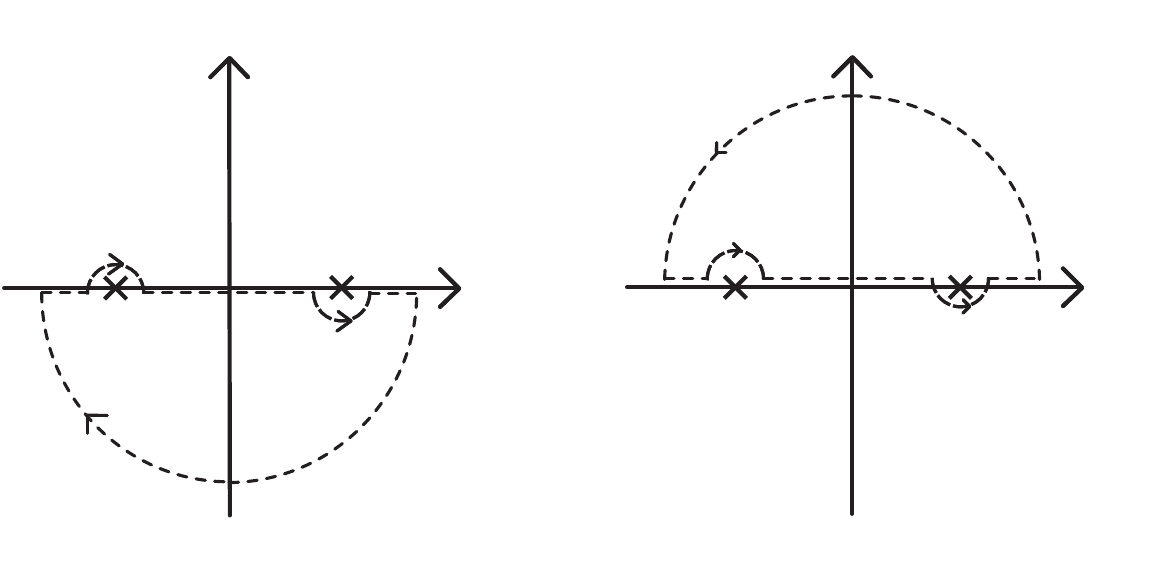}
 \put(74,50){\scriptsize $\mathrm{Im}(k_z)$}
  \put(15,50){\scriptsize $\mathrm{Im}(k_z)$}
  \put(100,25.3){\scriptsize $\mathrm{Re}(k_z)$}
    \put(41,25.3){\scriptsize $\mathrm{Re}(k_z)$}
        \put(19.5,31){\scriptsize $\gamma(\omega, k_x, k_y)$}
                \put(-5,31){\scriptsize $-\gamma(\omega, k_x, k_y)$}
\end{overpic} 
	\caption{Integration contours for plane-wave expansion of the outgoing field. a) $z< - z_0$; b) $z>z_0$.}
	\label{fig:Plane-wave_contours}
\end{figure}
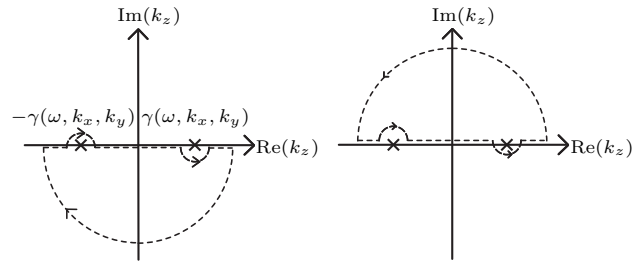

To obtain the electric and magnetic fields as functions of $\left( \omega, \mathbf{p} \right)$ in the source-free region (i.e., outside the antenna), we take the inverse wavenumber transform of (\ref{Maxsoln}). We perform the $k_z$ integral by evaluating the residues of the two poles, closing the contour in the upper-half plane for $z>z_0$, and in the lower-half plane for $z<z_0$, illustrated in Fig.~\ref{fig:Plane-wave_contours}:
\begin{align} \label{EHpw}
    & \left[ \begin{array}{c} \mathbf{E}(\omega,\mathbf{p}) \\ \mathbf{H}(\omega,\mathbf{p})  \end{array} \right]  
    = \int_{-\infty}^{\infty} \int_{-\infty}^{\infty} \frac{d k_x \, d k_y}{(2 \pi)^2} \frac{\imagunit }{2 \gamma(\omega,k_x,k_y)} \nonumber \\
     &\quad \cdot  \left. \left[ \begin{array}{c} \left( \frac{\kappa^2 \mathbf{I}_3 - \mathbf{k} \mathbf{k}^{\Ttran}}{-\imagunit\omega \epsilon_0} \right) \\
(\imagunit\mathbf{k} \times)
\end{array} \right]  \mathbf{J}(\omega,\mathbf{k}) e^{\imagunit \mathbf{k}^{\Ttran} \mathbf{p}} \right|_{k_z = \mathrm{sgn}(z) \cdot \gamma(\omega,k_x,k_y) }  , \nonumber \\
& \quad \quad |z|>z_0 .
\end{align}
This formula embodies the most important principle of wave propagation: For \emph{any} compact space-time distribution of electric current, the resulting external EM field comprises a superposition of outgoing plane-waves \cite{Hansen_Yaghjian}.
The plane-waves are of two types: \emph{ordinary} (\emph{propagating}) for $k_x^2 + k_y^2 < \kappa^2$ where $k_z$ is positive-real and \emph{evanescent} (\emph{inhomogeneous}) for $k_x^2 + k_y^2 > \kappa^2$ where $k_z$ is positive-imaginary and the wave decays exponentially fast in $|z|$. The evanescent waves carry only reactive power in the $z$-direction.
The difference between these types was previously illustrated in Fig.~\ref{fig:wavenumber_support}, where we concluded that only the ordinary type provides spatial DoF that can be used to carry data to the radiative near-field and far-field of the source.

Every $\left( k_x,k_y \right)$ represents a plane-wave 
$e^{\imagunit \left( k_x x + k_y y \pm \gamma z \right)}$ propagating in the $\pm z$ directions. For each of these plane-waves, the wavenumber vector, $\mathbf{k}$, and the electric and magnetic fields are mutually orthogonal.\footnote{Here we mean ``complex-orthogonal'', i.e., $\mathbf{a}^{\Ttran} \mathbf{b} = 0$, not $\mathbf{a}^{\Htran} \mathbf{b} = 0$. }

The external outgoing field depends only on $\mathbf{J}(\omega,k_x,k_y, \pm \gamma(\omega,k_x,k_y))$, implying that more than one current source distribution can generate the \emph{same} external field. It is the variability in the external field that can be used to carry data, and thereby determine the spatial DoF.

\subsubsection{The plane-wave representation of the spherical wave}

Consider the Helmholtz equation (i.e., the temporal Fourier transform of the wave equation), driven by a spatial impulse,
\begin{equation} \label{Fequ}
    \left( \nabla^2 + \kappa^2 \right) F(\omega,\mathbf{p}) = -4 \pi \delta (x) \delta(y) \delta(z) .
\end{equation}
It is not yet obvious, but the solution is the spherical wave, $F(\omega,\mathbf{p}) =\frac{e^{\imagunit\kappa |\mathbf{p}|}}{|\mathbf{p}|}$. As before, we take spatial Fourier transforms of both sides of (\ref{Fequ}), algebraically solve for $F(\omega,\mathbf{k})$, and then take inverse wavenumber transforms to obtain the plane-wave representation called the \emph{Weyl integral} \cite{Weyl}:
\begin{align} \label{Weyl}
    F(\omega,\mathbf{p}) &= \int_{- \infty}^{\infty} \int_{- \infty}^{\infty} \frac{d k_x \, d k_y}{(2 \pi)^2} \frac{\imagunit 2 \pi}{\gamma(\omega,k_x,k_y)} \nonumber \\
    & \quad \cdot e^{\imagunit \left( k_x x + k_y y + \gamma(\omega,k_x,k_y) |z|   \right)} .
\end{align}
To show that (\ref{Weyl}) is indeed representing a spherical wave,  we switch to cylindrical coordinates, and use the fact that (\ref{Fequ}) is spherically symmetric, implying that  $F(\omega,\mathbf{p})$ is also spherically symmetric:
\begin{align}
F(\omega,\mathbf{p}) &= F(\omega,0,0,|\mathbf{p}|) \nonumber \\
&= \int_0^{\infty} \int_0^{2 \pi} \frac{ k_r d k_r \, d \phi}{2 \pi} \frac{\imagunit }{\sqrt{\kappa^2 - k_r^2}} \cdot e^{\imagunit  \sqrt{\kappa^2 - k_r^2} |\mathbf{p}| } \nonumber \\
&=  \left. -  \frac{e^{\imagunit  \sqrt{\kappa^2 - k_r^2} |\mathbf{p}| } }{|\mathbf{p}|}\right|_0^{\infty} = \frac{e^{\imagunit \kappa |\mathbf{p}| } }{|\mathbf{p}|}, 
\end{align}
where the choice of integration contour is an exercise in the application of the Cauchy integral theorem: the contour cannot cross the branch cut associated with the square-root singularity.
Hence, we have established that the spherical wave $\frac{e^{\imagunit \kappa |\mathbf{p}| } }{|\mathbf{p}|}$ can be expanded as the integral in \eqref{Weyl} over all plane-waves, both propagating and evanescent.

It is well-known that many wavelengths away, a spherical wave looks like a plane-wave \emph{locally}. When the observer is sufficiently small to observe a plane-wave, we say that it is in the far-field.
It is much less obvious that the spherical wave can exactly be represented by a superposition of plane-waves. The local behavior of the spherical wave in the far-field can be inferred by application of the method of stationary phase: for $|\mathbf{p}| \gg \lambda$, the phase of the integrand of (\ref{Weyl}) oscillates violently, and the only significant contribution to the integral occurs in the vicinity of $\mathbf{k} = \kappa \mathbf{p}/|\mathbf{p}|$.

We note that (\ref{Maxsoln}) is a product, in the frequency/wavenumber domain, of three terms, equivalent to convolutions in the space/time domain. The first two terms comprise the Green's function
\begin{equation} \label{Greens_k}
    \mathbf{G}(\omega,\mathbf{k}) =  \frac{1}{\mathbf{k}^{\Ttran} \mathbf{k} - \kappa^2 } \cdot \left[ \begin{array}{c} \left( \frac{\kappa^2 \mathbf{I}_3 - \mathbf{k} \mathbf{k}^{\Ttran}}{-\imagunit\omega \epsilon_0} \right) \\
\left( \imagunit\mathbf{k} \times \right)
\end{array} \right].
\end{equation}
Given the Fourier transform relation, $\frac{1}{\mathbf{k}^{\Ttran} \mathbf{k} - \kappa^2 } \leftrightarrow \frac{e^{\imagunit \kappa |\mathbf{p}| } }{4 \pi |\mathbf{p}|}$, we directly obtain the Green's function in the space/frequency domain as 
\begin{equation} \label{Gpart}
    \mathbf{G}(\omega,\mathbf{p}) = \left[ \begin{array}{c} \mathbf{G}_E(\omega,\mathbf{p}) \\ \mathbf{G}_H(\omega,\mathbf{p})
    \end{array} \right] ,
\end{equation}
where
\begin{align}
 &\mathbf{G}_E(\omega,\mathbf{p}) = \frac{1}{-\imagunit \omega \epsilon_0} \left( \kappa^2 \mathbf{I}_3 + \nabla \nabla^{\Ttran}  \right) \frac{e^{\imagunit \kappa |\mathbf{p}| } }{4 \pi |\mathbf{p}|} \nonumber \\
    & = \frac{1}{-\imagunit \omega \epsilon_0 } \left[ \begin{array}{ccc}  
\frac{\partial^2}{\partial x^2} + \kappa^2 & \frac{\partial^2}{\partial x \partial y} & \frac{\partial^2}{\partial x \partial z} \\
\frac{\partial^2}{\partial y \partial x} & \frac{\partial^2}{\partial y^2} + \kappa^2 & \frac{\partial^2}{\partial y \partial z} \\
\frac{\partial^2}{\partial z \partial x} & \frac{\partial^2}{\partial z \partial y} & \frac{\partial^2}{\partial z^2} + \kappa^2
\end{array} \right] \frac{e^{\imagunit \kappa |\mathbf{p}| } }{4 \pi |\mathbf{p}|} , \nonumber
\end{align}
and
\begin{align} \label{G_H}
     &\mathbf{G}_H(\omega,\mathbf{p}) = \left[ \nabla \times \right] \frac{e^{\imagunit \kappa |\mathbf{p}| } }{4 \pi |\mathbf{p}|}  = \left[ \begin{array}{ccc}  
0 & - \frac{\partial}{\partial z} & \frac{\partial}{\partial y} \\
\frac{\partial}{\partial z} & 0 & - \frac{\partial}{\partial x} \\
- \frac{\partial}{\partial y} & \frac{\partial}{\partial x} & 0
\end{array}  \right] \frac{e^{\imagunit \kappa |\mathbf{p}| } }{4 \pi |\mathbf{p}|} . \nonumber
\end{align}
For a particular electric current distribution, $\mathbf{J}(\omega,\mathbf{p})$, we can perform a spatial convolution with the Green's function to obtain the electric and magnitude fields as superpositions of spherical waves and their first and second spatial derivatives.

We have now described two solution techniques for Maxwell's equations. The first one in (\ref{EHpw}) represents the electric and magnetic fields as a superposition of plane-waves. The second one in (\ref{Green's}) and (\ref{Gpart}) represents the fields as a superposition of spherical waves. The spherical-wave representation is useful for computing antenna self-impedances. In general, the plane-wave representation is vastly superior. Firstly, the outgoing fields can be efficiently computed via 2D discrete Fourier transform (DFTs). Secondly, plane-waves, unlike spherical waves, are origin-free so we must only characterize their respective strength.
Thirdly, there is a particularly simple description of the generation of plane-waves by Cartesian-grid antenna arrays.
Finally, propagation in a horizontally-stratified medium can be solved by expanding the source distribution in plane-waves, propagating the constituent plane-waves through the parallel layers of the medium, and finally integrating over horizontal wavenumber to obtain the EM field within each layer.

\subsection{Mutual- and self-impedance}

The creation of an electric current distribution entails the exertion of power (both real and reactive) to drive the current against an electric field which itself arises from the combination of the same current distribution and other current distributions. The computation of this power yields either the self-impedance of an antenna or the mutual impedance between two antennas. The instantaneous power associated with the interaction of a current density and an electric field is
\begin{equation} \label{powerEJ}
    p(t) = - \int  \mathbf{J}^{\Ttran} (t,\mathbf{p}) \mathbf{E}(t,\mathbf{p}) d \mathbf{p} .
\end{equation}
An antenna, located at the position $\mathbf{p}_1$, has the associated electric current density
\begin{equation} \label{Jant}
    \mathbf{J}_1 (t,\mathbf{p}) = \mathrm{Re} \left( I_1(\omega) \mathbf{s}(\mathbf{p}- \mathbf{p}_1) e^{-\imagunit \omega t} \right),
\end{equation}
where $\mathbf{s}(\mathbf{p})$ is a real-valued function of space that describes the shape of the antenna's current distribution. The convolution of the current distribution with the Green's function is equal to an electric field, $\mathbf{E}_1(t,\mathbf{p})$. A second antenna, located at the position $\mathbf{p}_2$, creates a current distribution, $\mathbf{J}_2(t,\mathbf{p})$, with an expenditure of instantaneous power due to the mutual coupling, $p_{21}(t) = - \int \mathbf{J}_2^{\Ttran} (t,\mathbf{p}) \mathbf{E}_1(t,\mathbf{p}) d \mathbf{p}$. This integral yields the mutual impedance between the two antennas:
\begin{align} \label{mut_imp_p}
  &  Z(\omega,\mathbf{p}_1 - \mathbf{p}_2) \nonumber \\
   & = - \int \int  \mathbf{s}^{\Ttran}(\mathbf{p}) \mathbf{G}_E (\omega,\mathbf{p} - \acute{\mathbf{p}} + \mathbf{p}_1 - \mathbf{p}_2 ) \mathbf{s}(\acute{\mathbf{p}})  d \mathbf{p} \, d \acute{\mathbf{p}}.
\end{align}
For example, an incremental vertical electric dipole of length $L_0$ is characterized by $\mathbf{s}(\mathbf{p}) = L_0 \mathbf{e}_z \delta(x) \delta(y) \delta(z)$, and the mutual impedance is
\begin{align} \label{Edipole}
    Z(\omega,\mathbf{p}_1 - \mathbf{p}_2) &= - L_0^2 \mathbf{e}_z^{\Ttran} \mathbf{G}_E (\omega,\mathbf{p}_1 - \mathbf{p}_2 ) \mathbf{e}_z \nonumber \\
    &=  \frac{L_0^2}{\imagunit \omega \epsilon_0} \left[ \frac{\partial^2}{\partial z^2} + \kappa^2 \right] \frac{e^{\imagunit \kappa |\mathbf{p}| } }{4 \pi |\mathbf{p}|} .
\end{align}
For $\mathbf{p}_1 = \mathbf{p}_2$, (\ref{Edipole}) yields the self-impedance; the imaginary part is infinite, but the real part (that figures in the computation of transmitted power) is finite.

An incremental current loop (magnetic dipole) with area $A_0$, oriented in the $z$-direction, is characterized by $\mathbf{s}(\mathbf{p}) = A_0 \delta(x) \delta(y) \delta(z) \left[ \frac{\partial}{\partial y} \, -\frac{\partial}{\partial x} \ \, 0 \right]^{\Ttran}$,
for which the mutual impedance is
\begin{equation} \label{Hdipole}
    Z(\omega,\mathbf{p}_1 - \mathbf{p}_2) = \frac{A_0^2}{\imagunit \omega \epsilon_0} \left[ \frac{\partial^2}{\partial x^2} + \frac{\partial^2}{\partial y^2} \right] \frac{e^{\imagunit \kappa |\mathbf{p}| } }{4 \pi |\mathbf{p}|} .
\end{equation}

\begin{figure}[t!]
	\centering 
	\begin{overpic}[trim={5mm 0mm 15mm 10mm},width=0.8\columnwidth,tics=10]{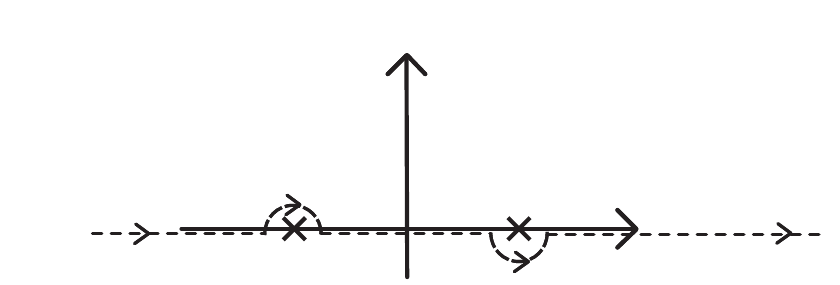}
  \put(47,34.5){\small $\mathrm{Im}(k_z)$}
  \put(88,10){\small $\mathrm{Re}(k_z)$}
        \put(58,13.5){\small $\gamma(\omega, k_x, k_y)$}
                \put(21.5,13.5){\small $-\gamma(\omega, k_x, k_y)$}
        \end{overpic} 
	\caption{The only contribution to the real part of the self-impedance are the half-residues associated with the poles for $k_x^2 + k_y^2 < \kappa^2$.}
	\label{fig:Real_Self_Imped_contour}
\end{figure}
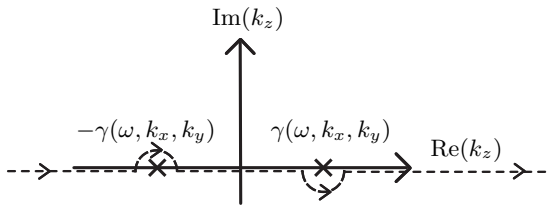

The expression for the mutual impedance in (\ref{mut_imp_p}) can be written in the wavenumber domain as 
\begin{equation} \label{mut_imp_k}
    Z(\omega,\mathbf{p}) = \int \frac{d \mathbf{k}}{(2 \pi)^3} \mathbf{S}^{\Htran}(\mathbf{k}) \mathbf{G}_E(\omega,\mathbf{k}) \mathbf{S}(\mathbf{k}) e^{\imagunit  \mathbf{k}^{\Htran} \mathbf{p}} .
\end{equation}
For $|\mathbf{p}|$ greater than the diameter of the antenna, we can extract the two residues to obtain the plane-wave representation, which can be computed via 2D DFT, as
\begin{align} \label{mut_inf_res}
    Z(\omega,\mathbf{p}) &= \int \int \frac{d k_x \, d k_y}{(2 \pi)^2} \frac{1}{2 \gamma} \mathbf{S}^{\Htran}(\mathbf{k}) \mathbf{G}_E(\omega,\mathbf{k}) \nonumber \\
    & \cdot  \mathbf{S}(\mathbf{k}) \left. e^{\imagunit \left( k_x x + k_y y + k_z |z| \right)} \right|_{k_z = \gamma(\omega,k_x,k_y)} .
\end{align}
For $\mathbf{p}= \mathbf{0}$, (\ref{mut_imp_k}) appears to give a purely imaginary-valued self-impedance, because $\mathbf{G}_E$ (\ref{Greens_k}) is imaginary-valued for real-valued $\mathbf{k}$. In fact, this does not happen, because the $k_z$ contour has to be indented; the sole contributions to the real part of self-impedance are the half-residues associated with the propagating
plane-waves as shown in Fig.~\ref{fig:Real_Self_Imped_contour}, which becomes
\begin{align} \label{self_imp_real}
    \mathrm{Re} \left( Z(\omega,\mathbf{0}) \right) &= \underset{k_x^2 + k_y^2 < \kappa^2}{\int \int} \frac{d k_x \, d k_y}{(2 \pi)^2} \nonumber \\
    &\cdot \left. \frac{\mathbf{S}^{\Htran}(\mathbf{k}) \left( \kappa^2 \mathbf{I}_3 - \mathbf{k} \mathbf{k}^{\Ttran} \right)\mathbf{S}(\mathbf{k})  }{2 \omega \epsilon_0 \gamma} \right|_{k_z = \gamma(\omega,k_x,k_y)} .
\end{align}

\subsection{Applications of EM theory to communications}
\label{subsec:EMtheory_applications}

By utilizing the exact EM models, we can gain insights into the design of UM-MIMO communication systems. We will revisit the DoF concept and its connection to polarization, take a closer look at the evanescent waves, and finally discuss how to model thermal noise.

\subsubsection{Polarization and degrees-of-freedom}
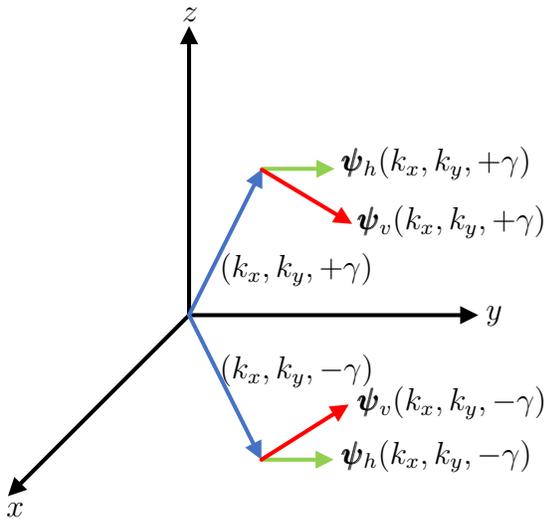
\begin{figure}[t!]
	\centering 
	\begin{overpic}[trim={5mm 0mm 15mm 10mm},width=0.8\columnwidth,tics=10]{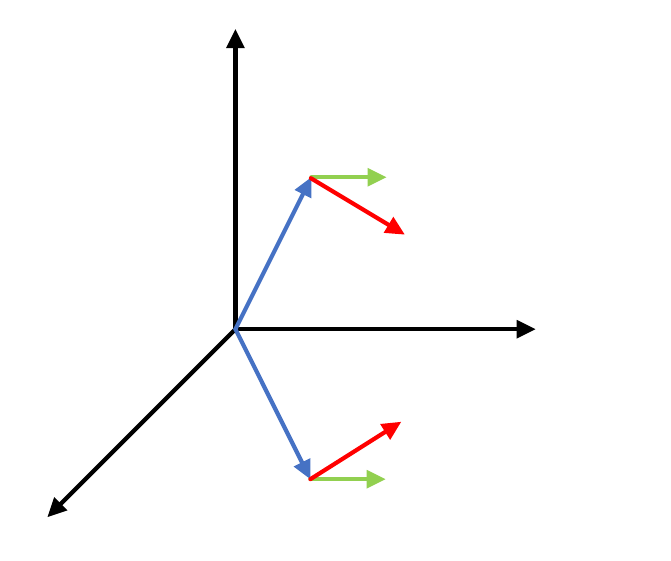}
  \put(36,97){\large $z$}
  \put(93,41){\large $y$}
    \put(3,4){\large $x$}
        \put(43,49){\large $(k_x, k_y,+\gamma)$}
                \put(43,30){\large $(k_x, k_y,-\gamma)$}
                      \put(66,69){\large $\pmb{\psi}_h(k_x, k_y,+\gamma)$}
                        \put(69,58){\large $\pmb{\psi}_v(k_x, k_y,+\gamma)$}
 \put(66,14){\large $\pmb{\psi}_h(k_x, k_y,-\gamma)$}
                        \put(69,24){\large $\pmb{\psi}_v(k_x, k_y,-\gamma)$}\end{overpic} 
	\caption{For every pair of horizontal wavenumbers $(k_x,k_y)$ there are four linearly independent plane-waves: in either the plus- or minus-$z$ directions an $h$-wave whose $E$-field lies in the horizontal plane, $\pmb{\psi}_h$, and a $v$-wave whose $E$-field lies in a vertical plane, $\pmb{\psi}_v$.}
	\label{fig:Polarization_Coordinates}
\end{figure}

The plane-wave expansion in (\ref{EHpw}) implies that, for every $( \omega, k_x, k_y )$, there are two plane-waves having wavenumber vector $\mathbf{k} = \left[ k_x \, k_y \, \, \pm \gamma  \right]^{\Ttran}$. Recall that the wavenumber vector and the electric and magnetic field vectors are mutually orthogonal. Consequently, for each of the $\pm z$-waves, the electric and magnetic field vectors are confined to a two-dimensional subspace. One way to characterize the subspace is a set of three mutually orthogonal unit vectors, which form a unitary matrix 
\begin{align} \label{PSI}
\mathbf{\Psi} (\mathbf{k}) &= \left[ \begin{array}{ccc} \frac{\mathbf{k}}{\kappa} & \pmb{\psi}_v &  \pmb{\psi}_h  \end{array} \right] \nonumber \\
&= \left[ \begin{array}{ccc} +\frac{k_x}{\kappa} & +\frac{k_{x} k_{z}}{k_{r} \kappa} &   -\frac{k_{y}}{k_{r}} \\
+\frac{k_{y}}{\kappa} & +\frac{k_y k_{z}}{k_{r} \kappa}  &    +\frac{k_{x}}{k_{\mathrm{r}}} \\
+\frac{k_{z}}{\kappa} & -\frac{k_{\mathrm{r}}}{\kappa} &  0
\end{array} \right] ,
\end{align}
where $k_r = \sqrt{k_{x}^2 + k_{y}^2}$ and $k_z = \pm \sqrt{\kappa^2 - k_r^2}$.
The first column vector is the normalized wavenumber vector, the second lies in a vertical plane, and the third lies in the horizontal, $(x,y)$, plane. We can re-write the plane-wave representation in (\ref{EHpw}) in terms of horizontally and vertically polarized plane-waves by projecting both the current distribution and the $E$ and $H$ fields onto the orthogonal unit vectors (note that $\mathbf{E}$ and $Z_0 \mathbf{H}$ have the same physical units, where $Z_0 = \sqrt{\frac{\mu_0}{\epsilon_0}}$ represents the characteristic impedance, thereby  simplifying the expression):
\begin{align} \label{EHpolar}
   & \left[ \begin{array}{c} \mathbf{E}(\omega,\mathbf{p}) \\ Z_0 \mathbf{H}(\omega,\mathbf{p})  \end{array} \right] = \int_{-\infty}^{\infty} \int_{-\infty}^{\infty} \frac{d k_x \, d k_y}{(2 \pi)^2} \frac{ - \mu_0 }{2 \gamma} \nonumber  e^{\imagunit \left( k_x x + k_y y + \gamma |z| \right)} \nonumber \\
    & \cdot \left\{ \begin{array}{l} 
    \left. \left( A_h^+ \left[ \begin{array}{c} \pmb{\psi}_h \\ - \pmb{\psi}_v \end{array} \right] + A_v^+ \left[ \begin{array}{c} \pmb{\psi}_v \\  \pmb{\psi}_h \end{array} \right] \right) \right|_{k_z = + \gamma}, \ z>+z_0 \\
   \left. \left( A_h^- \left[ \begin{array}{c} \pmb{\psi}_h \\ - \pmb{\psi}_v \end{array} \right] + A_v^- \left[ \begin{array}{c} \pmb{\psi}_v \\  \pmb{\psi}_h \end{array} \right] \right) \right|_{k_z = - \gamma}, \ z<-z_0
    \end{array} \right. ,
\end{align}
where the \emph{polarization amplitudes}, $A_h^+, A_v^+, A_h^-, A_v^-$, are related to the current density by
\begin{equation} \label{Ahvpm}
\left[ \begin{array}{c} 
A_h^{\pm}(\omega,k_x,k_y) \\ A_v^{\pm}(\omega,k_x,k_y)
\end{array} \right] = \left( \left[ \begin{array}{c}
\pmb{\psi}_h^{\Ttran}(\mathbf{k}) \\ \pmb{\psi}_v^{\Ttran}(\mathbf{k})
\end{array} \right] \mathbf{J}(\omega,\mathbf{k}) \right) _{k_z = \pm \gamma} .
\end{equation}
The polarization coordinate system is illustrated in Fig.~\ref{fig:Polarization_Coordinates}.

When spatial DoF were discussed in Section~\ref{sec:spat_deg_freedom}, we only considered a single polarization. We will now revisit this concept by considering the extent to which an array of antennas can control the spectrum of polarimetric plane-waves.
Consider a $N \times N$ UPA in the $(x,y)$-plane with antenna spacing $\Delta$. The resulting current density is
\begin{align} \label{Jarray}
    & \mathbf{J} (\omega,\mathbf{p}) = \sum_{n_x=0}^{N-1} \sum_{n_y=0}^{N-1} I_{n_x,n_y}(\omega) \nonumber \\
   & \cdot\mathbf{s} \left( x - \Delta \left[ n_x - \frac{N-1}{2} \right],y - \Delta \left[ n_y - \frac{N-1}{2} \right],z \right) .
\end{align}
The currents that drive the antennas are made equal to an inverse 2D DFT:
\begin{align} \label{IinvDFT}
    I_{n_x,n_y}(\omega) &= \frac{1}{N} \sum_{m_x=0}^{N-1} \sum_{m_y=0}^{N-1} \hat{I}_{m_x,m_y}(\omega) e^{- \frac{\imagunit \pi (N-1) (m_x + m_y)}{N}} \nonumber \\ &\cdot e^{\frac{\imagunit 2 \pi \left(m_x n_x + m_y n_y  \right)}{N}} .\end{align}
We substitute (\ref{IinvDFT}) into (\ref{Jarray}), and then take the wavenumber Fourier transforms to obtain
\begin{align} \label{Jkarray}
     \mathbf{J}(\omega,\mathbf{k}) &= \frac{ \mathbf{S}(\mathbf{k})}{N} \sum_{m_x=0}^{N-1} \sum_{m_y=0}^{N-1} \hat{I}_{m_x,m_y}(\omega) \nonumber \\
    & \quad \cdot \frac{\sin \left[  \frac{(N-1) \Delta}{2} \cdot \left( k_x - \frac{2 \pi m_x}{{N \Delta}} \right) \right]} 
    { \sin \left[ \frac{\Delta}{2} \cdot \left( k_x - \frac{2 \pi m_x}{{N \Delta}} \right) \right] } \nonumber \\
    & \quad \cdot \frac{\sin \left[  \frac{(N-1) \Delta}{2} \cdot \left( k_y - \frac{2 \pi m_y}{{N \Delta}} \right) \right]} 
    { \sin \left[  \frac{\Delta}{2} \cdot \left( k_y - \frac{2 \pi m_y}{{N \Delta}} \right) \right] } .
    \end{align}
  We note that these Dirichlet kernel functions are periodic in the wavenumber, with a period equal to $\frac{2 \pi}{\Delta}$. Furthermore, the $N^2$ Dirichlet-products are mutually orthogonal as well. To avoid aliasing the propagating plane-waves, the spacing must be no more than half of a wavelength, so $\Delta \leq \frac{\lambda}{2}$.

  Associated with each DFT current coefficient, $\hat{I}_{m_x,m_y}(\omega)$ 
  is a bundle of plane-waves, centered at $\left( k_x, k_y \right) = \left( \frac{2 \pi m_x}{N \Delta},\frac{2 \pi m_y}{N \Delta} \right)$, occupying a square region $\frac{2 \pi}{N \Delta}$ on a side.\footnote{When the range is comparable to the size of the array, the Dirichlet kernels behave, in the distributional sense, as Dirac delta functions, and every DFT current coefficient is associated with a single discrete plane-wave.} 
  For sufficiently large $| \mathbf{p} |$ (i.e., in the far-field), the dominant contribution to the field is a single DFT coefficient such that $\left( \frac{x}{| \mathbf{p} |},  \frac{y}{| \mathbf{p} |}\right) = \left( \frac{2 \pi m_x}{\kappa N \Delta} , \frac{2 \pi m_y}{\kappa N \Delta} \right)$.
  
  Recall that the propagating plane-waves correspond to $k_x^2 + k_y^2 \leq \kappa^2$, or $m_x^2 + m_y^2 \leq \left(  \frac{N \Delta}{\lambda} \right)^2$, while the others are evanescent. The number of propagating plane-waves is approximately $\pi\left(  {N \Delta}/{\lambda} \right)^2$, which was stated in \eqref{eq:sampling_theorem_space_LHLV} as the spatial DoF of a UPA.

The preceding calculation has profound implications for planar and volumetric arrays. Firstly, the activity of halving the spacing between the antennas while simultaneously quadrupling the number of antennas does not enable the UPA to create any additional propagating plane-waves. A UPA with $\lambda/2$-spacing is sufficient to control all spatial DoF, in the sense of giving full control of all possible propagating plane-waves, subject to the wavenumber resolution of the array. This is consistent with the discussion in Section~\ref{subsec:MIMO-DOF}.

Secondly, although the considered UPA can create $\pi\left(  \frac{N \Delta}{\lambda} \right)^2$ propagating plane-waves, there is only one DoF per plane wave so we cannot distinguish between the four waves sharing the same horizontal wavenumbers $k_x,k_y$ (i.e., $\pm z$, and the two polarizations).
A different kind of array design is required to obtain four DoF per plane-wave, thereby quadrupling the total spatial DoF. Two possible designs are:
\begin{itemize}
\item Four parallel planar arrays, separated in $z$ by at least $\lambda/2$, with two arrays employing, for example, vertical electric dipoles, and two arrays employing vertical magnetic dipoles.
        
\item A single polarimetric planar array that simultaneously employs, for example, $x$- and $y$-electric dipoles, and $x$- and $y$-magnetic dipoles. Not all combinations of the six types of antennas are admissible. For example, a vertical magnetic dipole combined with two horizontal electric dipoles is linearly redundant in view of the identity $\mathbf{H} \propto \nabla \times \mathbf{E}$ \cite{Marzetta2002}.
\end{itemize}
Except as noted above, the expansion of a planar array into a volume array does not yield additional DoF.

Antenna polarization features have been exploited in wireless communication systems for many years \cite{ LiEtAlQuasiOrthogonalSpaceTime2012,KimEtAlLimitedFeedbackBeamformingSystems2010}, and even predate spatial multiplexing \cite{amitay1984linear}.

\subsubsection{Are evanescent waves of any use?}

Conventionally, DoFs are counted as the number of linearly independent propagating plane-waves that can be created. This excludes evanescent waves, based on the fact that they decay exponentially fast in the $z$-direction, and carry only reactive power in the $z$-direction. 
There are two scenarios, however, which may contradict this popular notion.

The first scenario is an extreme near-field operation (i.e., in the reactive near-field region), where the evanescent wave could be a significant component of the EM field. Just as an increased array aperture extends the Fraunhofer distance so that many practical communication situations can take place in the radiative-near-field, it also expands the reactive near-field so it can be used for some short-range systems.

The second scenario builds on super-directivity. A super-directive array (for example a planar $xy$-array) has sub-wavelength spacing \cite{Schelkunoff1943},\cite{Uzkof1946}, deliberately to create strong mutual coupling. The evanescent waves decay exponentially in the $z$-direction, but not in the $x-y$-directions \cite{Marzetta2019Asilomar}. Moreover, the evanescent waves can carry real power in the $x-y$ directions. A linear array of $M$ antennas (say, along the $x$-axis), operating in end-fire mode, has a limiting gain of $M^2$ as $\Delta \rightarrow 0$, compared with the gain of $M$ for $\Delta = \lambda/2$. The super-directive array is an old concept but has never been realized on a large scale because the antennas have to be driven by numerically large currents which create unacceptable $I^2 R$ losses (unless super-conductive antennas are used), and an enormous local reactive field.

\subsubsection{Dense scattering propagation}
    The plane-wave expansion turns out to be the ideal theoretical tool for handling scattering propagation \cite{Marzetta2002,Pizzo_Marzetta2020}. Consider a transmit array and a receive array embedded in a scattering environment. As before, we can expand the transmitted field into outgoing plane-waves, characterized for every horizontal wavenumber pair, $(k_{\mathrm{T}x},k_{\mathrm{T}y})$, by four polarization amplitudes, $\left\{A_{\mathrm{T}h}^\pm(\omega,k_{\mathrm{T}x},k_{\mathrm{T}y}) , A_{\mathrm{T}v}^{\pm}(\omega,k_{\mathrm{T}x},k_{\mathrm{T}y}) \right\}$. The transmit plane-waves interact with the scatterers to produce the incoming field that is measured by the receive array, which itself comprises a superposition of plane-waves. The received plane-waves are characterized by their own polarization amplitudes, $\left\{A_{\mathrm{R}h}^\pm(\omega,k_{\mathrm{R}x},k_{\mathrm{R}y}) , A_{\mathrm{R}v}^{\pm}(\omega,k_{\mathrm{R}x},k_{\mathrm{R}y}) \right\}$. The most general linear relation between the transmitted field and the received field is through the action of a $4 \times 4$ scattering kernel, $\mathbf{K} (\omega,k_{\mathrm{R}x},k_{\mathrm{R}y},k_{\mathrm{T}x},k_{\mathrm{T}y})$,
    \begin{align} \label{scatK}
        &\mathbf{A}_\mathrm{R}(\omega,k_{\mathrm{R}x},k_{\mathrm{R}y}) = \int_{-\infty}^{\infty}\int_{-\infty}^{\infty} d k_{\mathrm{T}x} \,d k_{\mathrm{T}y} \nonumber \\
        &\mathbf{K} (\omega,k_{\mathrm{R}x},k_{\mathrm{R}y},k_{\mathrm{T}x},k_{\mathrm{T}y}) \mathbf{A}_\mathrm{T}(\omega,k_{\mathrm{T}x},k_{\mathrm{T}y}) , 
        \end{align}
where
        \begin{align}
        &\mathbf{A}_\mathrm{T}(\omega,k_{\mathrm{T}x},k_{\mathrm{T}y})= \left[ \begin{array}{c}
        A_{\mathrm{T}h}^+ (\omega,k_{\mathrm{T}x},k_{\mathrm{T}y}) \\
        A_{\mathrm{T}v}^+ (\omega,k_{\mathrm{T}x},k_{\mathrm{T}y}) \\
        A_{\mathrm{T}h}^- (\omega,k_{\mathrm{T}x},k_{\mathrm{T}y}) \\
        A_{\mathrm{T}v}^- (\omega,k_{\mathrm{T}x},k_{\mathrm{T}y})
        \end{array} \right], \nonumber \\
        &\mathbf{A}_\mathrm{R}(\omega,k_{\mathrm{R}x},k_{\mathrm{R}y})= \left[ \begin{array}{c}
        A_{\mathrm{R}h}^+ (\omega,k_{\mathrm{R}x},k_{\mathrm{R}y}) \\
        A_{\mathrm{R}v}^+ (\omega,k_{\mathrm{R}x},k_{\mathrm{R}y}) \\
        A_{\mathrm{R}h}^- (\omega,k_{\mathrm{R}x},k_{\mathrm{R}y}) \\
        A_{\mathrm{R}v}^- (\omega,k_{\mathrm{R}x},k_{\mathrm{R}y})
        \end{array} \right] .
    \end{align}
    This formulation is both physically and mathematically exact.

    Particular assumptions concerning the scattering kernel result in a spatially-stationary stochastic model for the propagation. First, assume that neither transmitted nor received evanescent waves contribute significantly to the propagation, i.e., the scattering kernel vanishes if either $k_{\mathrm{T}x}^2 + k_{\mathrm{T}y}^2 > \kappa^2$ or $k_{\mathrm{R}x}^2 + k_{\mathrm{R}y}^2 > \kappa^2$. Second, assume that the sixteen elements of the scattering kernel are complex Gaussian distributed, and independent among themselves and independent over the four wavenumbers. Then the Green's function for the propagation is spatially stationary in both transmit and receive coordinates. This is the generalization of the spectral representation for a stationary Gaussian random process to spatial random fields. This formulation results in the most general physically tenable (e.g., satisfying Maxwell's equations in free-space) spatially-stationary stochastic model for propagation.

    Isotropic scattering results in a spatial autocorrelation function for the Green's function which is proportional to $\mathrm{sinc}(\kappa |\mathbf{p}_\mathrm{R}|) \cdot \mathrm{sinc}(\kappa|\mathbf{p}_\mathrm{T}|)$ which is equivalent to the Clarke model \cite{Clarke1968}. This is as close to i.i.d. Rayleigh fading as could ever exist, and features the spatial correlation behaviors shown in Fig.~\ref{fig:eigenvalue} when the antenna spacing is smaller than $\lambda/2$ or the array is planar.

\subsubsection{Rayleigh-Jeans-Clarke model for ambient thermal noise}
The additive noise in the receiver array can also be spatially correlated. Classical statistical mechanics, when combined with elementary wave propagation theory, yields a space/time model for thermally induced ambient noise \cite{Marzetta_Hansen_RJC}. We begin by considering a resonant chamber, such as a lossless copper box. The interior of the box supports a countably infinite number of EM standing-wave (Sturm-Liouville) normal modes whose tangential electric fields vanish on the boundaries of the box \cite{Morse_Feshbach}. According to the Equipartition Theorem, every energy-storage mode has expected thermal energy equal to $\frac{k_{\mathrm{B}} T}{2}$, where $k_{\mathrm{B}}$ is Boltzmann's constant and $T$ is absolute temperature \cite{Joos}. The superposition of the modes results in a Gaussian space/time stochastic process which is stationary in time, but non-stationary in space due to the boundary conditions. As the size of the box grows large compared with the wavelength, the random EM field becomes stationary in space as well as time. At a given point in space, the temporal power-density spectrum is proportional to $k_{\mathrm{B}}T \omega^2$, called the \emph{Rayleigh-Jeans spectrum} \cite{Joos}. At a particular temporal frequency, the spatial autocorrelation function is proportional to $\mathrm{sinc}(\kappa |\mathbf{p}|)$, which is the Clarke autocorrelation \cite{Clarke1968}. This \emph{Rayleigh-Jeans-Clarke} spectrum describes dark-sky noise, and it represents the minimum EM noise that a receive antenna would be subject to.

\subsubsection{Shannon capacity of a wireless link in a resonant chamber}
As a case study, consider a transmit antenna and a receive antenna operating inside a lossless resonant chamber \cite{Singh_Marzetta2023}. The $2 \times 2$ impedance matrix (whose entries are obtained from EM theory) is purely imaginary, and it has a countably infinite number of simple poles on the real-$\omega$ axis. The transmit antenna is driven by a current source, subject to bandwidth and power constraints. The receive antenna is connected to a load resistor, $R_{\mathrm{L}}$, which in turn is connected to the infinite impedance input of a voltage amplifier. There are two sources of noise in the receiver: the additive amplifier noise, and the Johnson noise of the load resistor, whose voltage spectral density is equal to $2 k_{\mathrm{B}} T R_{\mathrm{L}}$.\footnote{The thermally-induced standing-waves in the resonant chamber do \emph{not} constitute noise in addition to the resistor Johnson noise. To include them would constitute a form of double-counting.}

The exercise of computing the channel capacity yields two surprises:
1) For a fixed transmit power, the capacity increases without bound as the load resistance, $R_{\mathrm{L}}$, approaches infinity, despite the Johnson noise spectral density becoming infinite.
2) The Shannon-optimum allocation of transmit power versus frequency avoids placing power in the vicinity of the resonant frequencies (e.g., the system poles).

The conclusion is that EM theory governs the channel modeling in wireless communications, including the available spatial DoF, interaction between the antennas and scattering environment, array design, and fundamental thermal noise. We will take a closer look at several of these aspects in the remainder of this paper. The general way to expand the DoF is to increase the array size, but one can also exploit polarization and adapt the arrays to a specific scattering environment.


\section{Circuit Theory for Physical Channel Modeling}
\label{sec:physically-consistent-channel}

The last section provided a comprehensive overview of the physics governing wireless transmission through Maxwell’s equations, along with an introduction to the lumped impedance representation of antennas. Nevertheless, a fundamental challenge in designing advanced transmission techniques, such as Massive MIMO systems, stems from the intricate relationship between array signal processing quantities (e.g., signals from the analog-to-digital converters (ADCs) or to the digital-to-analog converters (DACs)) and the resultant fields. This complexity is heightened by mutual coupling effects among the antenna elements, making the connection between these processing quantities and the realized fields complicated \cite{Ivrlac_Nossek}. 

The objective of this section is to describe the analytical tools necessary to establish this connection and formulate an end-to-end model of such a communication link. This model from \cite{MezghaniEtAlReincorporatingCircuitTheory2023} encompasses both the antenna and radio-frequency (RF) frontend and leverages fundamental physics, such as the superposition principle, the conservation of power, and reciprocity. Throughout this section, for the sake of simplicity in notation, we may occasionally omit the frequency variable~$\omega$. 

\begin{figure}[!t]
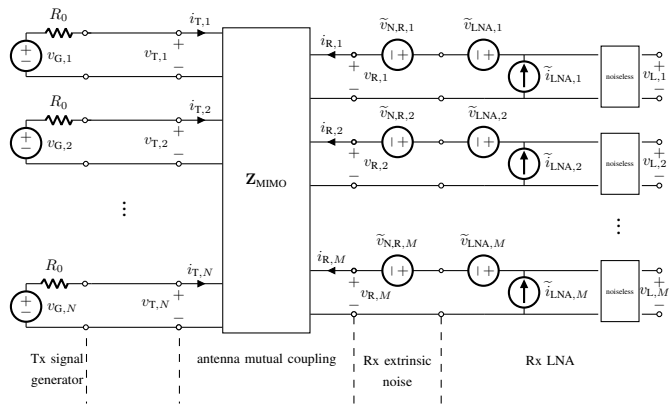

         \include{figures/fig9a.tex} \vspace{-4mm}
         \caption{A linear multiport model of a MIMO communication system showing the signal generation, antenna mutual coupling, and noise from both extrinsic sources (i.e., picked up by the antennas) and intrinsic sources (i.e., generated by low-noise amplifiers and local circuitry).}
         \label{fig:MIMO-communication-circuit}
\end{figure}

A complete MIMO transceiver system can be modeled as a noisy multiport circuit with the mutual MIMO impedance matrix $\vect{Z}_{\rm MIMO}$ from (\ref{ImpMatrixPart}) as illustrated in Fig.~\ref{fig:MIMO-communication-circuit}. In most practical  systems, the back-scattering effects between the transmit and receive antennas can be neglected and the unilateral approximation, where only the forward channel impedance is taken into account, can be made:
\begin{equation}
    \vect{Z}_{\rm MIMO}(\omega) \approx \left[ \begin{matrix} 
    \vect{Z}_{\rm T}(\omega) & \vect{0} \\
\vect{Z}_{\rm RT}(\omega) & \vect{Z}_{\rm R}(\omega) 
    \end{matrix} \right],
\end{equation}
where $\vect{Z}_{\rm T}(\omega)$ is the mutual impedance matrix of the transmitting array and  $\vect{Z}_{\rm R}(\omega)$ is the mutual impedance matrix of the receiving array, while the propagation channel is represented by the transimpedance $\vect{Z}_{\rm RT}(\omega)$. In the far-field, the latter matrix can be related to the open circuit radiation responses of both arrays and the coefficients of the channel directions
\begin{equation}
    \vect{Z}_{\rm RT}(\omega) = \sum_{k} g(\varphi_k,\theta_k,\omega) \vect{s}_{\rm R}^{\rm OC}(\varphi_k,\theta_k,\omega) \left(\vect{s}_{\rm T}^{\rm OC}(\varphi_k,\theta_k,\omega)\right)^{\Ttran}.
\end{equation} 
In Fig.~\ref{fig:MIMO-communication-circuit}, the ambient thermal noise as well as the thermal noise due to the antenna losses are represented by the vector $\bm{\widetilde{v}}_{\textrm{N,R}}$.
Each transmit amplifier is represented as a Th\'evenin equivalent source with a generator open circuit voltage $v_{{\rm G},n}$ and internal resistance $R_0$. Each low-noise amplifier (LNA) is modeled as a noisy linear two-port network with two equivalent input noise sources. The noiseless part of the LNA can be assumed to have open-circuit input and unit voltage gain, without affecting the communication performance.

Using basic circuit analysis \cite{Ivrlac_Nossek}, we can establish the input-output relationship of the MIMO communication system between the input voltage $\bm{v}_{\textrm{G}}(\omega)$ and the output voltage $\bm{v}_{\textrm{L}}(\omega)$ as 
\begin{equation}\label{eq:in-out-relationship}
    \bm{v}_{\textrm{L}}(\omega) = \vect{H}(\omega)\,\bm{v}_{\mathrm{G}}(\omega) + \vect{n}(\omega),
\end{equation}
where
\begin{subequations}
    \begin{align}
    \vect{H}(\omega) &= \vect{Z}_{\mathrm{RT}}(\omega) \,\Big(\vect{Z}_{\textrm{T}}(\omega) + R_0\,\mathbf{I}_N \Big)^{-1}, \\
    \vect{n}(\omega) &= \bm{\widetilde{v}}_{\textrm{N,R}}(\omega) + \bm{\widetilde{v}}_{\textrm{LNA}}(\omega) + \vect{Z}_{\rm R} \bm{\widetilde{i}}_{\textrm{LNA}}(\omega).
    \label{eq:noise-mimo}
    \end{align}
\end{subequations}

\noindent The characterization of the circuit model in Fig.~\ref{fig:MIMO-communication-circuit} requires the specification of the circuit structure of the joint impedance matrix $\vect{Z}_{\text{MIMO}}$ along with the statistical signal and noise properties.

\subsection{Antenna circuit models  and their  key properties}

\begin{figure}[!t]
    \centering
    \input{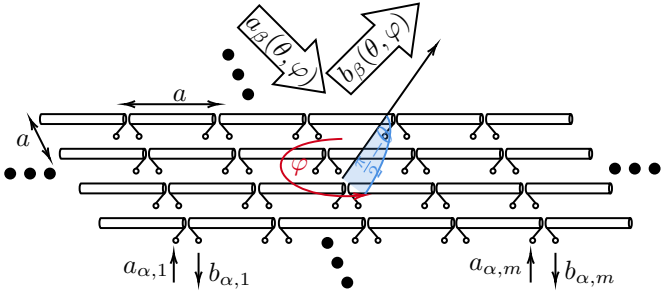}
\caption{Antenna scattering with $m$ accessible ports.}
\label{antenna_scattering}
\end{figure}

An antenna can be viewed as a wave transformer that converts (single-mode) guided waves at the terminal ports into EM fields that propagate in free space and vice versa. 
The EM properties of an antenna array are thus characterized by its radiating/receiving patterns, the space-side scattering pattern, and the electrical multi-port behavior of its terminals \cite{Kerns1960TheoryOD,MezghaniEtAlReincorporatingCircuitTheory2023} as depicted in Fig.~\ref{antenna_scattering}. For simplicity, we simplify the EM field, typically a complex vector quantity, as a complex-valued scalar quantity (e.g., considering a single polarization as we did in Section~\ref{sec:spat_deg_freedom}). At each accessible port ($m$th port), the forward and backward traveling wave phasors along the antenna feed line are denoted by $a_{\alpha,m}$ and $b_{\alpha,m}$, respectively, which are related to the port current and voltages as

\begin{equation}
\begin{aligned}
    \B{a}_\alpha&=\frac{\B{v}+R_0\B{i}}{2\sqrt{R_0}}= \frac{\bm{v}_{\mathrm{G}}}{2\sqrt{R_0}}, \\
    \vect{b}_\alpha&=\frac{\B{v}-R_0\B{i}}{2\sqrt{R_0}}.
\end{aligned}
\label{wave_rel_v_i}
\end{equation}
In the far-field, the angular transmission characteristic of a specific polarization (i.e., the complex embedded pattern under reference resistance termination) is defined when the antenna port $m$ is connected to a wave-source amplitude $a_{\alpha,m}$ (in $\sqrt{\text{W}}$) at port $m$ while all other ports are terminated with $R_0$:
\begin{equation}
   {\rm s}_m(\varphi,\theta)=\lim\limits_{r \rightarrow \infty} -{\imagunit} \,r \,e^{\imagunit \kappa r} \,\frac{E^{(m)}(\varphi,\theta,r)}{a_{\alpha,m}} \sqrt{\frac{1}{Z_0}},
\end{equation}
where $Z_0 =\sqrt{\mu_0/\epsilon_0}$ is the characteristic impedance of free space and $E^{(m)}(\varphi,\theta,r)$ is the corresponding generated electrical far-field.
The complex patterns of the embedded elements, denoted as ${\rm s}_m(\varphi,\theta)$, are aggregated to construct the characteristic response vector $\vect{s}_{\alpha\beta}(\varphi,\theta)$ for receiving and transmitting.
Considering a single polarization, for simplicity, the overall description of the antenna array can be expressed through the linear scattering representation on the terminal side
\begin{equation}
 \vect{b}_\alpha = \vect{S}_{\alpha\alpha} \boldsymbol{a}_\alpha + \int_{0}^{\pi} \int_{-\pi}^{\pi} \!\!  \vect{s}_{\alpha\beta}(\varphi,\theta)  \,a_\beta(\varphi,\theta) \,   
\sin(\theta)\, d \varphi\, d \theta,
\label{S_parameters}
\end{equation}
where $ \B{a}_\alpha\triangleq [a_{\alpha,1},\ldots, a_{\alpha,M}]^{\Ttran}$ and $\vect{S}_{\alpha\alpha}$ is the scattering matrix. For the space-side, the angular spectra of the outgoing propagating wave phasors $b_\beta(\varphi,\theta)$ are expressed as linear functions of the incoming wave $a_\beta(\varphi,\theta)$ ($=g(\varphi,\theta)\cdot s$ for a single source in space) as well as the port incident phasor  
\begin{equation}
\begin{aligned}
 b_\beta(\varphi,\theta) \,=~&   \vect{s}^{\Ttran}_{\alpha\beta}(\varphi,\theta)  \B{a}_\alpha  + \\
 &\hspace{-0.3cm}\int_{0}^{\pi} \! \!\!  \int_{-\pi}^{\pi} \!\!\! {\rm s}_{\beta\beta}(\varphi,\theta,\varphi',\theta')  \,a_\beta(\varphi',\theta') \,   
\sin(\theta') \,d \varphi' \,d\theta',
\end{aligned}
\end{equation}
where ${\rm s}_{\beta\beta}(\varphi,\theta,\varphi',\theta')$ is the wave back-scattering characteristic. Ideally, ${\rm s}_{\beta\beta}(\varphi,\theta,\varphi',\theta')$ is effectively negligible or is just considered as part of the fixed environment. 
Alternatively, one can use port current and voltage phasors to describe the antenna instead of wave phasors.
Substituting (\ref{wave_rel_v_i}) into (\ref{S_parameters}), 
we get the alternative impedance-based representation
\begin{equation}
 \B{v} = \vect{Z} \B{i} + \int_{0}^{\pi} \int_{-\pi}^{\pi} \!\!  \vect{s}_{\alpha\beta}^{\rm OC}(\varphi,\theta)  \,a_\beta(\varphi,\theta) \,   
\sin(\theta)\, d \varphi\, d \theta,
\label{Z_parameters}
\end{equation}
where the impedance matrix is then related to the S-matrix via the relation 
\begin{equation}
\vect{Z}=R_0({\bf I}_M - \vect{S}_{\alpha\alpha})^{-1}({\bf I}_M + \vect{S}_{\alpha\alpha}),
\end{equation}
and the open circuit embedded radiation pattern is related to the terminated radiation pattern as
\begin{equation}
    \vect{a}^{\rm OC}_{\alpha\beta}(\varphi,\theta)=\frac{1}{\sqrt{R_0}} (\vect{Z}+R_0{\bf I}_M)\,\vect{s}_{\alpha\beta}(\varphi,\theta).
\end{equation}

Assuming lossless antennas with the property 
\begin{equation}
\begin{aligned}
\vect{S}_{\alpha\alpha}\vect{S}_{\alpha\alpha}^{\Htran}+ \underbrace{\int_{0}^{\pi} \!\!  \int_{-\pi}^{\pi} \!\!\! \vect{s}_{\alpha\beta}(\varphi,\theta)    \vect{s}^{\Htran}_{\alpha\beta}(\varphi,\theta)
\sin(\theta) d \varphi d\theta}_{{\triangleq}\vect{B}} = {\bf I}_M,
\end{aligned}
\label{matrix_B}
\end{equation}
then a relationship between the embedded pattern coupling matrix $\vect{B}$ (also called radiation matrix) and the S-matrix $\vect{S}_{\alpha\alpha}$ is given by  
\begin{equation}
    \vect{B}={\bf I}_M-\vect{S}_{\alpha\alpha}\vect{S}_{\alpha\alpha}^{\Htran}=\vect{U}\B{\Lambda}\vect{U}^{\Htran}.
\end{equation}
Due to reciprocity (i.e.,  $\vect{S}_{\alpha\alpha}=\vect{S}_{\alpha\alpha}^{\Ttran}$), we can obtain when all eigenvalues of $\vect{B}$ are distinct 
\begin{equation}
    \vect{S}_{\alpha\alpha}=\vect{U}\diag(e^{\imagunit \alpha_1},\ldots,e^{ \imagunit\alpha_M})({\bf I}_M-\B{\Lambda})^{1/2}\vect{U}^{\Ttran},
\label{s_matrix}
\end{equation}
with arbitrary phases $\alpha_1,\ldots, \alpha_M$ (which are equal if the antennas are assumed to have identical embedded patterns). It is worth noting that the port scattering matrix $\vect{S}_{\alpha\alpha}$, the impedance matrix $\vect{Z}$, and the pattern coupling matrix $\vect{B}$ have the same eigenvectors but with different eigenvalues.

Antenna mutual coupling, primarily occurring in the reactive near-field, can be characterized by considering the complex far-field pattern, up to a diagonal complex rotation. While this observation seems to be counter-intuitive, it follows from the uniqueness theorem in electromagnetism, wherein boundary conditions are specified at infinity. It is crucial to emphasize that the required patterns for this derivation are the embedded patterns as opposed to the isolated patterns which relativizes the usefulness of this observation.

If the antenna is lossy, then (\ref{matrix_B}) becomes an inequality. A simple way to account for losses is to introduce the loss factor $\eta \leq 1$ which defines the antenna efficiency. Accordingly, (\ref{s_matrix}) becomes
\begin{equation}
    \vect{S}_{\alpha\alpha}=\vect{U}\diag(e^{\imagunit \alpha_1},\ldots,e^{\imagunit \alpha_M})\left(\eta{\bf I}_M-\B{\Lambda}\right)^{1/2}\vect{U}^{\Ttran}.
\end{equation}

Generally, characterizing the circuit behavior of large antenna arrays across different frequencies through the embedded pattern is challenging computationally as well as experimentally. In addition, the current distribution is not known in advance to solve (\ref{EHpw}) in a straightforward manner. Instead, some boundary conditions are to be imposed depending on the antenna structure. In the following, two main analytical techniques for characterizing the mutual coupling effects are discussed.

\subsubsection{Infinite arrays}

\begin{figure}[!t]
    \centering
    \tikzset{every picture/.style={line width=0.75pt}} 

\begin{tikzpicture}[x=0.75pt,y=0.75pt,yscale=-0.7,xscale=0.7]


\draw   (96.66,33.97) -- (223.36,34.03) .. controls (224.27,34.03) and (225,36.49) .. (225,39.52) .. controls (225,42.55) and (224.26,45) .. (223.36,45) -- (96.65,44.93) .. controls (95.75,44.93) and (95.01,42.48) .. (95.01,39.45) .. controls (95.01,36.42) and (95.75,33.97) .. (96.66,33.97) .. controls (97.57,33.97) and (98.3,36.42) .. (98.3,39.45) .. controls (98.3,42.48) and (97.56,44.93) .. (96.65,44.93) ;
\draw   (231.64,34) -- (358.35,34.07) .. controls (359.25,34.07) and (359.99,36.52) .. (359.99,39.55) .. controls (359.99,42.58) and (359.25,45.03) .. (358.34,45.03) -- (231.64,44.97) .. controls (230.73,44.97) and (230,42.51) .. (230,39.48) .. controls (230,36.45) and (230.74,34) .. (231.64,34) .. controls (232.55,34) and (233.29,36.46) .. (233.29,39.48) .. controls (233.29,42.51) and (232.55,44.97) .. (231.64,44.97) ;
\draw   (366.65,34.03) -- (493.35,34.1) .. controls (494.26,34.1) and (494.99,36.56) .. (494.99,39.59) .. controls (494.99,42.61) and (494.25,45.07) .. (493.35,45.07) -- (366.64,45) .. controls (365.74,45) and (365,42.55) .. (365,39.52) .. controls (365,36.49) and (365.74,34.03) .. (366.65,34.03) .. controls (367.56,34.04) and (368.29,36.49) .. (368.29,39.52) .. controls (368.29,42.55) and (367.55,45) .. (366.64,45) ;
\draw    (135,40) -- (168,40) ;
\draw [shift={(170,40)}, rotate = 180] [color={rgb, 255:red, 0; green, 0; blue, 0 }  ][line width=0.75]    (10.93,-3.29) .. controls (6.95,-1.4) and (3.31,-0.3) .. (0,0) .. controls (3.31,0.3) and (6.95,1.4) .. (10.93,3.29)   ;
\draw    (539.99,39.59) -- (568,39.97) ;
\draw [shift={(570,40)}, rotate = 180.79] [color={rgb, 255:red, 0; green, 0; blue, 0 }  ][line width=0.75]    (10.93,-3.29) .. controls (6.95,-1.4) and (3.31,-0.3) .. (0,0) .. controls (3.31,0.3) and (6.95,1.4) .. (10.93,3.29)   ;
\draw    (223.36,45) -- (220,55) ;
\draw    (231.64,44.97) -- (235,55) ;
\draw    (359.36,44) -- (356,54) ;
\draw    (367.64,43.97) -- (371,54) ;
\draw  [dash pattern={on 4.5pt off 4.5pt}]  (494.99,39.59) -- (539.99,39.59) ;
\draw  [dash pattern={on 4.5pt off 4.5pt}]  (72.51,39.74) -- (95.01,39.45) ;
\draw   (353.17,57) .. controls (353.17,55.34) and (354.51,54) .. (356.17,54) .. controls (357.82,54) and (359.17,55.34) .. (359.17,57) .. controls (359.17,58.66) and (357.82,60) .. (356.17,60) .. controls (354.51,60) and (353.17,58.66) .. (353.17,57) -- cycle ;
\draw   (368,57) .. controls (368,55.34) and (369.34,54) .. (371,54) .. controls (372.66,54) and (374,55.34) .. (374,57) .. controls (374,58.66) and (372.66,60) .. (371,60) .. controls (369.34,60) and (368,58.66) .. (368,57) -- cycle ;
\draw   (217.17,57.98) .. controls (217.17,56.33) and (218.51,54.98) .. (220.17,54.98) .. controls (221.82,54.98) and (223.17,56.33) .. (223.17,57.98) .. controls (223.17,59.64) and (221.82,60.98) .. (220.17,60.98) .. controls (218.51,60.98) and (217.17,59.64) .. (217.17,57.98) -- cycle ;
\draw   (232,57.98) .. controls (232,56.33) and (233.34,54.98) .. (235,54.98) .. controls (236.66,54.98) and (238,56.33) .. (238,57.98) .. controls (238,59.64) and (236.66,60.98) .. (235,60.98) .. controls (233.34,60.98) and (232,59.64) .. (232,57.98) -- cycle ;
\draw    (110,16) -- (110,32) ;
\draw [shift={(110,34)}, rotate = 270] [color={rgb, 255:red, 0; green, 0; blue, 0 }  ][line width=0.75]    (10.93,-3.29) .. controls (6.95,-1.4) and (3.31,-0.3) .. (0,0) .. controls (3.31,0.3) and (6.95,1.4) .. (10.93,3.29)   ;
\draw    (110,60) -- (110,47) ;
\draw [shift={(110,45)}, rotate = 90] [color={rgb, 255:red, 0; green, 0; blue, 0 }  ][line width=0.75]    (10.93,-3.29) .. controls (6.95,-1.4) and (3.31,-0.3) .. (0,0) .. controls (3.31,0.3) and (6.95,1.4) .. (10.93,3.29)   ;
\draw    (227,20) -- (358,20) ;
\draw [shift={(360,20)}, rotate = 180] [color={rgb, 255:red, 0; green, 0; blue, 0 }  ][line width=0.75]    (10.93,-3.29) .. controls (6.95,-1.4) and (3.31,-0.3) .. (0,0) .. controls (3.31,0.3) and (6.95,1.4) .. (10.93,3.29)   ;
\draw [shift={(225,20)}, rotate = 0] [color={rgb, 255:red, 0; green, 0; blue, 0 }  ][line width=0.75]    (10.93,-3.29) .. controls (6.95,-1.4) and (3.31,-0.3) .. (0,0) .. controls (3.31,0.3) and (6.95,1.4) .. (10.93,3.29)   ;

\draw (206,49) node [anchor=north west][inner sep=0.75pt]   [align=left] {-};
\draw (240,49) node [anchor=north west][inner sep=0.75pt]   [align=left] {\mbox{+}};
\draw (220,60) node [anchor=north west][inner sep=0.75pt]   [align=left] {$\displaystyle V_{m-1}$};
\draw (342,48) node [anchor=north west][inner sep=0.75pt]   [align=left] {-};
\draw (376,48) node [anchor=north west][inner sep=0.75pt]   [align=left] {\mbox{+}};
\draw (355.17,60) node [anchor=north west][inner sep=0.75pt]   [align=left] {$\displaystyle V_{m}$};
\draw (141,47.4) node [anchor=north west][inner sep=0.75pt]    {$I( z)$};
\draw (551,44.4) node [anchor=north west][inner sep=0.75pt]    {$z$};
\draw (114,14.4) node [anchor=north west][inner sep=0.75pt]    {$2a$};
\draw (286,3.4) node [anchor=north west][inner sep=0.75pt]    {$\Delta $};

\end{tikzpicture} \vspace{-6mm}
    \caption{Infinite uniform connected co-linear array.}
    \label{fig:colinear}
\end{figure}
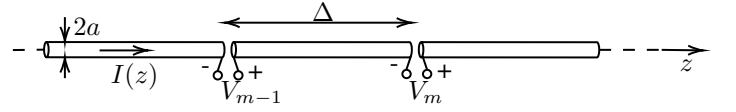

Infinite periodic arrays are generally more tractable to analyze due to the identical radiation properties of the elements and the absence of edge effects. As an example, consider the infinite co-linear one-dimensional array in Fig.~\ref{fig:colinear}, which can be treated as an infinitely long dipole with multiple feeds.
This setup serves as a valuable approximation for large arrays and is mathematically manageable in numerous instances, thanks to the periodic structure. 
The feed points are periodic infinitesimal gaps distributed along the thin wire. To find the corresponding admittance matrix, we first examine the current distribution on a linear cylindrical antenna center-driven by a delta function generator. For a hollow-cylindrical antenna with a radius  $a$ and infinite length,
the boundary condition in the polar coordinates system along the center-driven wire is 
\begin{equation}
E_z(\rho=a,z)   = -V \delta(z),
  \label{Ez_eq}     
\end{equation} 
where $V$ is the voltage maintained at the driving point $z = 0$. 
By solving (\ref{Maxsoln}) in the Fourier domain of $z$ under the above boundary condition, we obtain the solution (which is a particular solution of the Pocklington's integral equation \cite{Pocklington_1897})
\begin{equation}
y(z)=\frac{I(z)}{V}= \frac {2\kappa}{\pi Z_ {0}}  \int_{-\infty}^{\infty}  \frac{e^{\imagunit \alpha z}}{\beta^{2}J_ {0}(\beta a)H_ {0}^{(2)}(\beta a )} d \alpha,
\label{current_dist_exact}  
\end{equation}
where $\beta= \sqrt{\kappa^2-\alpha^2}$, while $J_ {0}(\cdot)$ and $H_ {0}^{(2)}(\cdot )$ are the zero-order Bessel and Hankel functions. 
The resulting magnetic field reads as (c.f. (\ref{Green's})) 
\begin{equation}
\begin{aligned}
& H_ {\varphi, \rm far-field }(\varphi,\theta,r)= \\
& \sin (\theta)  e^ {-\imagunit \kappa r}  \frac {\imagunit \kappa }{4\pi r}  \int _ {-\infty }^ {\infty}\!\! \frac {I(z) }{2\pi } \! \int _ {0}^ {2\pi } \! e^ {-\imagunit \kappa (z\cos (\theta) +a\sin (\theta) \cos( \varphi) )} d \varphi d z \\
&=  \frac {\imagunit V e^{-\imagunit \kappa r} }{\pi rZ_0 \sin 
(\theta) H_ {0}^{(2)} (a \kappa \sin (\theta) )}.
\end{aligned}
\end{equation}
Hence, the short-circuit pattern for an excitation port at $z=0$ is expressed as 
\begin{equation}
\begin{aligned}
   {\rm s}_{\alpha\beta}^{\rm SC}(\varphi,\theta)&=-\imagunit \,r \,e^{\imagunit \kappa r} \,\frac{H_ {\varphi, \rm far-field }(\varphi,\theta,r)}{V} \sqrt{Z_0} \\
   &=\frac { 1 }{\pi \sqrt{Z_0} \sin 
(\theta) H_ {0}^{(2)} (a \kappa \sin (\theta) )} .
\end{aligned}
\end{equation}

To derive the impedance description of the infinite co-linear array, we uniformly discretize the admittance function in (\ref{current_dist_exact}) with a sampling spacing of $\Delta$. The periodic admittance spectrum can subsequently be inverted to form the impedance function
\begin{equation}
  z[m\Delta]= \frac {Z_0\Delta^2}{8\pi \kappa}  \int_{0}^{\frac{2\pi}{\Delta}} {\frac{e^{\imagunit \alpha m\Delta}}{\sum\limits_{\ell=-\infty}^{\infty}\frac{1}{\beta_\ell^{2}J_ {0}(\beta_\ell a)H_ {0}^{(2)}(\beta_\ell a )}}} d \alpha,
\end{equation}
where $\beta_\ell=\sqrt{\kappa^2-(\alpha-\frac{2\pi \ell}{\Delta})^2}$. 
In addition, the open-circuit embedded  pattern is obtained similarly as  
\begin{equation}
\begin{aligned}
 &{\rm s}_{\alpha\beta}^{\rm OC}(\varphi,\theta)=  \frac{{\rm s}_{\alpha\beta}^{\rm SC}(\varphi,\theta) Z_0 \Delta}{
 \sum\limits_{\ell=-\infty}^{\infty} \frac{4\kappa}{\beta_\ell^{2}J_ {0}(\beta_\ell a)H_ {0}^{(2)}(\beta_\ell a )}
 },~ {\scriptstyle \beta_\ell=\sqrt{\kappa^2-(\kappa \cos(\theta)-\frac{2\pi \ell}{\Delta})^2}}. 
\end{aligned}
\end{equation}
The impedance description of the infinite array is more appropriate than the admittance one to approximate the actual finite case since open-circuiting the edge portions of the infinite array forces the edge currents to decay rapidly to zero rather than slowly in the shorted array case.

\subsubsection{Array of identical minimum scattering antennas}
\label{msa_antenna}

\noindent The radiation patterns of embedded elements usually differ from the patterns emitted by an element when other elements in the array are absent, known as isolated element patterns, due to scattering effects from nearby elements. Finding the impedance matrix $\vect{Z}$ directly from the isolated far-field radiation pattern is however possible in the case of minimum scattering antennas \cite{Wasylkiwskyj_1970} based on the result in (\ref{mut_inf_res}).  Minimum scattering antennas are
invisible (or radio-transparent) under certain reactive termination. In the canonical case, this happens when the antenna port is open-circuited.
This generally means that the antenna elements, embedded into the array, induce in the structure individual current distributions within non-overlapping spheres when fed with current sources. 
In such a case, the embedded open circuit patterns are equivalent to the isolated pattern. Minimum scattering antennas also have other properties such as identical radiation and scattering patterns that are symmetrical with respect to the origin \cite{Wasylkiwskyj_1970}. By changing the variables $k_x$ and $k_y$ to the angular coordinates,  (\ref{mut_inf_res}) can be rewritten in terms of normalized isolated radiation pattern (assumed to be known for complex angles) \cite{Wasylkiwskyj_1970}
\begin{equation}
\frac{Z(\omega,\mathbf{p})}{{\mathrm{Re}}\left(Z(\omega,\mathbf{0})\right)}= 2\int_0^{2 \pi}\!\!\! \int_{0}^{\frac{\pi}{2}+\imagunit\infty} e^{-\imagunit \kappa \vect{e}^{\rm T}_r\vect{p} }  |{\rm s}^{\rm iso}(\varphi,\theta)|^2 \sin( \theta) d \theta d\varphi, 
\label{iso_pattern_integ}
\end{equation}
where the $\theta$ integration is taken along the real axis, $0\leq \theta \leq \frac{\pi}{2}$, and then along the line ${\rm Re}(\theta)=\frac{\pi}{2}$, $0\leq {\rm Im}(\theta) < \infty $ ,  
with the normalized isolated angular pattern ${\rm s}^{\rm iso}( \varphi,\theta)$
\begin{equation}
 \int_0^{2 \pi}\int_{0}^\pi   |{\rm s}^{\rm iso}(\varphi,\theta)|^2 \sin (\theta) d \theta d\varphi=1. 
\end{equation}
It is worth mentioning that the result in (\ref{iso_pattern_integ}) is also valid in the non-minimum scattering case if the elements are sufficiently spaced ($>\lambda$) \cite{Wasylkiwskyj_1970_2}. It might be, however,  inaccurate for closely spaced elements.

\subsection{Signal power}
The characterizing of signal power is a fundamental aspect in the analysis and optimization of communication systems. While in communication theory, mainly the Euclidean norm of signals is emphasized, reflecting their abstract representations and mathematical properties,  UM-MIMO systems necessitate an understanding of physical power involving different physical quantities such as voltages and currents. Here, two different power metrics can be relevant in the realm of physical-consistent modeling and analysis: the radiated power and the maximum available power spectral densities obtained by taking the signal duration $T_0$ to infinity:
\begin{align}
p_{\rm avail}(\omega)&=\frac{1}{2} \lim_{T_0\to\infty}\frac{1}{T_0} \,\mathbb{E}\left\{\|\B{a}_\alpha(\omega)\|_2^2\right\} \nonumber\\
&= \lim_{T_0\to\infty}\frac{1}{T_0} \frac{\mathbb{E}\left\{\|\bm{v}_{\mathrm{G}}(\omega)\|_2^2\right\}}{8R_0}, 
\end{align}
\begin{align}
p_{\rm T}(\omega)&=\frac{1}{2} \lim_{T_0\to\infty}\frac{1}{T_0} \,\mathbb{E}\left\{ \B{a}_\alpha^{\Htran}(\omega) \vect{B}(\omega)\B{a}_\alpha(\omega) \right\} \nonumber\\
&\underset{\rm \underset{antenna}{lossy}}{\leq}  \frac{1}{2} \lim_{T_0\to\infty}\frac{1}{T_0} \,\mathbb{E}\left\{ \B{i}_{\rm T}^{\Htran}(\omega){\rm Re}(\vect{Z}_{\rm T}(\omega))\B{i}_{\rm T}(\omega) \right\} \nonumber\\
&\underset{\rm \underset{Mismatch}{Imped.}} 
{\leq} p_{\rm avail}(\omega).
\end{align}
The radiated power density $p_{\rm T}(\omega)$ is usually restricted by regulatory measures and interference policies, whereas the available power $p_{\rm avail}(\omega)$ is constrained by device/technology limitations.
In the case of perfect matching between the generator and antenna impedances ($\vect{Z}_{\rm T}=R_0\vect{I}$) along with a lossless array structure, these two powers are equivalent. In advanced antenna systems, however,  impedance mismatching, structural losses, and technology limitations introduce additional model complexities that require careful characterization of the different power limitations and their relevance. At the same time, it enables flexible signal design as the antenna structure can serve as spatial filters for blocking unwanted distortion \cite{Mezghani_prec_2022}.

\subsection{Noise modeling}

\subsubsection{Background noise of antennas}
The multi-port network $\vect{Z}_\text{MIMO}$ is only composed of passive components which can be assumed to have the same absolute temperature $T$ of the environment \cite{Ivrlac_Nossek}. Therefore, the noise of the joint impedance matrix $\vect{Z}_\text{MIMO}$ originates solely from the thermal agitation of the electrons flowing inside its all passive components, a.k.a, thermal noise at the equilibrium temperature $T$ \cite{nyquist1928thermal}. Due to the unilateral approximation, the transmit side noise is neglected. In Fig.~\ref{fig:MIMO-communication-circuit}, the receive noise voltages $\widetilde{\bm{v}}_{\textrm{N,R}}(\omega)$ model the background noise of receive antennas as well as the resistive losses. When the mutual coupling is taken into account within the transmit/receive arrays, the correlation between the $m$th and $m^{\prime}$th receive noise voltages $\widetilde{v}_{\textrm{N,R},m}(\omega)$ and $\widetilde{v}_{\textrm{N,R},m^{\prime}}(\omega)$ taking the signal duration $T_0$ to infinity \cite{nyquist1928thermal} is
\begin{equation}
\label{eq:noise-voltage-auto-correlation}
\begin{aligned}
       & \lim_{T_0\to\infty}\frac{1}{T_0}\,\mathbb{E}\big\{\widetilde{v}_{\textrm{N,R},m}(\omega)\,\widetilde{v}^*_{\textrm{N,R},m^\prime}(\omega)\big\} ~=~ 4\,k_{\mathrm{B}}\,T\,\mathrm{Re}\left(Z_{\textrm{R},mm^\prime}\right) \\
    &    \hspace{4cm}\quad \forall \,m,\,m^\prime \in [1,\dots,M].
 \end{aligned}
\end{equation}

\subsubsection{The receive LNA model} The LNAs are modeled as independently noisy frequency flat devices with unit gain. For the $m$th amplifier, the spectral second-order statistics of the noise voltage, $\widetilde{v}_{\text{LNA},m}(\omega)$, and current, $\widetilde{i}_{\text{LNA},m}(\omega)$, generated inside the LNA are determined using the truncated Fourier transform:
\begin{subequations}\label{eq:noise-autocorrelation-amplifier}
\begin{align}
  &  \lim_{T_0\to\infty}\frac{1}{T_0} \,\mathbb{E}\left\{|\widetilde{v}_{\text{LNA},m}(\omega)|^2\right\} = 4\,k_{\mathrm{B}}\,T\,R_{\textrm{v,LNA}}, \\
  &     \lim_{T_0\to\infty}\frac{1}{T_0} \,\mathbb{E}\left\{|\widetilde{i}_{\text{LNA},m}(\omega)|^2\right\} = 4\,k_{\mathrm{B}}\,T\,G_{\textrm{i,LNA}},\\
 &      \lim_{T_0\to\infty}\frac{1}{T_0} \,\mathbb{E}\left\{\widetilde{i}_{\text{LNA},m}(\omega) \widetilde{v}_{\text{LNA},m}(\omega)^* \right\} = 4\,k_{\mathrm{B}}\,T\,\beta_{\textrm{LNA}}.
\end{align}
\end{subequations}
Additionally, at the $m$th receive antenna, the LNA noise voltage, $\widetilde{v}_{\text{N},\text{LNA},m}(\omega)$, is uncorrelated with the receive noise voltage, $\widetilde{v}_{\text{N,R},m}(\omega)$. 
 Using (\ref{eq:noise-mimo}) and (\ref{eq:noise-voltage-auto-correlation})--(\ref{eq:noise-autocorrelation-amplifier}), the noise correlation matrix is obtained as:
\begin{equation}\label{eq:noise-correlation}
    \vect{R}_{\vect{n}} = 4\,k_{\mathrm{B}} \,T \left[\mathrm{Re}\left((1+\beta_{\textrm{LNA}}){\vect{Z}_{\textrm{R}}}\right)+ R_{\textrm{v,LNA}} \vect{I}_M + G_{\textrm{i,LNA}}\vect{Z}_{\textrm{R}}\vect{Z}_{\textrm{R}}^{\Htran}\right].
\end{equation}
Generally, noise is spatially correlated due to mutual coupling effects. However, if the noise from the LNAs dominates or the antenna structure is lossy ($\vect{Z}_{\textrm{R}}$ dominated by resistive losses), then the conventional uncorrelated noise assumption becomes more justified.   


\section{Antenna near-field and polarization modeling}
\label{sec:polarization-modeling}

Maxwell's equations govern not only how fields propagate through the environment but also how antennas respond to fields and excitations. In prior works on wireless communication and array processing, the antenna is generally treated as an isotropic point source with linear polarization. While such a point source does not exist in practice, this model is appropriate under far-field and narrowband assumptions. As the array sizes grow larger and antennas grow more complex, however, their near-field characteristics and polarization become increasingly important. In this section, we overview prior work on realistic array modeling and a novel EM-based approach developed in \cite{CastellanosHeathElectromagneticManifoldCharacterizationAntenna2023}.

Physically consistent antenna and channel models are critical in isolating the characteristics of the array from the channel, understanding antenna effects on MIMO communications, and designing systems that achieve high capacity in specific environments.
There are several different aspects to this.
Firstly, the radiative near-field of an antenna array extends much further than that of an individual antenna, as we elaborated on in Section~\ref{sec:near-field}. Hence, we need models that account for the amplitude and phase variations observed at different antenna elements due to impinging spherical wavefronts. 
Secondly, the antenna polarization must be captured since it affects both the channel modeling and spatial DoF. The fundamental EM theory for polarized antennas was provided in Section~\ref{subsec:EMtheory_applications}, but practical antenna hardware has impairments. 
Each antenna is designed for a specific polarization but reacts to both the intended polarization, known as the co-polarization, and the orthogonal cross-polarization. The ratio between the strength of these two polarizations is known as the cross-polarization discrimination and determines how well a practical antenna can isolate co-polarized signals. The antenna polarization can be modeled by accounting for both the co-polarization and cross-polarization gain patterns in the signal model \cite{BhaOesHea:A-New-Double-Directional-Channel-Model:10, CastellanosHeathLinearPolarizationOptimizationWideband2023,FriedlanderPolarizationSensitivityAntennaArrays2019}.

State-of-the-art array models aim to analytically characterize all antenna features simultaneously. The array manifold model in  \cite{EfstathopoulosManikasExtendedArrayManifolds2011} used a generalized array response vector model that accounts for arbitrary locations, spherical propagation, and gain patterns. An EM-based array model proposed in \cite{FriedlanderExtendedManifoldAntennaArrays2020} leveraged a Hertzian dipole framework for far-field array propagation. This framework is extended in \cite{CastellanosHeathElectromagneticManifoldCharacterizationAntenna2023} to capture mutual coupling, polarization, and near-field propagation. We will provide an overview of this model below.

\subsection{Electromagnetic-based array model}

Electromagnetically, an antenna is a device that converts electrical currents to fields. Electrical signals are fed into the antenna, inducing a current distribution throughout the structure. Let $\cV$ denote the volume containing the antenna. For any point $\bp \in \cV$, we let $\bJ(\bp)$ be the current density along the antenna when it is excited by a unit current (we omitted the time index for simplicity). The radiated electric and magnetic fields can be computed from $\bJ(\bp)$ using the magnetic vector potential $\bA(\bp)$  computed as \cite{StutzmanThieleAntennaTheoryAndDesign2012}
\begin{equation}
  \label{eq:vector-potential}
  \bA(\bp) = \mu_0 \int_\cV \frac{e^{\imagunit \kappa \, \vert \bp - \bp' \vert }}{4 \pi \vert \bp - \bp' \vert} \bJ(\bp') d \bp'.
\end{equation}
The electric $\bE(\bp)$ and the magnetic field $\mathbf{H}(\bp)$ are given by Maxwell's equations in \eqref{Maxwelltxyz}.
The current distribution along the antenna fully determines the radiated field patterns. Unfortunately, this current distribution can be difficult to obtain in a tractable form that allows easy computation of the magnetic vector potential.

To simplify the calculation of the radiated fields, we can apply a discretization technique to model the antenna as a number of easily characterized segments. EM computational software discretizes antenna volumes into smaller portions to solve complicated problems. We incorporate a similar approach by partitioning $\cV$ into $K$ non-overlapping pieces $\{\cV_k \}_{k=1}^K$ to obtain
\begin{equation}
  \label{eq:disc-vector-potential}
    \bA(\bp) = \mu_0 \sum_{k=1}^K \int_{\cV_k} \frac{e^{\imagunit \kappa \, \vert \bp - \bp' \vert }}{4 \pi \vert \bp - \bp' \vert} \bJ(\bp') d \bp'.
  \end{equation}
  We assume that the discretization is fine enough such that the following two hold:
1) The current distribution over the $k$th segment is equal to a constant $\bJ_k$ (i.e., $\bJ(\bp) = \bJ_k$ for $\bp \in \cV_k$); 2) Each segment behaves as a point source (i.e., $\bp - \bp' = \bp_k$ for $\bp' \in \cV_k$).

  Under these assumptions, \eqref{eq:disc-vector-potential} can be approximated as
  \begin{align}
    \label{eq:sum-vec-pot}
    \bA(\bp) & \approx \mu_0 \sum_{k=1}^{K} \int_{V_k} \frac{e^{\imagunit \kappa \, \vert \bp_k \vert}}{4 \pi \vert \bp_k \vert } \bJ_k d \bp'. \\
    & = \mu_0 \sum_{k=1}^K  \frac{e^{\imagunit \kappa \, \abs{ \bp_k}}} {4 \pi \abs{\bp_k}} \bJ_k \abs{\cV_k},
  \end{align}
  where $\abs{\cV_k}$ denotes the volume of $\cV_k$. The product of the current distribution and the volume is known as the moment, and will be denoted as $\bmm_k = \bJ_k \abs{\cV_k}$. If we define the magnetic vector potential of the $k$th segment as
    $\bA_k(\bp) = \frac{\mu_0 \bmm_k}{4 \pi \abs{\bp_k}} e^{\imagunit \kappa \, \abs{\bp_k}}$, then we can compute \eqref{eq:sum-vec-pot} as
  \begin{align}
    \label{eq:sum-pot-seg}
      \bA(\bp) \approx \sum_{k=1}^K \bA_k(\bp).
  \end{align}
  In essence, the discretization procedure has converted the antenna into an array of point sources, each with magnetic vector potential $\bA_k$. The overall fields can therefore be found from the superposition of the fields from the extended array.

  The simplest physically consistent radiating structure is a point source with constant current, known as a Hertzian dipole. Each segment of the antenna can be modeled as a Hertzian dipole with the current determined from the discretization procedure. The radiated field of a Hertzian dipole can be obtained in closed form following the derivation in \cite{StutzmanThieleAntennaTheoryAndDesign2012,CastellanosHeathElectromagneticManifoldCharacterizationAntenna2023}. Let $\bp$ have a spherical representation with radial distance $r$, azimuth angle $\phi$, and elevation angle $\theta$ as
\begin{equation}
  \label{eq:point-sph}
  \bp = r[\cos (\phi) \cos(\theta), \, \sin( \phi) \cos(\theta), \, \sin( \theta)]^{\Ttran}.
\end{equation}
We define the unit vectors of the spherical orthonormal basis at $\bp$ as
\begin{align}
  \label{eq:unit-sph-vecs}
  \bur(\bp) & = \left[ \cos (\phi) \cos(\theta), \, \sin (\phi) \cos(\theta), \, \sin (\theta) \right]^{\Ttran}, \nonumber \\
  \bue(\bp) & = \left[ -\sin (\phi), \, \cos (\phi), \, 0 \right]^{\Ttran},  \nonumber \\
    \bua(\bp) & = \left[ -\cos (\phi) \cos(\theta), \, -\sin(\phi) \cos(\theta), \, \sin (\theta) \right]^{\Ttran}.
  \end{align}
The electric field of a Hertzian dipole decays with distance along each spherical coordinate. We define the decaying amplitudes of the radial and angular components as $\arad(\bp)$ and $\aang(\bp)$, which are given by 
  \begin{align}
    \label{eq:rad-amp}
    &\arad(\bp) = \frac{e^{-\imagunit \kappa r}}{\imagunit \omega \e0 2 \pi}\left(\frac{1}{r^3} + \frac{\imagunit \kappa}{r^2} \right), \\
    \label{eq:ang-amp}
    &\aang(\bp) = -\frac{e^{-\imagunit \kappa r}}{\imagunit \omega \e0 4 \pi} \left(\frac{1}{r^3} + \frac{\imagunit \kappa}{r^2} - \frac{\kappa^2}{r} \right).
  \end{align}
By further defining the $3 \times 3$ matrix $\bT(\bp)$ as the dipole field transform
  \begin{equation}
    \label{eq:dipole-field-transform}
        \bT(\bp) = \left[
      \begin{array}{ccc}
        \arad(\bp) \bur(\bp), & \aang(\bp) \bua(\bp), & \aang(\bp) \bue(\bp)
      \end{array}
    \right]^{\Ttran},
  \end{equation}
  the electric field of a Hertzian dipole at the origin with moment $\bmm$ is given by
  \begin{equation}
    \label{eq:dipole-field}
    \bE_{\sf dip}(\bp) = \bT(\bp) \bmm.
  \end{equation}
  Two important properties captured by the Hertzian dipole model that are not seen in an isotropic point source are the polarization and the near-field pattern. The Hertzian dipole has a polarization that is dependent on the orientation of the moment vector. For example, a moment vector that is aligned with the $z$-axis will exhibit a vertically polarized field. The amplitude variations in the field as a function of distance are also incorporated into the field expressions through $\arad(\bp)$ and $\aang(\bp)$.

  The complete antenna response can be obtained by superposing the fields from Hertzian dipoles located at each segment. To translate the Hertzian dipole response from the origin to the center of the $k$th segment, we need to account for the dependence of the spherical basis on the position. Letting $\bQ(\bp) = [\bur(\bp), \, \bua(\bp), \, \bue(\bp)]$ be the rotation matrix at $\bp$, the field at $\bp$ from a dipole with moment $\bmm_k$ located at the $k$th segment is
  \begin{equation}
    \label{eq:dipole-field-translated}
    \bE_{{\sf dip}, k}(\bp) = \bQ(\bp) \bQ(\bp_k) \bT(\bp_k) \bmm_k.
  \end{equation}
  The two rotations serve to express the radiated field in spherical coordinates with respect to the origin. Letting $\bR(\bp, \bp_k) = \bQ(\bp) \bQ(\bp_k)$, the antenna response becomes
  \begin{equation}
    \label{eq:extended-ant-model}
    \bE_{\sf ant}(\bp) = \sum_{k=1}^K \bR(\bp, \bp_k)  \bT(\bp_k) \bmm_k.
  \end{equation}
  The polarization and near-field radiation of the antenna are characterized by the combined effect of all of the dipole segments. 

  The framework can be applied to an antenna array. Consider an array with $N$ antenna elements excited by the length $N$ transmit signal $\bx$, where the $n$th antenna is discretized into $K_n$ segments. Under the same assumptions regarding the size of the discretization, the $k$th segment of the $n$th antenna can be associated with a point $\bp_{n,k}$ and a moment vector $\bmm_{n,k}(\bx)$. We note that because of mutual coupling, the current distribution in all of the antennas will depend on the excitation signal $\bx$. The array radiated field can be computed as
  \begin{equation}
    \label{eq:extended-array-model}
    \bE_{\sf arr}(\bp) = \sum_{n=1}^N \sum_{k=1}^{K_n} \bR(\bp, \bp_{n,k})  \bT(\bp_{n,k}) \bmm_{n,k}(\bx).
  \end{equation}
  We emphasize that the model still functions in the same way as in the single-antenna case, thus, the total radiated field is calculated by treating the array structure as a larger array of Hertzian dipoles. Linearity can also be applied to further simplify \eqref{eq:extended-array-model}. Let $\bee_n$ denote the $N \times 1$ vector with $1$ on the $n$th entry and zeros elsewhere and consider the $3 \times N$ matrix $\bM_{n,k}$ with $\ell$th column given by $\bmm_{n,k}(\bee_\ell) $.
  It can be shown that
  \begin{equation}
    \label{eq:1}
    \bE_{\sf arr}(\bp) = \sum_{n=1}^N \sum_{k=1}^{K_n} \bR(\bp, \bp_{n,k})  \bT(\bp_{n,k}) \bM_{n,k} \bx.
  \end{equation}
  This isolates the role of the transmit signal by incorporating mutual coupling between the array elements into $\bM_{n,k}$.

\begin{figure}
    \centering
	\begin{overpic}[width=\columnwidth,tics=10]{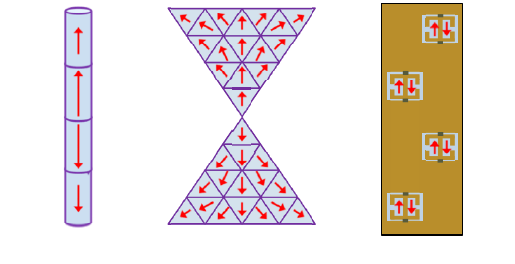}
  \put(10,0){Dipole}
  \put(42,0){Bowtie}
        \put(71,0){Meta-antenna}
        \end{overpic} 
    \caption{Examples of the EM characterization of antennas based on discretization. Each antenna is partitioned into a large number of segments, each of which is treated as a Hertzian dipole. The radiated field from each antenna comes from the combined fields of the dipoles.}
    \label{fig:em-antenna-model}
  \end{figure}

  We end the section with a few notes on the usefulness of this representation for an antenna array. One of the key benefits is the flexibility in characterizing arbitrary arrays since we did not make any assumptions about the structure or shape. Any type of antenna that can be meshed in the considered manner can be approximated by the sum of the Hertzian dipoles, as shown in Fig.~\ref{fig:em-antenna-model}. In addition, the use of Hertzian dipoles as the fundamental building blocks of each antenna means that properties such as the polarization and near-field radiation patterns of the entire array are natively incorporated into the linear model.

\section{Antenna Array Design for UM-MIMO}
\label{sec:antenna-array-design}

\begin{figure*}[t]
\centering{\includegraphics[width=1.9\columnwidth,keepaspectratio]{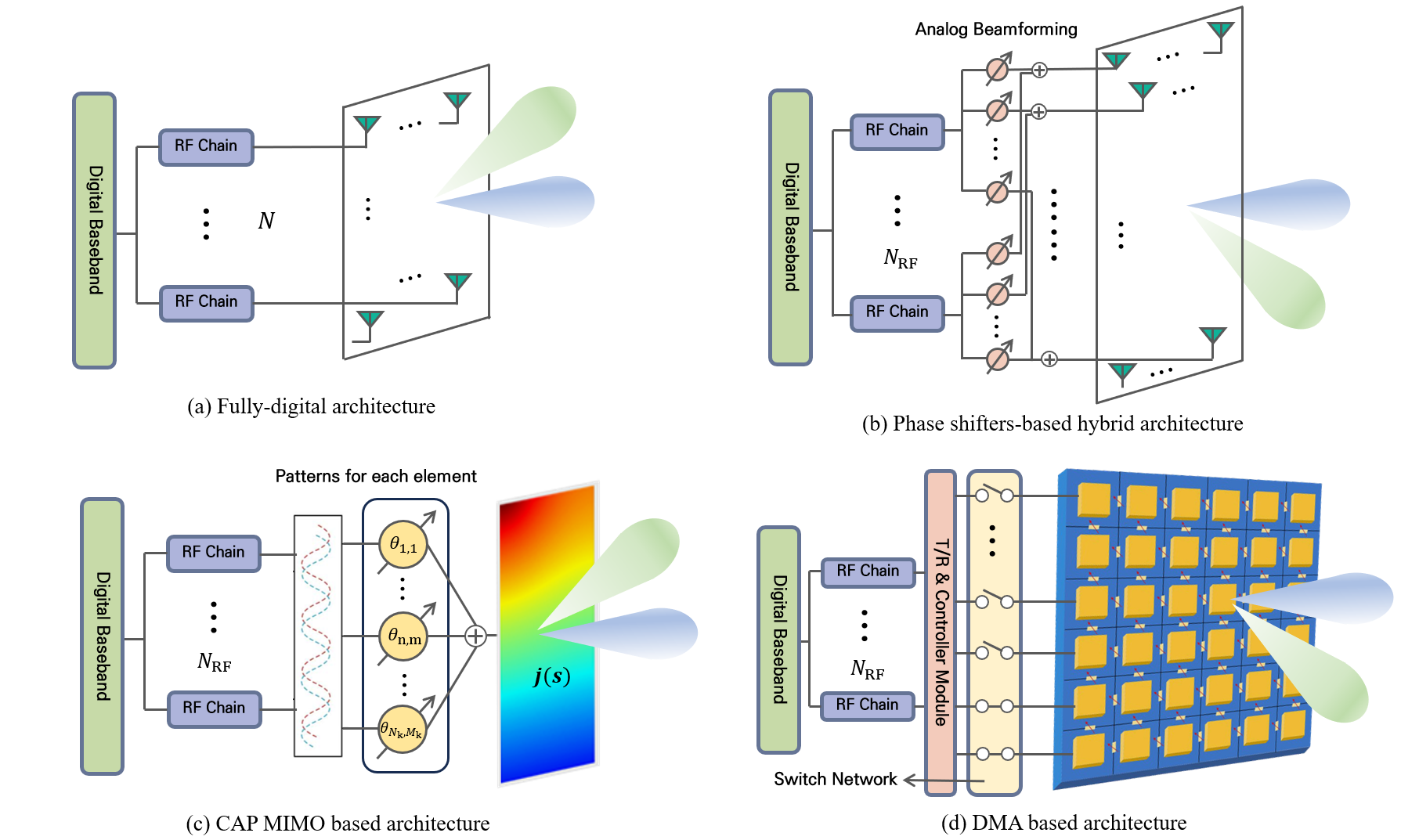}}
	\vspace{-2pt}
	\caption{The four classes of UM-MIMO antenna array architectures. a) Fully-digital architecture, b) Hybrid architecture, c) CAP MIMO-based architecture, d) DMA-based architecture.}
	\label{fig:sec8:21} 
\end{figure*}

The beamfocusing and massive spatial multiplexing characteristics of UM-MIMO systems were discussed in Section~\ref{sec:near-field}, but the ability to control them depends on the signal processing capabilities of the antenna array~\cite{e}. 
The achievable SE varies between hardware architectures that have the same form factor.
While half-wavelength antenna spacing is the norm in conventional Massive MIMO systems, a key feature of UM-MIMO is placing an extremely large number of antennas in a small aperture area, using spacings much smaller than $\lambda/2$. 
Although this feature does not inherently increase spatial DoF, this approach allows for several candidate architecture designs for spatial efficiency and new functionalities to meet the extreme requirements of 6G. These designs range from thousands of discrete antenna elements to methods using metasurfaces, and even approaches that make apertures effectively continuous.

There are many implementation challenges related UM-MIMO arrays. Firstly, large and dense array structures introduce issues related to mutual coupling, polarization discrimination, channel estimation, and near-field propagation, which have been touched upon earlier in this paper.
Moreover, the large number of antennas and reduced spacing, coupled with novel signal processing techniques, can significantly increase hardware costs, energy consumption, and complexity in the production of UM-MIMO systems.
The efficient realization of such high-density arrays has spurred the development of advanced antenna and device designs for 6G, including using metamaterials for antenna design or developing flexible fluid antenna systems. This section explains key antenna array technologies for 6G UM-MIMO systems, discussing how these technologies can be designed, their implementation methods, and potential application areas.

\subsection{Uniform array-based radiation architecture}

One might implement a UM-MIMO array by following the conventional approach with a UPA with discrete elements as shown in Fig.~\ref{fig:sec8:21}(a). While conventional Massive MIMO at the BSs typically involves around 64 antennas, UM-MIMO is envisioned to include many more antennas in the array~\cite{MyersHeathInfocus2022}. This is made possible by making the array aperture larger and possibly by reducing the spacing between antennas. In~\cite{g}, a BS equipped with 40\,000 antennas was proposed, and \cite{h} investigated various antenna spacing values, such as $\lambda/6$ and $\lambda/15$. Additionally, structures composed of thousands of patch antennas have been researched since 5G, including how the antenna size affects the EM modeling in large arrays~\cite{i}.

Massive MIMO for the $3.5$\,GHz band has been implemented in 5G using the fully digital beamforming architecture shown in Fig.~\ref{fig:sec8:21}(a).
In this structure, each antenna is connected to a dedicated RF chain, allowing the transceivers to generate any desirable superposition of near-field and far-field beams, thus offering high spatial flexibility~\cite{j}. This is the ideal implementation from an SE perspective, but not from a complexity viewpoint.
A fully digital UM-MIMO architecture requires the use of a large number of RF chains, leading to substantially higher design complexity, energy consumption, and cost.
These issues might grow faster than linearly with the number of antennas; for example, the computational complexity for interference-suppression beamforming grows cubically with the number of antennas \cite{massivemimobook}.
In principle, a fully analog architecture could be utilized that has a single RF chain that connects to the antennas through phase-shifters, but this implementation lacks the ability of spatial multiplexing---the main motivation behind UM-MIMO.
Consequently, intermediate methods have been proposed that divide large arrays into several analog sub-arrays~\cite{k}. This is specifically implemented through hybrid processing, which combines the analog and digital architectures and uses fewer RF chains than the number of antennas, realized through the connection of phase shifters and viewed as a promising method to reduce hardware complexity~\cite{l, l2}. This architecture is illustrated in Fig.~\ref{fig:sec8:21}(b).
These implementation simplifications can have little impact on the SE when there are more RF chains than strong multipath clusters in the propagation environment, which is particularly the case in line-of-sight scenarios.

\begin{figure}[!t]
	\centering{\includegraphics[width=\columnwidth,keepaspectratio]{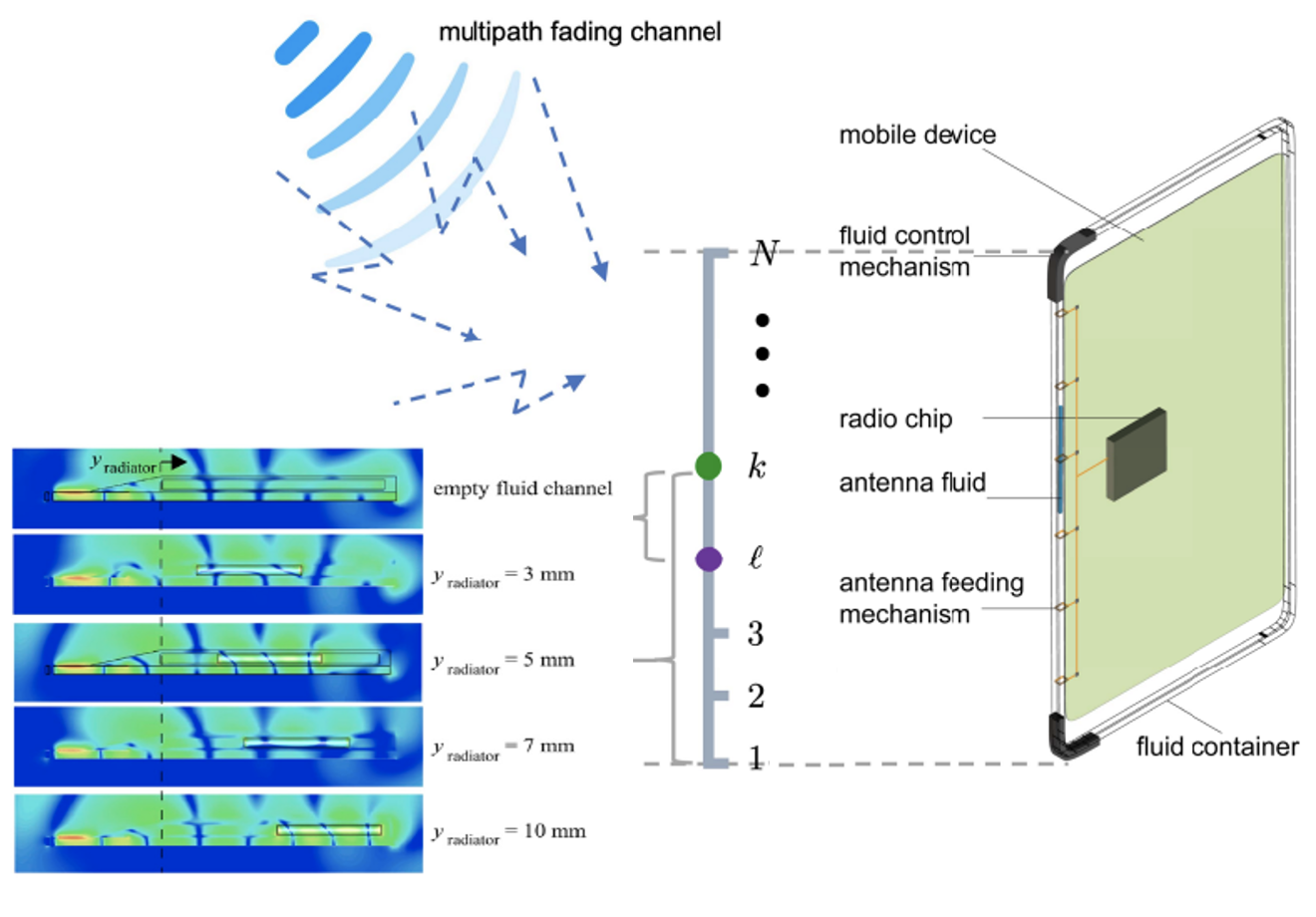}}
	\vspace{-5pt}
	\caption{Fluid antenna systems and $E$-field distribution of the surface wave when the fluid element changes its location\cite{wong}.}
	\label{fig:fluid} 
\end{figure}

\subsection{Continuous aperture MIMO system}

To achieve full control over the array aperture, continuous-aperture MIMO (CAP MIMO) architectures are gaining attention. CAP MIMO is a novel implementation that modulates information directly in the form of EM waves through a continuous antenna surface, ideally possessing a spatially continuous EM volume composed of an infinite number of infinitesimally small antennas as shown in Fig.~\ref{fig:sec8:21}(c). This is also being researched under the name of holographic MIMO~\cite{n, Pizzo_Marzetta2020} and differs from traditional discrete antenna arrays by effectively using a continuous antenna aperture, thereby offering transmission efficiency and application flexibility. Although the spatial DoF is the same as for a discrete array of the same size, channel properties such as the singular values of the channel matrix can be improved within those dimensions.

CAP-based UM-MIMO systems have the capability to generate and control all current distributions on a spatially continuous surface, and can directly modulate artificially constructed EM waves to radiate into space. This can also be seen as a specific case of point-antenna-based UPA UM-MIMO from a mathematical perspective, by considering the limit where the antenna element sizes approach zero.
This can be approximately realized through the development of metamaterials and highly flexible reconfigurable antennas, fully leveraging the physical properties of space propagation to achieve high SE and energy efficiency.
In~\cite{m}, an antenna structure with a circular CAP plane of 10\,m radius was introduced. The work in \cite{o} presented a lens antenna with a spatially continuous aperture, where hundreds of antenna elements were placed along the focal arc of an EM lens with a radius of 5\,m.

The signal processing for discrete MIMO arrays mainly consists of linear algebra operations, but these turn into continuous operations when using CAP MIMO. Designing continuous beampattern functions for CAP MIMO is typically a non-convex problem, and it is nontrivial to make use of conventional MIMO methods. \cite{p} introduced a pattern division multiplexing (PDM) technique that addresses these challenges by transforming the design of continuous pattern functions into a finite orthogonal basis-based projection length design. Subsequently, the EM performance comparison between CAP MIMO and conventional discrete MIMO systems was demonstrated using a non-asymptotic approach~\cite{q}. This technique provides an efficient means to simultaneously serve multiple UEs and demonstrates the potential of CAP MIMO systems in meeting the diverse performance requirements of future networks.

\subsection{Fluid antenna system}

Traditional array designs consist of highly conductive and static metal elements at fixed locations, whose join radiation pattern is controlled by modifying the current distribution over the elements.
A fundamentally different approach is fluid antenna systems (FAS) that can dynamically change shapes and positions of the radiating elements~\cite{fas_cl1, fas_cl2,fas_cl3}. FAS represents an antenna design capable of reconfiguring characteristics such as shape, position, polarization, and radiation pattern, based on controllable conductive or dielectric elements. The flexibility and benefits over traditional solid materials have led to the emergence of numerous fluid antennas in recent years~\cite{r, r2, r3}.
As seen in Fig.~\ref{fig:fluid}, the fluid element can move the antenna to one of $N$ fixed positions (referred to as ports) within a predefined space. At any given moment, the FAS can explore the spatial fading characteristics by moving to a spatial location with better signal quality to avoid deep fades. To accomplish this, the FAS must have a fine spatial resolution, allowing the fluid to optimize for each port, thus requiring a large $N$ value. Although the spatial footprint of FAS is small, $N$ can be very large, allowing for diversity across numerous spatially correlated ports.
From a communication perspective, FAS behaves as an antenna selection system, and can thus achieve diversity gains that can be comparable to those obtained with maximum ratio processing \cite{wong2, wong3}.

\begin{figure}[!t]
	\centering{\includegraphics[width=.8\columnwidth,keepaspectratio]{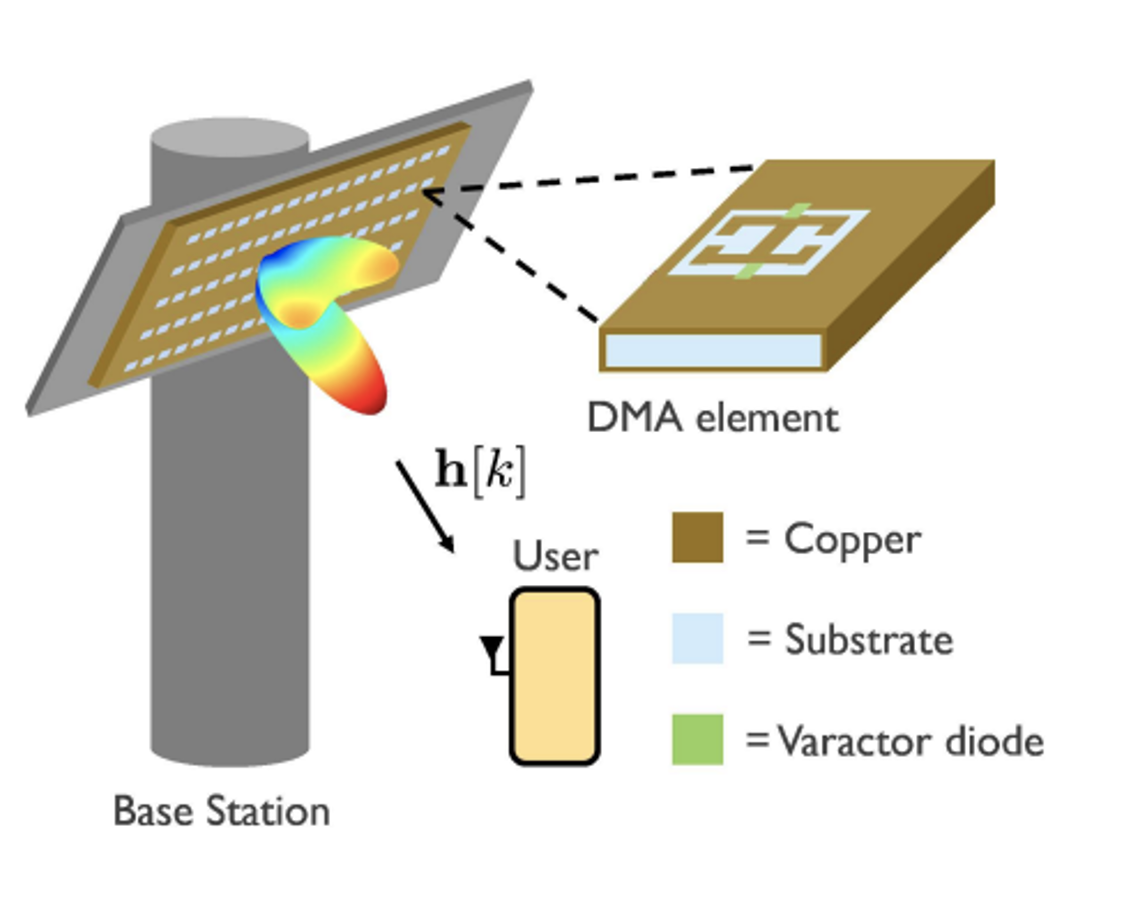}}
	\vspace{-5pt}
\caption{Various metasurface antenna designs. A DMA-based system that uses metasurface-based antenna arrays~\cite{y}.}
	\label{fig:meta} 
\end{figure}

To achieve multiplexing gains, the FAS can make use of multiple fluid antenna elements (each with $N$ different ports), which can move independently within a specific non-overlapping space.
The number of elements must be equal to or larger than the desired number of layers. The joint optimization of the elements can be utilized to design a MIMO channel matrix with advantageous properties for a multiplexing perspective. 
The concept of reconfigurable antennas has been considered in the communication literature, which contains algorithms for performance optimization \cite{BoeBer:Performance-Study-of-Pattern:08,PiazzaEtAlDesignAndEvaluationReconfigurable2008,SayRag:Maximizing-MIMO-Capacity:07,HasBahCet:Downlink-Multi-User-MIMO:18}.
FAS is a potent technology for realizing these visions.

From an implementation perspective, FAS can be fabricated using liquid metals like eutectic gallium-indium~(eGaIn). The movement of the fluid can be electrically controlled using methods like electrowetting~(EW) or RF MEMS, as used in displays, and must have a device structure optimized for the desired radiation pattern~\cite{EW, EW2}. One of the major bottlenecks of FAS is response time, as the time it takes for the fluid to move to the desired position significantly impacts communication performance. While devices with sufficient response speed have not yet been presented, future improvements are expected through creative geometric methods. Additionally, some fluids used in FAS may cause undesirable chemical side effects, which should be noted. While FAS offers significant efficiency in terms of space and cost, stability issues must also be considered.

\subsection{Metasurface-inspired antenna system}

The implementation of UM-MIMO using hybrid architectures might lead to lower complexity and energy consumption than a fully digital architecture but still requires plenty of analog circuits due to the addition of numerous phase shifters. In this context, it has been recognized that metasurfaces, capable of effectively controlling EM properties, can replace traditional analog beamforming structures~\cite{s}. Metasurfaces, comprising dense arrays of reconfigurable elements smaller than the wavelength, can precisely and dynamically manipulate EM waves, actively participating in improving signal transmission and reducing interference. Research analyzing the Shannon entropy of digitally-coded metasurface communication systems has also been presented~\cite{ref}.

Metasurfaces have been considered in the EM literature for decades, and the development can be divided into generations. Metasurface~1.0 was based on homogeneous periodic structures, Metasurface~2.0 allows spatial modulation with variable amplitude and phase. Further advancements have led to Metasurface~3.0, capable of real-time modulation in both time and space and featuring electrically large arrays that adapt to environmental changes~\cite{t}. 

Within the latest generation, the term dynamic metasurface antennas~(DMA) has emerged as a promising way of implementing UM-MIMO arrays. This architecture is illustrated in Fig.~\ref{fig:sec8:21}(d). Comprising densely packed arrays of reconfigurable metadevices, DMAs utilize the tuning characteristics of each unit cell to provide analog signal processing functions without dedicated analog circuits as shown in Fig.~\ref{fig:meta}. By controlling the amplitude and phase of the DMA elements, they enable low-power precoding structure that does not require phase shifters~\cite{u}.

Furthermore, by exploiting the properties of metadevices to create an effective medium at much smaller scales than the wavelength, DMAs can design densely packed antenna arrays with spacings smaller than half a wavelength, enhancing beamfocusing with higher efficiency. A major challenge in implementing DMA-based communication is the Lorentzian-limited beamforming weights in the analog precoding process. Studies have been conducted to optimally adjust these limited weights~\cite{v, v2}, with \cite{x} demonstrating that the proposed approach can reduce the number of RF chains while achieving most of the beamforming gain.
The paper~\cite{w} mathematically analyzed how large-scale MIMO systems using fully digital, hybrid, and DMA architectures impact beamforming capabilities in the radiative near-field. Additionally, \cite{y} integrated DFT codebooks with DMAs in low-power MIMO systems, showing that transmitter systems using developed DMAs can surpass conventional analog beamforming systems in terms of SE and energy efficiency. An energy-efficiency analysis in \cite{CastellanosEtAlEnergyEfficientTriHybrid} also demonstrates the benefits of incorporating DMAs into traditional hybrid precoding architectures. These findings suggest that DMAs, capable of being packaged in a small physical area for a wide operating frequency band, hold significant potential as a design that can surpass conventional beamforming architectures.


\section{Interplay Between Array Design and Signal Processing}
\label{sec:array_implementation}

The last section described four general categories of array architectures, which are summarized in Fig.~\ref{fig:sec8:21}.
Architectures (a) and (b), fully digital and hybrid analog/digital, have a long history in the wireless infrastructure industry. On the other hand, (c) and (d), namely CAP and DMA, represent more modern approaches, and only time will reveal whether they will be implemented in BSs during the 6G life-cycle. 

We foresee that the first 6G systems will adopt fully digital and hybrid architectures, but such UM-MIMO implementations will face many practical challenges.  Improvements in hardware technology have without doubt made it possible to equip a BS with around 64 RF chains \cite{Vieria14}, but commercial BSs typically have 2-4 times more antenna elements than RF chains \cite{asplund2020advanced}.\footnote{Each RF chain is connected to a subarray with multiple antenna elements, to achieve a stronger vertical directivity than with a single element. Since no phase shifters are utilized, this implementation is still called fully digital.} 
Viewed from the perspective of the enormous number of antenna elements envisioned for UM-MIMO, it is likely that the number of RF chains will never catch up with the desired number of antenna elements at a BS. If so, a fully digital architecture with one element per RF chain will essentially be relegated to benchmark status only, and hybrid architectures are the practical ones to consider. However, a hybrid architecture is not limited to using analog phase shifters; rather, with the term \textit{hybrid} we refer to an architecture that somehow reduces the signal dimension between the antenna elements and the baseband processor.

It is essential to use the terms \textit{fully digital} and \textit{hybrid} judiciously. As discussed in Section \ref{sec:spat_deg_freedom}, the available spatial DoF, denoted as $\eta_{\rm 2D}$ in \eqref{eq:sampling_theorem_space_LHLV}, might be smaller than the number of deployed antenna elements. Consequently, a hybrid architecture can be lossless in terms of communication performance only if the number of RF chains is greater or equal to $\eta_{\rm 2D}$. Therefore, when referencing a fully digital architecture, it may implicitly refer to a hybrid one with $\eta_{\rm 2D}$ RF chains.

To identify an immediate issue with having $\eta_{\rm 2D}$ RF chains, consider a UM-MIMO array implemented on the jumbotron in a sports venue. Suppose the antenna elements cover an area $A$ of the jumbotron and the carrier frequency is $f_\mathrm{c}$.
It follows from \eqref{eq:sampling_theorem_space_LHLV} that $\eta_{\rm 2D}=A\frac{f_\mathrm{c}^2 \pi}{c^2}$.
For a bandwidth of $B$\,Hz, this produces, at the very minimum, $\eta_{\rm 2D}B$ samples per second as output from the UM-MIMO array. If we represent each sample with $b$ bits and forward all this information to a baseband unit (BBU), the rate at the input to the BBU is
\begin{equation}
    R_\mathrm{BBU}=ABb\frac{f_\mathrm{c}^2 \pi}{c^2}\,\, \textrm{bit/s}.
\end{equation}
To appreciate the scale of this rate, we consider the case of $B=100$\,MHz, $f_\mathrm{c}=3$\,GHz, $b=16$, and $A=10\,\text{m}^2$. This results in $R_\mathrm{BBU}\approx 5$\,Tbit/s. At mmWave frequencies with $10$ times more spectrum, the rate increases to $R_\mathrm{BBU}\approx 5000$ Tbit/s. These rates are simply not feasible to implement in the foreseeable future; hence, the design of high-performing but lossy hybrid architectures constitutes a promising research direction. 

The remainder of this section will describe ways to co-design antenna arrays with signal processing algorithms.
We will describe a modular design of hybrid UM-MIMO antennas and elaborate on the tradeoff between the number of BBU inputs and the processing at the antenna elements.

\subsection{Modular UM-MIMO design} \label{subsec:Rusekmodular}

We will now take a closer look at a UM-MIMO array and focus on its backplane; that is, the circuitry behind the antenna elements. Abstractly, we can view the UM-MIMO array as having several outputs, each connected to the BBU. When using a hybrid architecture, these outputs correspond to the RF chains. In a UM-MIMO setup, it is possible to define the density of these RF chains as $\mu \,\text{m}^{-2}$. Additionally, each RF chain may be linked only to a subset of the antennas within the array. We can then define an \emph{antenna panel} as a group of antennas, comprising an area $A_\mathrm{p}\,\mathrm{m^2}$ to which a given number $M_\mathrm{p}$ of RF chains are connected; clearly, $M_\mathrm{p}=A_\mathrm{p}\mu$. Panel-based implementations of UM-MIMO antennas are thoroughly discussed  in \cite{Callebaut23}.

\begin{figure}[!t]
	\centering
	\includegraphics[width=3in]{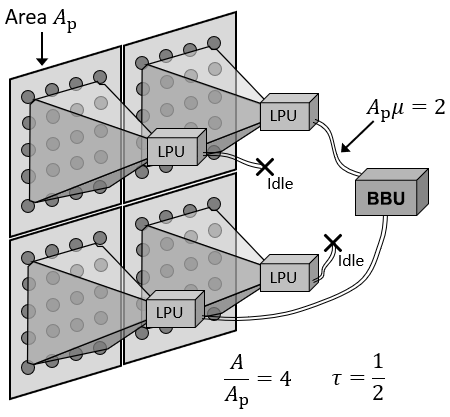}
	\caption{Example of a modular antenna design where only some panels are connected to the CPU at a given time instance.}
	\label{Rusek:fig1}
\end{figure}

For a given rate limitation at the input interface to the BBU, only a subset of the panels may forward their outputs to the BBU at a given time instance while the remaining are idle. If we let $\tau$ denote the fraction of the total antenna area, $A$, that is active, the total number of RF chains connected to the BBU becomes
\begin{equation} \label{eq:Rusek:1}
M = \frac{\tau A}{A_\mathrm{p}} M_\mathrm{p} = 
\frac{\tau A}{A_\mathrm{p}} A_\mathrm{p}\mu=A\tau\mu.
\end{equation}
A basic example is provided in Fig.~\ref{Rusek:fig1}, where four panels, each of area $A_\mathrm{p}$ are equipped with $M_\mathrm{p}=2$ RF chains. We consider $\tau=0.5$, which implies that at most $\tau A/A_\mathrm{p}=2$ panels may be active (i.e., forward their outputs to the BBU) at any given time. The total number of RF chains connected to the BBU at a given time becomes $M=M_\mathrm{p}\tau A/A_\mathrm{p}=4$. The connection between each panel and the BBU is digital and managed by a local processing unit (LPU); thus, there are no physical switches in the array but only a control mechanism that determines which LPUs forward signals to the BBU.

The preferable selection of the parameters $\tau$ and $\mu$ depends on the deployment scenario.
Choosing a very large $\mu$ (i.e., letting each panel have a large number of RF chains), implies that $\tau$ must be reduced to satisfy a given rate constraint on the BBU input. This may be favorable in case the UEs are believed to appear in hotspots (i.e., small subsets of the coverage area) so that they all have good propagation characteristics to the same panels. However, if the UEs are fairly uniformly distributed over the coverage area, then a small $\tau$ (i.e., only a few panels are active), may not be a good choice.

Ultimately, striking a good balance between $\tau$ and $\mu$ is not a problem that has an analytical solution, but rather requires extensive system simulations---possibly considering a digital twin of the intended deployment area. Let the metric(s) of interest be denoted by $\mathbf{c}$, and possibly be vector-valued. This may be the ergodic channel capacity, outage capacity, or any other metric of choice. What quantities/settings do $\mathbf{c}$ depend on? First of all, it strongly depends on the propagation environment and user distribution, which is why a numerical approach is required. Secondly, it depends on various fixed parameters such as the carrier frequency, bandwidth, available deployment area $A$ of the array, etc. Thirdly, it depends on the type of signal processing performed at the LPU. This can range from purely analog beamforming (e.g., using phase-shifters and signal combiners) to a fully digital LPU that can refine/compress the complex baseband samples.
Lastly, it depends on the quantities $\tau$, $\mu$, and $A_\mathrm{p}$; therefore, we express the metric as $\mathbf{c}(\tau,\mu,A_\mathrm{p})$ with the remark that all other properties discussed are held constant. Despite the fact that $A_\mathrm{p}$ does not show up in \eqref{eq:Rusek:1}, it does impact the overall performance as it impacts the quality of the LPU outputs. 

Rule-of-thumbs for selecting $\tau$ and $\mu$ in a line-of-sight-dominant propagation environment were obtained in \cite{Pereira22} by extensive simulations. These show a linear relation between the number of BBU inputs and the number of UEs and active panels. Further work is required to determine the generality of these results.

The signal processing that is performed at each LPU to limit the flow of data to the BBU can also be optimized; for example, to reduce the dimensionality while retaining most of the performance. This problem is studied in \cite{Vidal21,Vidal23} with a star-topology between the BBU and LPU, while sequential topologies were considered in \cite{Shaik2021a}.

\section{Final Remarks and Open Challenges}
\label{sec:conclusions}

Many new technology components are envisioned for 6G networks, but one that surely will play a key role is MIMO. 
An educated guess is that we will first see 6G MIMO in the upper mid-band (from 6-15 GHz), where 512 to 2048 antennas per array are within practical reach.
The success of the technology is tightly connected to scientific and hardware maturity, and many grand challenges remain to be tackled. We will end this paper by describing eight such challenges:

\begin{enumerate}

\item \textbf{Beamfocusing in realistic environments:} We showed how near-field effects improve the spatial multiplexing capabilities in line-of-sight scenarios. How prominent are these effects in more realistic environments?

\item \textbf{Fixed-complexity BBU processing:} The interface between antennas and BBU will be limited in practice, as well as the computational capabilities. How can LPUs reduce the signal dimensionality to harness the benefits of having many antennas with a fixed-complexity BBU?

\item \textbf{Energy-efficient operation:} The energy consumption of UM-MIMO will likely be higher than that of 5G MIMO, but the energy per bit can be substantially smaller when massive multiplexing is performed. However, intelligent sleep features are required to achieve energy efficiency when the traffic load is small (e.g., at night).
    
    \item \textbf{Channel estimation:} Channel coefficients must be estimated in every coherence block. Does this impose a fundamental limit on the useful spatial DoF? How does the limit depend on the available side-information in the estimator?
    
    \item \textbf{Array topology:} How to optimize the antenna number and spacing for a given area by considering mutual coupling and polarization effects? How to incorporate antenna reconfigurability and different kinds of array architectures? 

    \item \textbf{Distortion-aware processing:} Physical-consistent hardware and noise modeling, as well as RF impairments, make the end-to-end system model different than in textbooks. Can these ``distortions'' be overcome through digital processing?

    \item \textbf{Continuous processing:} The CAP and DMA architectures enable nearly continuous spatial processing over the aperture, but under specific phase and amplitude constraints. How can advanced spatial multiplexing methods (e.g., zero-forcing) be implemented in these cases, where the matrix operations have infinite dimensions?

    \item \textbf{Field trials:} Although near-field effects and their consequences can be simulated using EM-compliant models, true confidence in the technology is obtained through testbeds and measurements. UM-MIMO pushes the limits in terms of the cost and size of such field trials. Antenna designers and communication engineers must come together to overcome these challenges.
    
\end{enumerate}

\bibliographystyle{IEEEtran}

\bibliography{IEEEabrv,refs,refs_AM,refs_rc,refs_ant}

\end{document}